\documentclass[useAMS,usenatbib]{mn2e}

\usepackage{psfig, epsf, epsfig}

\title[Dust-regulated galaxy evolution]{Coevolution of dust, gas, and stars in galaxies - I.
Spatial distributions and scaling-relations of dust and molecular hydrogen}
\author[K. Bekki]
{Kenji Bekki${}^1$\thanks{E-mail:
bekki@cyllene.uwa.edu.au} \\
${}^1$ICRAR M468
The University of Western Australia
35 Stirling Hwy, Crawley
Western Australia 6009, Australia}

\begin{document}

\date{Accepted, Received 2005 February 20; in original form }

\pagerange{\pageref{firstpage}--\pageref{lastpage}} \pubyear{2005}

\maketitle

\label{firstpage}

\begin{abstract}

We investigate the  time evolution of dust properties,  molecular hydrogen 
(${\rm H}_2$) contents,
and star formation histories in galaxies
by using our original chemodynamical simulations.
The simulations include the formation of dust in the stellar winds of supernovae (SNe)
and asymptotic giant branch (AGB) stars, the growth and destruction processes
of dust in the interstellar 
medium (ISM),  the formation of polycyclic aromatic hydrocarbon (PAH) dust
in carbon-rich AGB stars, 
the ${\rm H_2}$ formation on dust grains,
and the ${\rm H_2}$ photo-dissociation due to far ultra-violet (FUV)
light  in a self-consistent manner. We focus mainly  on disk galaxies
with the total masses ranging from $10^{10} {\rm M}_{\odot}$ and $10^{12} {\rm M}_{\odot}$
in this preliminary study. 
The principle results are as follows: The  star formation histories of  disk galaxies
can be regulated  by the time evolution of interstellar dust,  mainly because the
formation rates of ${\rm H}_2$ can be controlled by dust properties.
The observed correlation between dust-to-gas-ratios ($D$) and gas-phase oxygen abundances
($A_{\rm O} \equiv 12+\log {\rm (O/H)}$) can be reproduced reasonably well
in the present models.
The disks show negative radial gradients (i.e., larger in inner regions)
of ${\rm H_2}$ fraction ($f_{\rm H_2}$), PAH-to-dust mass ratio ($f_{\rm PAH}$),
$D$, and $A_{\rm O}$ and these gradients evolve with time. 
The surface-mass densities of dust ($\Sigma_{\rm dust}$) are  correlated more strongly with
the total surface gas densities ($\Sigma_{\rm gas}$) than with those of ${\rm H_2}$
($\Sigma_{\rm H_2}$).
Local gaseous regions with higher $D$ are more likely to have higher $f_{\rm H_2}$
in individual disks  and total ${\rm H_2}$  masses ($M_{\rm H_2}$) correlate well
with total dust masses ($M_{\rm dust}$).
More massive disk galaxies are more likely to have higher $D$, $f_{\rm PAH}$, and $f_{\rm H_2}$
and smaller dust-to-stellar mass ratios ($R_{\rm dust}=M_{\rm dust}/M_{\rm star}$).
Early-type E/S0 galaxies formed by major galaxy merging  can have lower $R_{\rm dust}$
than isolated late-type disk galaxies.
 We also compare between galactic star  formation histories in 
the metallicity-dependent and dust-dependent star formation models and
find no major differences.
Based on these results, we discuss the roles of dust in chemical and dynamical evolution
of galaxies. 
\end{abstract}


\section{Introduction}

One of the many  important roles of interstellar dust in galaxies is the formation of molecular hydrogen
(${\rm H}_2$) on its surface (e.g., Gould \& Salpeter 1963; Hollenbach \& Salpeter 1971;
Cazaux \&  Tielens 2002). Giant molecular clouds (GMCs) composed of ${\rm H}_2$ 
are the major formation sites of stars in galaxies (e.g., Blitz 1993; Fukui \& Kawamura 2010).
Global star formation rates of galaxies are observed to be well correlated with
surface mass densities of interstellar gas 
(e.g., Schmidt 1959; Kennicutt 1998). Dust can originate from
stellar winds of AGB stars (e.g., Ferrarotti \& Gail 2006; Zhukovska et al. 2008)
and supernovae (e.g., Kozasa et al. 1991; Nozawa et al. 2003),
 and therefore the production rate of dust in a galaxy can
be determined by the star formation history that controls the formation rates
of AGB stars and supernovae. 
Thus,  the evolution processes of dust,  gas, and stars are mutually related,
and detailed investigation of these  coevolution processes 
can lead us to the better understanding of galaxy formation and evolution.

Recent observational studies by infrared space telescopes have  revealed
physical properties of dust,  their spatial distributions, and their correlations
with their host galaxy properties 
in nearby and distant galaxies (e.g., Draine et al. 2007; Meixner et al. 2010;
Roman-Duval et al. 2010;  Takagi et al. 2010;
Dunne et al. 2011; Kaneda et al. 2011; Cortese et al. 2012; 
Skibba et al. 2012; Smith et al. 2012).
For example, Draine et al. (2007) have investigated the total dust mass, the mass fraction
of the dust contributed by PAHs, and the correlation between gas-phase oxygen
abundance ($A_{\rm O}$=12 +log(O/H)) and the dust-to-gas ratio ($D$) for 65 nearby galaxies 
and found that the average PAH fraction in galaxies with $A_{\rm O}>8.1$ is 3.55\%.
Smith et al. (2012) have found a significant difference in the dust-to-stellar-mass
ratio between late-type spirals and S0s and thus provided a new clue to the 
origin of S0s.  Meixner et al. (2010) and Skibba et al. (2012) have derived
the detailed 2D maps of $D$, dust temperature, and PAH fraction in the Large Magellanic
Cloud (LMC).

Recent multi-wavelength observational studies of nearby galaxies
have investigated physical correlations between surface densities of
star formation ($\Sigma_{\rm SFR}$), neutral hydrogen (H~{\sc i};
$\Sigma_{\rm HI}$), and ${\rm H}_2$ ($\Sigma_{\rm H_2}$)
and discussed key parameters that control global  star formation
in galaxies (e.g., Bigiel et al. 2008; Leroy et al. 2008). 
These observation  have also provided data sets for radial gradients of
H~{\sc i} and ${\rm H}_2$ gas, which are quite useful for discussing
what can determine the H~{\sc i} and ${\rm H}_2$ gas mass fractions in galaxies.
High-resolution observational studies of GMCs for galaxies in the Local Group
(e.g., LMC and M33)
have provided vital clues to how individual GMCs composed
of ${\rm H}_2$  form from local H~{\sc i}
and how star formation proceeds within GMCs (e.g., Rosolowsky et al. 2003;
Kawamura et al. 2009).
A number of recent observational studies have investigated correlations between
properties of dust (e.g., $D$ and $\Sigma_{\rm dust}$)
and gas (e.g., chemical abundances and 
$\Sigma_{\rm H_2}$)  in galaxies (e.g., Leroy et al. 2011).

These observational results 
have raised the following three key questions  on the origin
of dust and gas in galaxies.
The first is  what physical processes are
responsible for the observed spatial variation of dust  properties  within galaxies
(e.g., radial variation of $D$ and PAH-dust mass fraction;  Meixner et al. 2010).
The majority of previous theoretical studies of dust formation and evolution in galaxies
are based on one-zone (or multi-zone) chemical evolution models 
(e.g., Dwek 1998, D98; Lisenfeld \& Ferrara 1998; Hirashita 1999; Edmunds 2001;
Calura et al. 2008; Piovan et al. 2011).
Although such previous models provided theoretical explanations
for a number of key observational results on dust properties in galaxies
(e.g., $A_{\rm O}-D$ relation), they did not allow astronomers to discuss
the spatial  distributions of dust properties and their correlations with other galaxy
properties (e.g., Hubble morphological types).
Therefore, it is largely unclear in these previous studies what physical mechanisms
are responsible for the observed spatial distributions of dust
properties in galaxies.

The second question pertains to the determinant factors for
${\rm H}_2$ properties and their correlations with galaxy properties (e.g., luminosities
and metallicities). 
Previously,  Elmegreen (1993) theoretically
discussed the importance of the pressure and radiation field
of the interstellar medium in the transition from H to ${\rm H}_2$  in galaxies.
Blitz \& Rosolowsky (2004) also investigated the importance of gas pressure
in ${\rm H - H_2}$ transition based on observational data from
nearby 28 galaxies.
Recent numerical simulations  have incorporated the ${\rm H}_2$ formation process
from H~{\sc i} 
(e.g., Pelupessy et al. 2006, P06;  Robertson \& Kravtsov 2008, RK08;
Gnedin et al. 2009; Christensen et al. 2012),
 which means that theoretical studies will soon contribute
to the solution of the above problem.
The ${\rm H}_2$ formation model dependent only on gaseous metallicities and
densities  proposed by 
Krumholz, McKee \& Tumlinson (2009, KMT09)
is practically  
useful for investigating ${\rm H}_2$ formation and star formation from ${\rm H}_2$
in galaxies
and thus used in recent semi-analytic models
(e.g., Fu et al. 2010; Lagos et al. 2012) and numerical simulations (Kuhlen et al. 2012, K12).
 However, these recent  ${\rm H}_2$ formation models assume that dust abundances
are linearly proportional to metallicities (i.e., dust-to-metal ratio, $D_{\rm z}$, 
is constant), which is inconsistent with 
observations
(e.g., Galamex et al. 2011) which show a large dispersion in $D$
for a given $A_{\rm O}$ (i.e., significantly different $D_{\rm z}$).
Furthermore, such an assumption of constant $D_{\rm z}$ is incompatible with
dust evolution models that predict
significant $D_{\rm z}$ evolution with time in galaxies
(e.g., Inoue 2003; Calura et al. 2008).
Thus we need to improve ${\rm H}_2$ formation models by incorporating 
dust formation and evolution in order to 
discuss the second question in a more quantitative manner.

\begin{table*}
\centering
\begin{minipage}{175mm}
\caption{A summary for recent observations to be compared with
the present simulations.}
\begin{tabular}{lll}
{ Properties }
& {  Physical meanings }
& { References \footnote{ `Prediction' means that only predicted properties
are presented: observational results are yet to be obtained. Only one
representative reference is given. A full list of relevant papers is given
in the main text.}} \\
$A_{\rm O} -D$  & Correlations between gas-phase oxygen abundance 
($A_{\rm O}$) and dust-to-gas ratios ($D$)   &  Galametz et al. (2011) \\
$A_{\rm O} -f_{\rm PAH}$  & Correlations between  
$A_{\rm O}$ and PAH-to-dust ratios ($f_{\rm PAH}$)   &  Draine et al.  (2007) \\
$f_{\rm H_2}-D$  & The dependences of  $f_{\rm H_2}$ on $D$ in local ISM
& Prediction \\ 
$\Sigma_{\rm dust}-\Sigma_{\rm gas}$  &  Correlations between local
surface densities of dust ($\Sigma_{\rm dust}$) and those of total gas
($\Sigma_{\rm gas}$) & Leroy et al. (2011) \\
$\Sigma_{\rm dust}-\Sigma_{\rm H_2}$  &  Correlations between local
surface densities of dust ($\Sigma_{\rm dust}$) and those of ${\rm H_2}$
($\Sigma_{\rm gas}$) & Leroy et al. (2011) \\
$dD/dR$  &  Radial gradients of $D$ 
&  Pappalardo et al. (2012) \\
$df_{\rm PAH}/dR$  &  Radial gradients of $f_{\rm PAH}$ 
& Meixner et al. (2010) \\
$M_{\rm dust}-M_{\rm gas}$  &  Correlations between total dust masses
($M_{\rm dust}$) and total gas masses  ($M_{\rm gas}$)
& Corbelli et al. (2012) \\
$R_{\rm dust}-M_{\rm star}$  &  Correlations between dust-to-star mass
ratios ($R_{\rm dust}$) and total stellar masses
($M_{\rm star}$)
& Cortese et al. (2012) \\
$R_{\rm dust}-\Sigma_{\rm star}$  &  The dependences of local
$R_{\rm dust}$ on local stellar surface densities ($\Sigma_{\rm star}$)
& Prediction \\
\end{tabular}
\end{minipage}
\end{table*}

The third question is on the origin of the observed correlations between
dust and galaxy properties (e.g., dust-to-stellar mass ratio along the Hubble
morphological types; Cortese et al. 2012)
and  between dust and gas properties (e.g.,  gas surface densities dependent on dust surface
densities;  Leroy et al. 2011).
In order to discuss this
third question in a quantitative manner,   theoretical studies need to 
model both (i) the time evolution of dust/gas contents and star formation rates
in galaxies and (ii) the dynamical evolution 
of galaxies in a self-consistent manner.
Only a few previous models included the formation and destruction
of dust and ${\rm H}_2$ formation on dust grains
in a self-consistent manner and thereby discussed
the time evolution of dust properties (e.g., Hirashita \& Ferrara 2002;
Yamasawa et al. 2011).
Since they did not explicitly include dynamical evolution
of galaxies in their models,   the above-mentioned correlations 
were not addressed at all.

Chemodynamical simulations 
are useful and powerful  tools for investigating  both the
spatial distributions
of stellar and gaseous  abundances
and galaxy scaling relations simultaneously 
(e.g., Theis et al. 1992; Bekki \& Shioya 1999; Kawata 2001;
Revaz \& Jablonka 2012; RJ12)
and thus 
should be also useful for 
theoretical studies on spatial distributions and scaling relations of 
dust and ${\rm H}_2$ properties: we can not discuss the above key
problems related to dust and ${\rm H}_2$ distributions without
performing chemodynamical simulations.
However, even the latest chemodynamical simulations with more sophisticated 
models for the time evolution of variously different chemical abundances
(e.g.,   Rahimi \& Kawata 2012, RK12; Bekki et al. 2012; B12)
did not explicitly include the formation and
evolution of dust. 
Although recent hydrodynamical simulations have incorporated ${\rm H}_2$ formation  on dust
grains (e.g.,P06), they did not incorporate the time evolution of dust properties
in a self-consistent manner.
{\it Thus it is high  time for us to develop a new chemodynamical model 
with dust and ${\rm H}_2$ formation
and thereby to try to answer the above three key questions.}

The three purposes of this paper are as follows. 
Firstly, we describe the details of the new simulation methods by which we can
investigate 3D distributions of dust, ${\rm H}_2$, and stars of galaxies
in a self-consistent manner.
Secondly, we present the preliminary results on the time evolution of dust and ${\rm H}_2$
contents and star formation histories of disk galaxies and their dependences on
the model parameters of the simulations.
Thirdly, we discuss recent observational results for
dust and ${\rm H}_2$ properties
of galaxies and their correlations with their host galaxy properties
based on the present simulation results; several  of these observations to be compared
with the present simulations are summarized in Table 1.
It should be stressed that these observations
of the  spatial distributions of dust and
${\rm H_2}$ and  their correlations with galaxy properties
can be appropriately addressed only by chemodynamical simulations with
dust evolution such
as the present one.
This is the first paper of a series of papers on the coevolution of dust, gas,
and stars in galaxies.
We therefore focus exclusively on the details of the new numerical methods
and the preliminary results.
We discuss 
other important issues such as the evolution of dust composition,  the important
influences of dust on star formation and chemical evolution histories of galaxies,
and the derivation of spectra energy distributions from the results of
chemodynamical simulations in our future papers.

The plan of the paper is as follows: In the next section,
we describe our new chemodynamical model with the formation and evolution
of dust and ${\rm H}_2$.  
In \S 3, we
present the numerical results
on the long-term evolution of physical properties
of dust and ${\rm H}_2$ and their correlations (e.g., $A_{\rm o}-D$ relation)
in disk galaxies.
In this section, we also discuss the dependences of the results on the adopted
model parameters.
In \S 4, we discuss the latest observational results on dust and ${\rm H}_2$
properties of galaxies derived mainly from the $Herschel$ and $Spitzer$ telescopes.
We also compare the predicted dust scaling relations with the corresponding 
observed relations
in this section.
We summarize our  conclusions in \S 5.

\section{The chemodynamical model}

We mainly investigate the time evolution of dust and ${\rm H_2}$ properties
in forming disk galaxies embedded in massive dark matter halos with their physical
properties (i.e., radial density profiles) consistent with predictions of
a Cold Dark Matter (CDM) cosmology (e.g., Navarro et al. 1996; NFW).
The present chemodynamical model incorporates chemical evolution
of variously different elements (e.g., He, C, N, O, Mg, and Ca),
chemical enrichment by  Type Ia, Type II,
and aspherical  supernovae (SNIa,  SNII, and ASN respectively)
and AGB stars,
supernova feedback effects of SNIa and SNII,
formation and destruction of dust,  ${\rm H_2}$ formation,
metallicity-dependent radiative cooling,  and ${\rm H_2}$-regulated star formation.
Therefore the present model
is a much improved version of those 
used in investigating chemodynamical evolution of galaxy mergers (Bekki \& Shioya 1999)
and disk galaxy evolution in groups (Bekki \& Couch 2011).

In order to perform  numerical simulations on GPU clusters,
we have revised our previous chemodynamical code (`GRAPE-SPH'; Bekki 2009)
that can be run on the special computer for gravitational dynamics (GRavity PipE;
Sugimoto et al. 1990). 
In the present paper, we describe only the key ingredients of the code;
the full details of the new code (including the code performance)
will be described in our forthcoming paper (Bekki 2013). 
Recent numerical simulations with ${\rm H_2}$-regulated star formation
(P06; R08; K12), the results of which are compared with those
of this paper,  do not include the important feedback effects of ASN  on
the interstellar mediums (ISM) of galaxies, which are investigated in B12.
We therefore disable the code's function of feedback effects and
chemical enrichment by ASN
in the present study.
For convenience,  physical meanings of symbols often used in
the present study are summarized in Table 2.

\subsection{Chemical enrichment}

Chemical enrichment through star formation and metal ejection from 
SNIa, II, and AGB stars is considered to proceed locally and inhomogeneously.
SNe and AGB stars are the production sites of dust, and some metals ejected from
these stars can be also accreted onto dust grains in the ISM of galaxies.
We investigate the time evolution of the 11 chemical elements of H, He, C, N, O, Fe,
Mg, Ca, Si, S, and Ba in order to predict both chemical abundances and dust properties
in the present study. The mean metallicity $Z$ for each $k$th stellar particle is 
represented by $Z_k$. The total mass of each $j$th ($j=1-11$) chemical component
ejected from each $k$th stellar particles at time $t$ is given as
\begin{equation}
\Delta z_{k,j}^{\rm ej}(t)=m_{\rm s, \it k} Y_{k,j}(t-t_k),
\end{equation}
where $m_{\rm s, \it k}$ is the mass of the $k$th stellar particle, $Y_{k,j}(t-t_k)$
is the mass of each $j$th chemical component ejected from stars per unit mass at 
time $t$, and $t_k$ represents the time when the $k$th stellar particle is 
born from a gas particle. $\Delta z_{k,j}^{\rm ej}(t)$ is given equally  to neighbor SPH gas
particles (with the total number of $N_{\rm nei, \it  k}$) 
located around the $k$th stellar particle.  Therefore, 
the mass increase of each $j$th chemical component for $i$th gas particle at time $t$
($\Delta z_{i,j}^{\rm ej}(t)$) is
given as 
\begin{equation}
\Delta z_{i,j}^{\rm ej}(t) = \sum_{k=1}^{N_{\rm nei, \it i}} 
m_{\rm s, \it k} Y_{k,j}(t-t_k)/N_{\rm nei, \it k},
\end{equation}
where $N_{\rm nei, \it i}$ is the total number of neighbor stellar particles whose metals
can be incorporated into the $i$th gas particle.

We consider the time delay between the epoch of star formation
and those  of supernova explosions and commencement of AGB phases (i.e.,
non-instantaneous recycling of chemical elements).
Therefore, the mass of each $j$th chemical component ejected from each
$i$th stellar particle is strongly time-dependent.
In order to  derive 
the mass-dependent lifetimes of stars that become SNe Ia, SNe II,
and AGB stars, we estimate 
the main-sequence
turn-off mass ($m_{\rm TO}$) for stellar particles.
We do so  by using the following formula 
(Renzini \& Buzzoni 1986):
\begin{equation}
\log m_{\rm TO}(t_{\rm s})
= 0.0558 (\log t_{\rm s})^2 - 1.338 \log t_{\rm s} + 7.764,
\end{equation}
where $m_{\rm TO}$  is in solar units and time $t_{\rm s}$ in years.

We adopt the `prompt SN Ia' model in which 
the delay time distribution (DTD)
of SNe Ia is consistent with  recent observational results by  extensive SN Ia surveys
(e.g.,  Mannucci et al. 2006; Sullivan et al. 2006).
In this prompt SN Ia mode, 
there is a time delay ($t_{\rm Ia}$) between the star formation
and the metal ejection for SNe Ia. We here adopt the following DTD
($g(t_{\rm Ia}$)) for 0.1 Gyr $\le t_{\rm Ia} \le$ 10 Gyr,
which is consistent with recent observational studies
on the SN Ia rate in extra-galaxies (e.g., Totani et al. 2008; Maoz et al. 2010,
2011):
\begin{equation}
g_{\rm Ia} (t_{\rm Ia})  = C_{\rm g}t_{\rm Ia}^{-1},
\end{equation}
where $C_{\rm g}$ is a normalization constant that is determined by
the number of SN Ia per unit mass  (which is controlled by the IMF
and the binary fraction for intermediate-mass stars
for  the adopted power-law slope of $-1$).
This adoption is pointed out to be necessary to explain the observed chemical properties 
of the LMC (Bekki \& Tsujimoto 2012)

The fraction of the stars that eventually
produce SNe Ia for 3--8$M_{\odot}$ has not been observationally determined
and thus is regarded as a free parameter, $f_{\rm b}$.
It is confirmed that  the present results on dust and ${\rm H_2}$ properties
do not depend so strongly on $f_{\rm b}$ for $0.03 \le f_{\rm b} \le 0.09$
in the present study.  We therefore show the results of the models with $f_{\rm b}=0.09$,
which is a reasonable value for investigating luminous disk galaxies like the Galaxy
(e.g., Tsujimoto et al. 2010).
The chemical yields adopted in the present study are the same as those used in
Bekki \& Tsujimoto (2012) except those from AGB stars.
We adopt the nucleosynthesis yields of SNe II and Ia from Tsujimoto et al. (1995; T95)
and AGB stars from van den Hoek \& Groenewegen (1997; VG97)
in order to estimate $Y_{k,j}(t-t_k)$ in the present study.

\begin{figure*}
\psfig{file=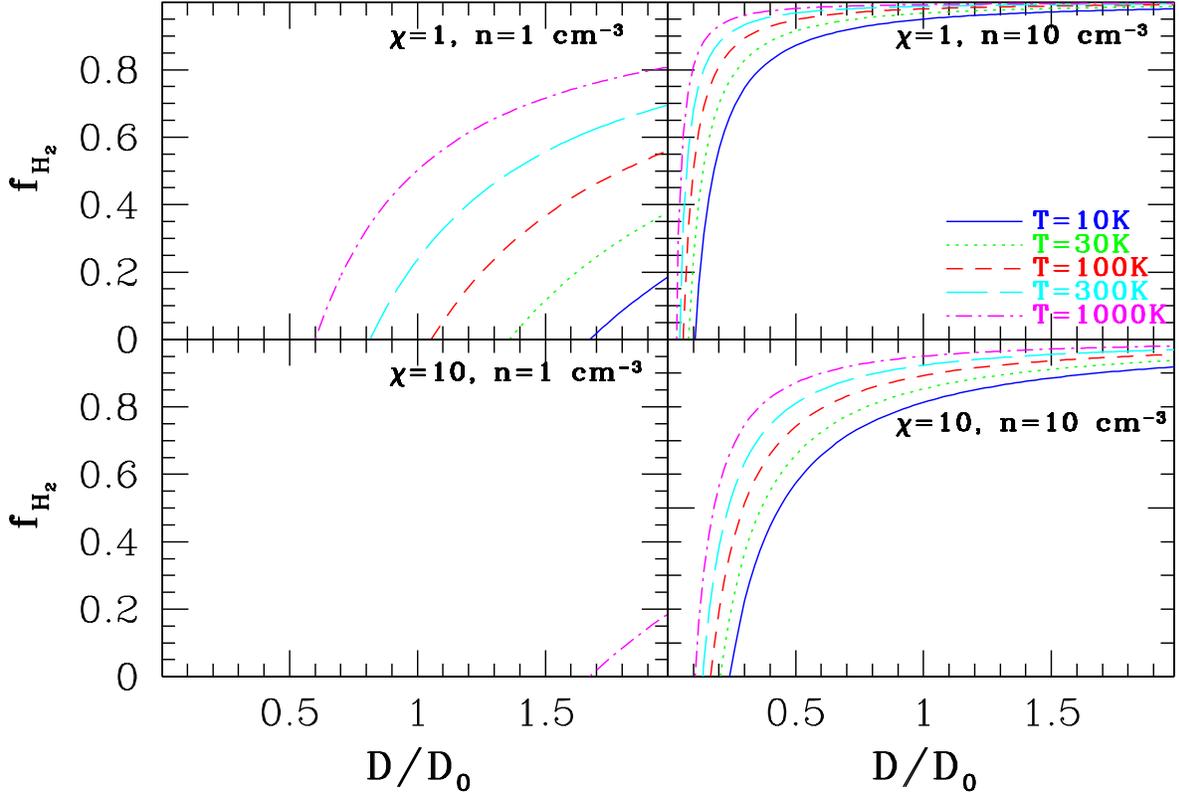,width=16.0cm}
\caption{
The ${\rm H_2}$ mass fraction ($f_{\rm H_2}$)
of a gas particle with the SPH smoothing length ($h$)
of 100pc 
as a function of $D$ (i.e., dust-to-gas ratio) for different $\chi$ 
(ISRF strength) and $n$ (hydrogen number density):
$\chi=1$ and $n=1$ cm$^{-3}$ (upper left), 
$\chi=1$ and $n=10$ cm$^{-3}$ (upper right), 
$\chi=10$ and $n=1$ cm$^{-3}$ (lower left), 
and $\chi=10$ and $n=10$ cm$^{-3}$ (lower right).
The constant $D_0$ is the dust-to-gas ratio of the solar neighborhood in the Galaxy. 
It should be noted here that  $f_{\rm H_2}$ depends on $h$ for a given $\chi$, $n$,
and $D$,  because hydrogen column densities ($N$, $N_1$, and $N_2$),
which are key parameters for $f_{\rm H_2}$, are calculated from $n$
($n_1$ and $n_2$)  and $h$: the results shown in this  figure are true only for $h=100$pc.
 Although $h$ is fixed for different $n$ and $\chi$ in this figure, $h$ is different
for different $n$ ($h \propto n^{1/3}$) and $\chi$ in real simulations.
Accordingly, this figure is used for an illustrative purpose (to demonstrate
the important role of $D$ in determining  $f_{\rm H_2}$.)
}
\label{Figure. 1}
\end{figure*}

\subsection{Dust model}

The present dust model is essentially the same as that adopted in the previous
multi-zone model by D98, which reproduced reasonably well the observed
chemical and dust properties of the Galaxy in a self-consistent manner.
The dust model consists of the following four components: (i) production in stellar winds
of SNe Ia and SNe II and AGB stars,  (ii) accretion of metals of ISM on dust grains,
(iii) destruction of dust by energetic SN explosions, and (iv) PAH formation.
The present model is somewhat idealized in that it does not include coagulation
of small dust grains and time evolution of dust sizes that have been
recently investigated in some one-zone models 
(e.g., Hirashita 2012). In the present paper, we focus exclusively on the most important
processes related to dust formation and evolution in galaxies.
The influences of dust coagulation and dust size evolution will therefore be investigated
in our forthcoming paper.

\begin{table}
\centering
\begin{minipage}{80mm}
\caption{Description of physical meanings for symbols often used in the present study.} 
\begin{tabular}{cc}
{Symbol}
& {Physical meaning}\\
$A_{\rm O}$  & gas-phase oxygen abundances  \\
$D$   &  dust-to-gas ratio \\
$f_{\rm H_2}$ & mass fraction of molecular hydrogen (${\rm H_2}$) \\
$f_{\rm PAH}$ & PAH-to-dust mass ratio \\
$\Sigma_{\rm H_2}$ & surface mass density of  ${\rm H_2}$  \\
$\Sigma_{\rm gas}$ &  surface mass density of gas (H~{\sc i}+${\rm H_2}$) \\
$\Sigma_{\rm Z}$ &  surface mass density of metals ($Z$)  \\
$\Sigma_{\rm dust}$ &  surface mass density of dust   \\
$\Sigma_{\rm PAH}$ &  surface mass density of PAH dust   \\
$\Sigma_{\rm star}$ &  surface mass density of stars   \\
$\mu_{\rm H_2}$ &  $\log_{10} \Sigma_{\rm H_2}$    \\
$\mu_{\rm dust}$ &  $\log_{10} \Sigma_{\rm dust}$    \\
$\mu_{\rm g}$ &  $\log_{10} \Sigma_{\rm gas}$    \\
$\mu_{\rm s}$ &  $\log_{10} \Sigma_{\rm star}$    \\
$R_{\rm dust}$ & mass ratio of dust to star ($M_{\rm dust}/M_{\rm star}$) \\
$F_{\rm dust}$  & initial dust-to-metal ratio in gas \\
$f_{\rm dust, m}$ & mass fraction of metals locked up in dust \\
$\rho_{\rm th}$ & threshold gas density for star formation \\
$\tau_{\rm acc}$ & dust accretion timescale   \\
$\tau_{\rm dest}$ & dust destruction  timescale   \\
$M_{\rm h}$ & initial dark halo mass   \\
\end{tabular}
\end{minipage}
\end{table}

\begin{table}
\centering
\begin{minipage}{80mm}
\caption{Description of the basic parameter values
for the fiducial model.}
\begin{tabular}{cc}
{Physical properties}
& {Parameter values}\\
{Total Mass \footnote{$M_{\rm h}=M_{\rm dm}+M_{\rm g}$, where
$M_{\rm dm}$ and $M_{\rm g}$ are the total masses of dark matter halo
and gas in a galaxy, respectively.}}
& $M_{\rm h}=10^{11} {\rm M}_{\odot}$  \\
{Structure \footnote{ The NFW profile with a virial radius ($r_{\rm vir}$)
and a $c$ parameter is adopted for the structure of dark matter halo.}}
& $r_{\rm vir}=120$ kpc,  $c=10$  \\
Gas fraction & $f_{\rm g}=0.1$     \\
Spin parameter &   $\lambda=0.038$  \\
Chemical yield  &  T95 for SN,  VG97 for AGB \\
Initial metallicity   &   ${\rm [Fe/H]_0}=-3$ \\
Dust formation  & $\tau_{\rm acc}=0.25$ Gyr, $\tau_{\rm dest}=0.5$ Gyr  \\
{PAH \footnote{$R_{\rm PAH}$ is the
mass fraction of PAH dust to total dust in the stellar ejecta
of C-rich AGB stars.}}   &  $R_{\rm PAH}=0.05$ \\
Dust yield  & D98-type \\
Initial dust/metal ratio  & 0.1  \\
Hubble-type & late-type disk (isolated) \\
{Feedback \footnote{$f_{\rm b}$ is the binary fraction of stars that
can finally explode as SNe Ia.}}  & SNIa ($f_{\rm b}=0.09$) and SNII \\
{SF \footnote{$\rho_{\rm th}$ is the threshold gas density for star formation
and interstellar radiation field (ISRF) is included in the estimation of 
${\rm H_2}$ mass fraction in this model.}}
& ${\rm H}_2$-dependent,  ISRF,  $\rho_{\rm th}=1$ cm$^{-3}$ \\
IMF & Salpeter ($\alpha=2.35$) \\
Particle number & $N_{\rm dm}=900000$, $N_{\rm g}=100000$ \\
Softening length  & $\epsilon_{\rm dm}=935$ pc, $\epsilon_{\rm g}=94$ pc \\
Gas mass resolution   & $m_{\rm g}=10^5 {\rm M}_{\odot}$ \\
\end{tabular}
\end{minipage}
\end{table}

\subsubsection{Production}

The total mass of $j$th component ($j$=C, O, Mg, Si, S, Ca, and Fe)
of dust from $k$th type of stars ($k$ = I, II, and AGB for SNe Ia, SNe II, and
AGB stars, respectively) is described as follows;
\begin{equation}
m_{\rm dust, \it j}^k= \delta_{\rm c, \it j}^k F_{\rm ej}(m_{\rm ej, \it j}^k),
\end{equation}
where $\delta_{\rm c, \it, j}^k$ is the condensation efficiency (i.e., the mass
fraction of metals that are locked up in dust grains) for each $j$th chemical component
from $k$th stellar type,
$F_{\rm ej}$ is the function that determines the total mass of metals that can be used 
for dust formation,
and $m_{\rm ej, j}^k$ is the mass of $j$th component
ejected from $k$th stellar type. 
The total mass of stellar ejecta is estimated by using stellar yield tables by T95 and VG97.
For stars with the initial masses ($m_{\rm I}$)
larger than $8 {\rm M}_{\odot}$,  $m_{\rm dust, \it, j}^{\rm II}$ is as follows:
\begin{equation}
m_{\rm dust, \it j}^{\rm II}= \left\{
\begin{array}{ll}
\delta_{\rm c, \it j}^{\rm II} m_{\rm ej, \it j}^{\rm II}
& \mbox{for other than O} \\
16 \sum_{l=1}^{n_l}  \delta_{\rm c, \it l}^{\rm II} 
m_{\rm ej, \it k}^{\rm II}/\mu_{l}   & \mbox{for O} \\ 
\end{array}
\right.
\end{equation}
In the above estimation of $m_{\rm dust, O}^{\rm II} $,
the summation is done for $l=$ Mg, Si, S, Ca, and Fe (i.e., $n_l=5$)
and $\mu_l$ is the mass of $l$th element in atomic mass units.
This formula for SNe II is used for SNe Ia in the present study.
$\delta_{\rm c, \it j}^{\rm I}$ and $\delta_{\rm c, \it j}^{\rm II}$  are
0.8 for $j$=Mg, Si, S, Ca, and Fe and 0.5 for C.

\begin{figure}
\psfig{file=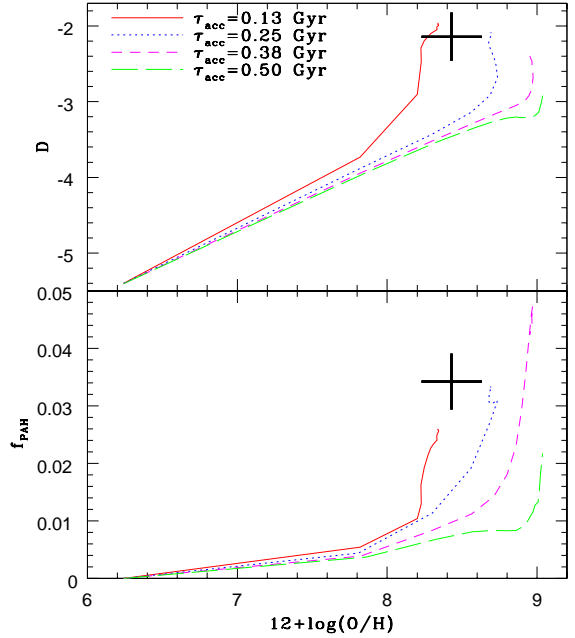,width=8.0cm}
\caption{
The time evolution of simulated  galaxies on the $A_{\rm O}-D$ (upper)
and $A_{\rm O}-f_{\rm PAH}$ planes (lower) for four representative
models with 
$\tau_{\rm acc}=0.13$ Gyr (solid, red),
$\tau_{\rm acc}=0.25$ Gyr (dotted, blue),
$\tau_{\rm acc}=0.38$ Gyr (short-dashed, magenta),
and $\tau_{\rm acc}=0.50$ Gyr (long-dashed, green).
The gas-phase oxygen abundance ($A_{\rm O}=12+\log {\rm (O/H)}$) is denoted as
$A_{\rm O}$ in this figure.
For these models,
$M_{\rm h}=10^{11} {\rm M}_{\odot}$ and $\lambda=0.038$ are adopted.
The large thick crosses indicate the observed location of the LMC
on the two planes for comparison.
The observational data are from Meixner et al. (2012) and Leroy et al. (2011), 
and the error bars of 0.2 dex are  shown for $A_{\rm O}$.
}
\label{Figure. 2}
\end{figure}

For stars with $m_{\rm I} \le 8 {\rm M}_{\odot}$ and $C/O>1$ in the ejecta,
$m_{\rm dust, \it, j}^{\rm AGB}$ is as follows:
\begin{equation}
m_{\rm dust, \it j}^{\rm AGB}= \left\{
\begin{array}{ll}
\delta_{\rm c,  C}^{\rm II} 
(m_{\rm ej,  C}^{\rm AGB}-0.75m_{\rm ej,  O}^{\rm AGB})
& \mbox{for C} \\
0  & \mbox{for other than C} \\ 
\end{array}
\right.
\end{equation}
The ejecta from these  C-rich stars is assumed to have
$\delta_{\rm c, C}^{\rm AGB}=1$ in the present study (i.e., the same value
adopted by D98).

For stars with $m_{\rm I} \le 8 {\rm M}_{\odot}$ and $C/O \le 1$ in the ejecta,
$m_{\rm dust, \it, j}^{\rm AGB}$ is as follows:
\begin{equation}
m_{\rm dust, \it j}^{\rm AGB}= \left\{
\begin{array}{ll}
0  & \mbox{for C} \\ 
\delta_{\rm c, \it j}^{\rm AGB} m_{\rm ej, \it j}^{\rm AGB}
& \mbox{for other than O \& C} \\
16 \sum_{l=1}^{n_l}  \delta_{\rm c, \it l}^{\rm AGB} 
m_{\rm ej, \it k}^{\rm II}/\mu_{l}   & \mbox{for O} \\ 
\end{array}
\right.
\end{equation}
In the above estimation of $m_{\rm dust, O}^{\rm AGB}$,
the summation is done for $l=$ Mg, Si, S, Ca, and Fe (i.e., $n_l=5$).
$\delta_{\rm c, \it j}^{\rm AGB}$  are
0.8 for Mg, Si, S, Ca, and Fe.
Although a range of $\delta_{\rm c, \it j}^k$ values could be adopted in the 
present study,  we investigate mainly the models with the parameter values adopted above.
These dust yield models  are referred to as `D98-type yield' for convenience in the present
study.
We discuss the total amount of dust and do not discuss the dust composition in detail.

 For comparison, 
we also investigate some models in which the mass fraction ($F_{\rm dust}$)
of dust among all gas (metals + dust)
ejected from  a star does not depend on the stellar mass.
These models are referred to as `uniform yield' for convenience.
We investigate models with uniform yield, mainly because several earlier  one-zone models
adopted such models for their investigation on the origin of the observed
$A_{\rm O}-D$ relation.
Although 
the dust mass fraction ($F_{\rm dust}$) is a (fixed) free parameter, we show the
models with $F_{\rm dust}=0.1$. This is because the models with $F_{\rm dust}=0.1$
can better explain the observed $D$ of galaxies. We mainly show the results of 
the models with D98-type yield rather than those with uniform one in the present study.

\subsubsection{Accretion}

Dust grains can grow by accretion of metals of ISM onto preexisting cores and this
accretion process is included in previous models (D98). Following D98, we consider
that the key parameter in dust accretion is the dust accretion timescale ($\tau_{\rm acc}$).
In the present study, this parameter can vary between different gas particles
and is thus represented by $\tau_{\rm acc, \it i}$ for $i$th gas particle.
The mass of $j$th component 
($j$=C, O, Mg, Si, S, Ca, and Fe) of dust for $i$th gas particle
at time $t$ ($d_{i,j}(t)$) can increase owing  to dust accretion processes.
The mass increase 
is described as 
\begin{equation}
\Delta d_{i,j}^{\rm acc}(t)=\Delta t_i (1-f_{\rm dust,\it i, j}) 
d_{i,j}(t) /\tau_{\rm acc, \it i},
\end{equation}
where $\Delta t_i$ is the individual time step width for the $i$th gas particle
and $f_{\rm dust, \it i, j}$ is the fraction of the $j$th chemical element that
is locked up in the dust. Owing to this dust growth, the mass of $j$th chemical
component that is {\it not} locked up in the dust ($z_{i,j}(t)$) 
can decrease, which is simply given as
\begin{equation}
\Delta z_{i,j}^{\rm acc}(t)=- \Delta t_i (1-f_{\rm dust,\it i, j}) 
d_{i,j}(t) /\tau_{\rm acc, \it i}
\end{equation}
As is clear in these equations, the total  mass of $j$th component in $i$th gas
particle ($m_{i,j}(t)$) is $z_{i,j}(t)+d_{i,j}(t)$. 
 It should be stressed that the notation for dust mass here is for
each particle whereas that for dust mass in \S 2.2.1 is for each star.

Although $\tau_{\rm acc, \it i}$ can be {\it locally} different and {\it time-dependent}
owing to complicated chemical and dynamical evolution of galaxies, we adopt
a fixed value for all gaseous particles in a simulation.  The fixed value (denoted
$\tau_{\rm acc}$) varies  between different models.  As described later,
we need to choose carefully this $\tau_{\rm acc}$ in order to  reproduce
the observed dust properties (e.g., $D$) in disk galaxies. Thus we avoid introducing
additional free parameters for  $\tau_{\rm acc,  \it i}$ possibly dependent on
time and the ISM properties in the present study.

\subsubsection{Destruction}

Dust grains can be destroyed though supernova blast waves 
in the ISM of galaxies (e.g., McKee 1989)
and the destruction process is parameterized by the destruction time scale
($\tau_{\rm dest}$) in previous one-zone models (e.g., Lisenfeld \& Ferrara 1998;
Hirashita 1999).  Following the previous models, 
the decrease  of the mass of $j$th component 
of dust for $i$th gas particle 
at time $t$ due to dust destruction process
is as follows
\begin{equation}
\Delta d_{i,j}^{\rm dest}(t)= - \Delta t_i  
d_{i,j}(t) /\tau_{\rm dest, \it i},
\end{equation}
where $\tau_{\rm dest, \it i}$ is the dust destruction timescale for $i$th particle.
The dust destroyed by supernova explosions can be returned back to the ISM,
and therefore the  mass  
of $j$th chemical
component that is not locked up in the dust 
increases  as follows:
\begin{equation}
\Delta z_{i,j}^{\rm dest}(t)= \Delta t_i 
d_{i,j}(t) /\tau_{\rm dest, \it i}
\end{equation}

Although $\tau_{\rm dest, \it i}$ can  possibly vary for different particles
owing to different physical conditions of the local ISM,
we adopt a fixed value (denoted as $\tau_{\rm dust}$) for {\it all} gas particles
in the present study.
We consider that a reasonable $\tau_{\rm dest}$ is 0.5 Gyr (D98) 
and show the results of the models
with $\tau_{\rm dest}=0.5$ Gyr:
As described later (\S 2.7),  only a narrow range of $\tau_{\rm dest}$ 
is allowed for reproducing observational results.
The dust destruction can occur only for gas particles that are located in the surrounding
regions of SNe Ia and SNe II
only for a timescale of $\tau_{\rm dest}$ in the present study. 

Thus the equation for the time evolution of $j$th component of metals
for $i$th gas particle  are given as
\begin{equation}
z_{i,j}(t+\Delta t_i)=z_{i,j}(t)+\Delta z_{i,j}^{\rm ej}(t)+\Delta z_{i,j}^{\rm acc}(t)
+\Delta z_{i,j}^{\rm dest}(t)
\end{equation}
Likewise, the equation for dust evolution is given as
\begin{equation}
d_{i,j}(t+\Delta t_i)=d_{i,j}(t)+\Delta d_{i,j}^{\rm acc}(t) 
+\Delta d_{i,j}^{\rm dest}(t)
\end{equation}
 Dust is locked up in stars as metals are done so, when gas particles are converted into
new stars. This means that star formation process itself has an effect
of destroying dust in the present study.

\subsubsection{PAH}
 
A growing number of observational studies on PAH properties have been accumulated
for galaxies within and beyond the Local Group (e.g, Draine et al. 2007; Meixner et al. 2010;
Takagi et al. 2010; Sandstrom et al. 2012).
It is accordingly timely for the present study to discuss the origin of the observed 
PAH properties in galaxies by using the new chemodynamical model.
The most  promising formation site of interstellar PAH dust is C-rich AGB stars,
though direct observation supporting PAH formation in stellar winds of AGB stars
is very weak (e.g., Tielens 2008 for a recent review). 
We consider that some fraction of carbon dust produced by C-rich AGB stars ($C/O>1$) 
can finally become PAH dust and thereby investigate the PAH properties in the present
study. The mass fraction of PAH dust to total carbon dust in the ejecta of C-rich AGB stars
is a parameter denoted as $R_{\rm PAH}$.
Using low-resolution simulations with different $R_{\rm PAH}$,
we find that $R_{\rm PAH}=0.05$ can better explain observations (described later in \S 2.7).
We therefore show the results of the models with
$R_{\rm PAH}=0.05$ in the present study.

\subsection{${\rm H_2}$ formation and dissociation}

Previous different numerical simulations  estimated the mass fractions 
of ${\rm H_2}$ 
to total hydrogen gas  ($f_{\rm H_2}$) in galaxies by using different ${\rm H_2}$
formation models (P06;  R08; K12). 
The model adopted in the present study is essentially similar to
that used in P06 in the sense that $f_{\rm H_2}$ is determined by 
local far-UV (FUV) radiation fields and  gas densities.
One of the important differences between P06 and the present study  is that the time evolution
of dust abundances and compositions are explicitly followed and used for
estimating ${\rm H_2}$ formation rates of local regions of galaxies in the present
study. ${\rm H_2}$ for each local region in a galaxy is determined by the
balance between ${\rm H_2}$ formation and ${\rm H}_2$ dissociation by FUV radiation.
The formulas used in Goldshmidt \& Sternberg (1995)
and Draine (2009) are adopted
in deriving  $f_{\rm H_2}$.

\subsubsection{${\rm H_2}$ formation on dust grains}

The ${\rm H_2}$ formation on dust grains via grain catalysis has been extensively
investigated by many authors  (e.g., Gould \& Salpeter 1963; Hollenbach \& Salpeter 1971;
Cazaux \& Tielens 2002). A key parameter for the process is the total grain geometric
cross section per H nucleon and is defined  as
\begin{equation}
\Sigma_{\rm gr} \equiv \frac{1}{n} \int \pi a^2 \frac{dn_{\rm gr}}{da} da,
\end{equation}
where $a$,  $n_{\rm gr}$, and $n$ are the sizes of dust grains,
the size distribution, and the hydrogen number density, respectively (Draine 2009).
$\Sigma_{\rm gr}$ is often denoted  in units of 
$10^{-21}$ cm$^2$ H$^{-1}$ (Spitzer 1978) for convenience (i.e., 
$\Sigma_{\rm gr}=\Sigma_{-21}$ $10^{-21}$ cm$^2$ H$^{-1}$).
The ${\rm H_2}$ formation rate is given as
\begin{equation}
\frac{dn_2}{dt} =  R_{\rm gr}n n_1,
\end{equation}
where $n_1$, $n_2$, and $R_{\rm gr}$ are the number density of atomic
hydrogen,  molecular hydrogen, and the effective ${\rm H_2}$ rate coefficient,
respectively.
$R_{\rm gr}$ is given as
\begin{equation}
R_{\rm gr}=\frac{1}{2}( \frac{8 {\rm k_B} T}{\pi {\rm m_H}} )^{1/2} \langle 
\epsilon_{\rm gr} \rangle \Sigma_{\rm gr},
\end{equation}
where ${\rm k_B}$, $T$, ${\rm m_H}$, and $\epsilon_{\rm gr}$
are the Boltzmann constant,  the temperature of gas, the mass of a hydrogen atom, 
and the formation efficiency averaged over the grain surface area, respectively. 
Observationally, $R_{\rm gr}$ is determined ($\sim 3 \times 10^{-17}$ cm$^3$ s$^{-1}$;
Jura 1975) so that the product of
$\langle \epsilon_{\rm gr} \rangle$ and $\Sigma_{\rm gr}$  can be 
estimated  ($\langle \epsilon_{\rm gr} \rangle \Sigma_{\rm gr} \approx 0.5$;
Draine 2009).  

We do not investigate the time evolution of the dust size distributions of galaxies
in the present study. We assume therefore that $\langle \epsilon_{\rm gr} \rangle$
is constant and thus $R_{\rm gr}$ is determined by the dust-to-gas ratio ($D$).
We adopt a reasonable value of $\langle \epsilon_{\rm gr} \rangle=0.5$ for classical silicate
and carbonaceous grains (Draine 2009), though
the adopted value could be a lowest possible value for the grains with the sizes larger
than $0.01 \mu$m (Draine 2009). 
$R_{\rm gr}$ for $i$th gas particle 
is determined as follows
\begin{equation}
R_{\rm gr,  \it i}=3.7 \times 10^{-17} ( \frac{T_i}{100 {\rm K}} )^{1/2} 
(\frac{D_i}{D_0}),
\end{equation}
where $T_i$ and $D_i$ 
are the temperature and the dust-to-gas ratio of $i$th gas particle,
respectively, and $D_0$ is the dust-to-gas ratio  for the Galaxy
(0.0064; Zubko et al. 2004).

\subsubsection{Formation/destruction equilibrium}

Goldshmidt \& Sternberg (1995) adopted a plane-parallel equilibrium cloud model
and thereby estimated the mass-ratios of ${\rm H_2}$ to H by solving the following
equation that describes the balance between ${\rm H_2}$ formation and dissociation
by FUV radiation fields:
\begin{equation}
R_{\rm gr}nn_1=\chi \zeta f_{\rm shield}(N_2) e^{-\tau} n_2,
\end{equation}
where $\zeta$ is the unattenuated ${\rm H}_2$ photo-dissociation rate for a unit
incident FUV photon flux within the 11.2 to 13.6 eV band, and $\chi$ is the
FUV intensity scaling factor relative to the unit FUV field,
$f_{\rm shield}(N_2)$ is the ${\rm H_2}$ self-shielding function,
$N_2$ ($N_1$) is the column density of ${\rm H_2}$ (H~{\sc i}),
and $\tau$ is the optical depth for FUV continuum and given
as $\tau=\sigma N_1 + 2\sigma N_2$, where $\sigma$ is the 
dust absorption cross section per hydrogen nucleus in the FUV wavelength range.

By substituting $\int n_1 dr$ and $\int n_2dr$ for $N_1$ and $N_2$, respectively,
in the above equation,  a separate differential equation can be derived as follows
\begin{equation}
R_{\rm gr}ndn_1=\chi \zeta f_{\rm shield}(N_2) e^{-\sigma(N_1+2N_2)} dn_2,
\end{equation}
where $n$(=$n_1+n_2$) is fixed.
By solving the above equation, 
the total ${\rm H_1}$ column density ($N_1^{\rm tot}$) can 
be derived and written as a function of $R_{\rm gr}$,
$\sigma$, $G_0$, $n$, and $\zeta$, and $\chi$ as follows
\begin{equation}
N_1^{\rm tot}=\frac{1}{\sigma} \ln(\sigma \alpha_0 G_0+1),
\end{equation}
where the dimensionless $\alpha_0$ is described as
\begin{equation}
\alpha_0 =  \frac{\chi \zeta}{R_{\rm gr}n},
\end{equation}
and $G_0$ is described as
\begin{equation}
G_0 = \int_0^{\infty}  f_{\rm shield}(N_2) e^{-2\sigma N_2} dN_2 .
\end{equation}
We can estimate $\chi$, $R_{\rm gr}$, $\sigma$, and $n$ for each gas particle
by using the local gas density, dust-to-gas ratio, and  interstellar radiation field (ISRF)
of the particle in the present simulations (described later).
Therefore, by estimating $G_0$ for a reasonable function of $f_{\rm shield}$,
we can derive $N_1$ and $N_2$ for
each gas particle (by using
$N=N_1+2N_2$ or $n=n_1+2n_2$). Draine \& Bertoldi (1996) derived an  approximation
for $f_{\rm shield}$ as follows
\begin{equation}
f_{\rm shield}=\frac{0.965}{(1+x/b_5)^2} + \frac{0.035}{(1+x)^{0.5}} 
\exp[-8.5\times 10^{-4}(1+x)^{0.5}],
\end{equation}
where $x=N_2/5\times 10^{14} {\rm cm}^{-2}$ and $b_5=b/{\rm km s^{-1}}$ ($b$ is the Doppler 
broadening parameter and equal to 3 km s$^{-1}$). Although this  is a reasonably accurate
approximation (Draine 2009),  we can not have an analytic formula for estimating
$G_0$ (as a function of $\sigma$), if we use this.
Therefore, we adopt the following approximation
from P06: 
\begin{equation}
f_{\rm shield}=(\frac{N_2}{N_{\rm ch}})^{-0.5},
\end{equation}
where $N_{\rm ch}$ is a characteristic column density of $N_{\rm ch}=1.75 \times 10^{11}$
cm$^{-2}$. Since $\int_0^{\infty} e^{-t} \frac{1}{\sqrt t} dt = \sqrt \pi$, we can
derive an analytic formula for $G_0$ in equation (23),
 which allows us to save computational time significantly
in the present simulations.

\begin{figure*}
\psfig{file=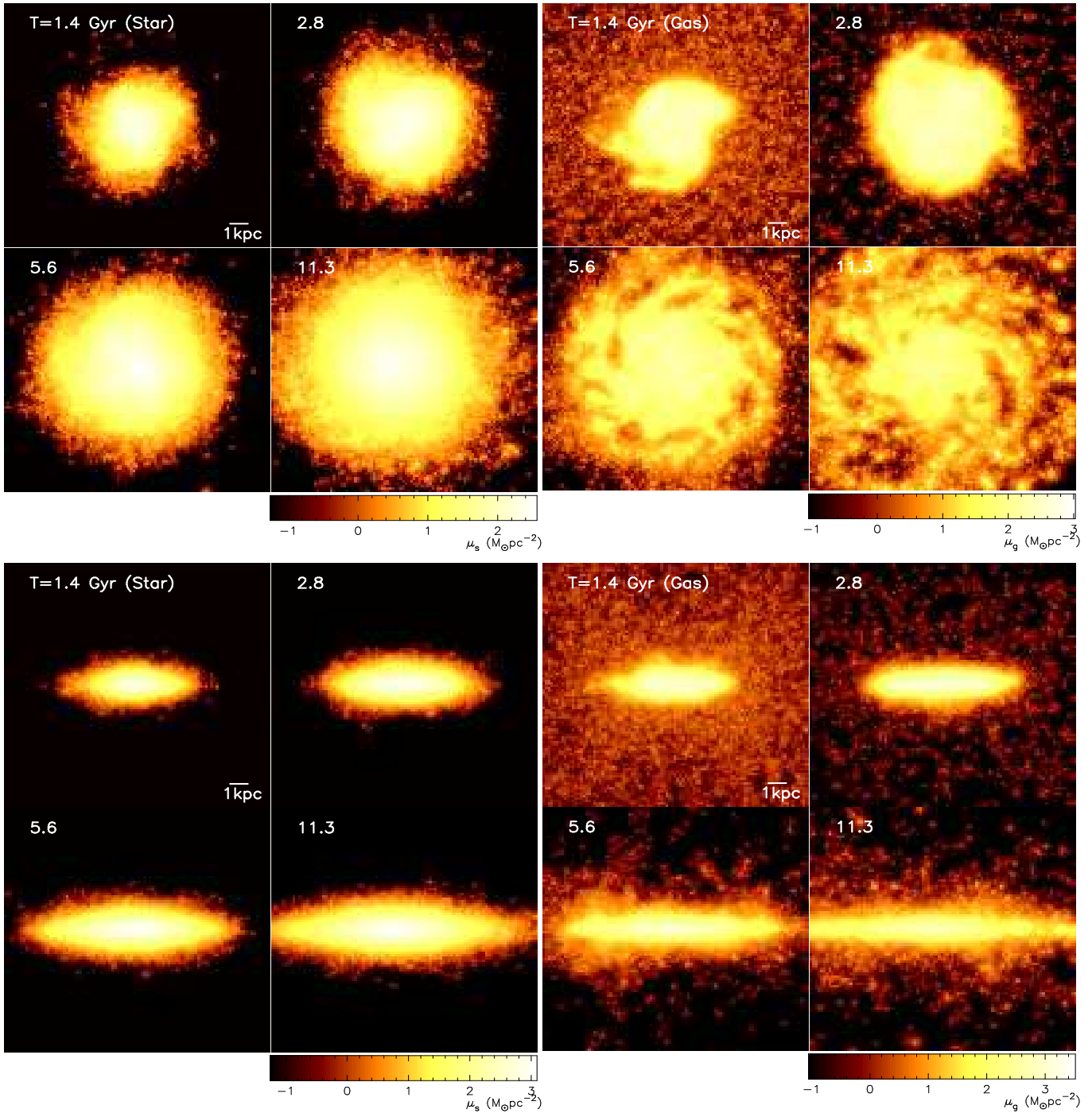,width=18.0cm}
\caption{
The time evolution of surface  mass densities  (in a logarithmic scale)
projected onto the $x$-$y$ (upper eight) and $x$-$z$ (lower eight) planes
for stars ($\mu_{\rm s}$, left) and gas ($\mu_{\rm g}$, right).
The time $T$ 
is given in units of Gyr and a thick bar indicates
1 kpc.  
}
\label{Figure. 3}
\end{figure*}

\begin{figure*}
\psfig{file=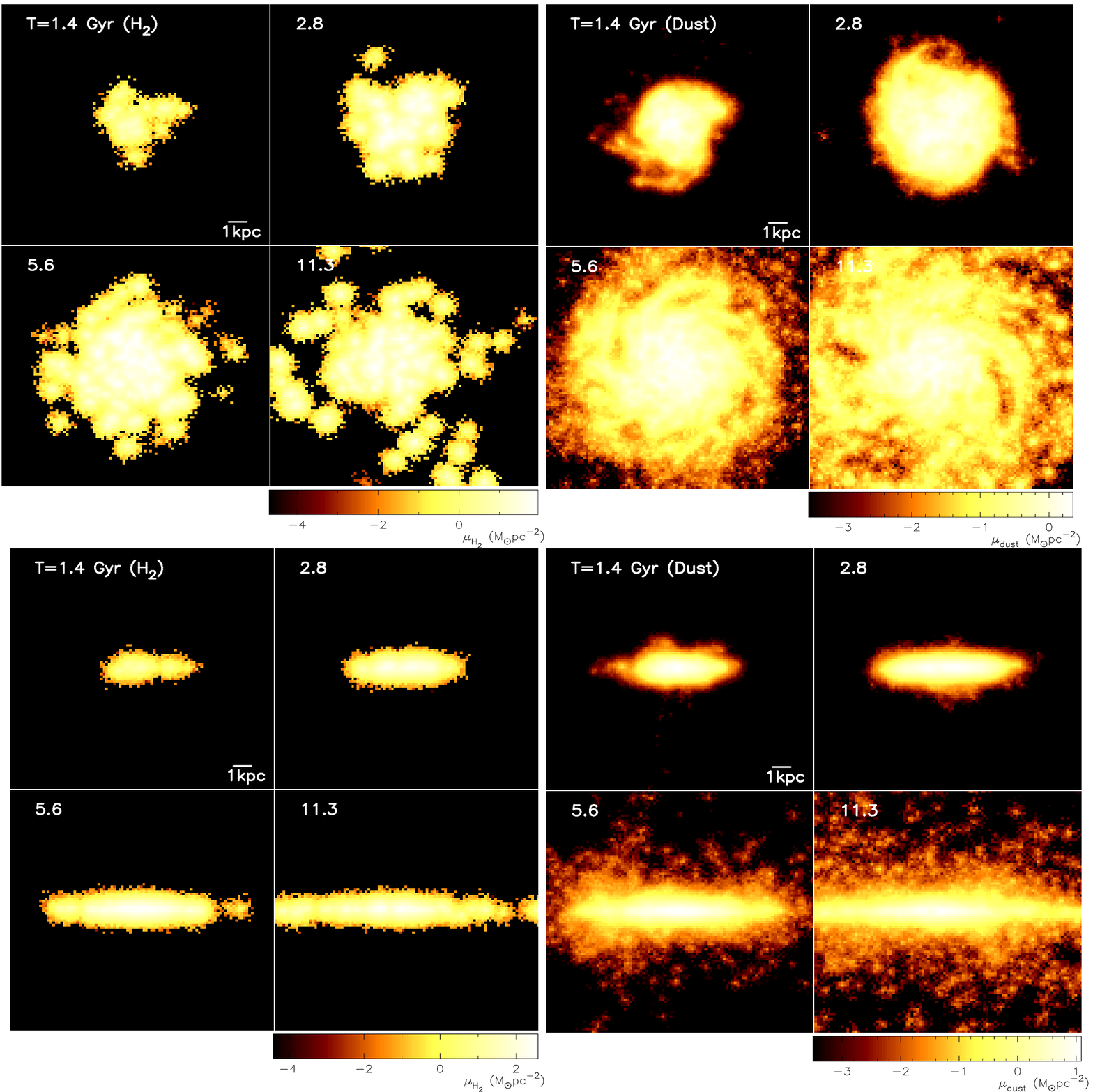,width=18.0cm}
\caption{
The same as Fig. 3 but for 
${\rm H_2}$  ($\mu_{\rm H_2}$, left) and dust ($\mu_{\rm dust}$, right).
}
\label{Figure. 4}
\end{figure*}

\subsubsection{ISRF}

In order to estimate $\chi$ for each gas particle, we use the spectral energy distributions
(SEDs) for stellar populations
with different ages and metallicities derived by Bruzual \& Charlot (2003; BC03).
We first  find stellar particles that are located within $\epsilon_{\rm g}$ (i.e., gravitational
softening length for gas particles)  of a gas particle,
then derive the total flux at 1000\AA $\,$ 
around the gas particle
from the sum of the SEDs of the stellar particles by using 
the SEDs of BC03.  Finally, we estimate the FUV flux density around the gas particle
for the local volume ($V_{\rm ISRF} \propto k_{\rm ISRF}^3\epsilon_{\rm g}^3$,
where $k_{\rm ISRF}=1$) at each time step.
We confirm that the present results do not depend on $k_{\rm ISRF}$ and
discuss the results of the models with different $k_{\rm ISRF}$ in Appendix A.
For consistency with the present IMF adopted for chemical evolution and star formation histories
of galaxies,
we use the SEDs of BC03 for the Salpeter IMF
($\alpha=2.35$).

\subsubsection{$f_{\rm H_2}$}

We can estimate $N_{1,i}$ (and $N_{2,i}$) and thus $f_{\rm H_2, \it i}$ for 
$i$th  gas particle
by substituting
the values of the temperature ($T_i$), dust-to-gas ratio ($D_i$), 
ISRF ($\chi_i$), and hydrogen gas density ($\rho_{\rm H}$)
in these equations (19), (21), (22), (23), and (25).
Since the ${\rm H}_2$ formation rate 
$R_{\rm gr, \it i}$ and
dust absorption cross section ($\sigma_i$) depend  on
$D_{\rm i}$ (e.g., $\sigma_i=1.9 \times 10^{-21} D_i/D_0$ cm$^2$),
$f_{\rm H_2, \it i}$ depends strongly on dust properties in the present study.
In converting $N_{1,i}$ into $n_{1,i}$ for $i$th gas particle,
we use the local smoothing length ($h_i$)
of the particle (i.e., $N_{1,i}=\int_{-h_i}^{h_i} n_{1,i} dr \approx 2n_{1,i}h_i$).
Fig. 1 shows an example of $D$-dependence of $f_{\rm H_2}$  
for  a gas cloud size ($h$) of 100pc, $\chi = 1$ or 10, $n=1$ or 10 cm$^{-3}$ and
five different $T$ (=10, 30, 100, 300, and 1000 K).

\subsubsection{Z-dependent ${\rm H_2}$ and star formation}

Recent numerical simulations of galaxy formation
and evolution  with ${\rm H}_2$-regulated star formation (e.g., P06 and K12)
assumed that dust abundances ($D$) of galaxies are linearly proportional to $Z$
and the proportionality constant does not vary with time and location in galaxies.
For convenience, the ${\rm H_2}$ and SF models adopted in the present study
and that in P06 and K12 are referred to as
$D-$dependent and $Z$-dependent, respectively,  for clarity.
An important question here is how different
the predicted properties of galaxies are between numerical simulations with Z-dependent
(e.g., P06 \& K12) and the present
$D$-dependent ${\rm H_2}$ and 
star formation models.

We  investigate this
question by adopting the same star formation recipe (i.e., Z-dependent) used in
KMT09 for our disk galaxy formation model with $M_{\rm h}=10^{11} {\rm M}_{\odot}$
(the fiducial model). In estimating dust optical depth ($\tau_{\rm c}$)
for local regions in a galaxy,
K12 used the cell's column density in their adaptive mesh refinement simulation code.
Since our code is not a mesh code,  we estimate $\tau_{\rm c}$ for each local region
in a simulated galaxy
by using the smoothing
length of a SPH particle at the region.
Since our main focus is not the Z-dependent model,
we describe the results of only  one Z-dependent  model.

\begin{figure}
\psfig{file=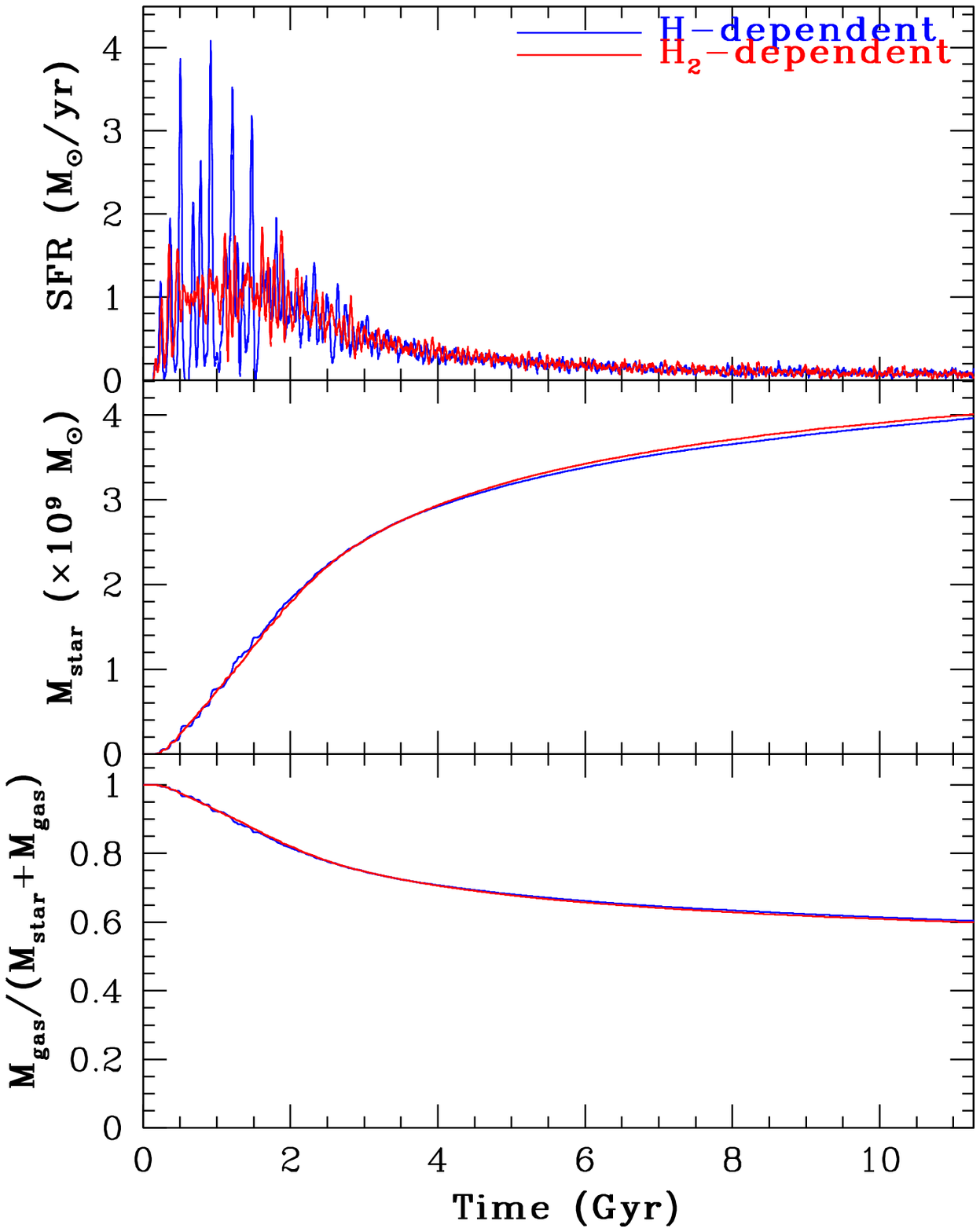,width=8.0cm}
\caption{
The time evolution of star formation rates (SFRs, top),
total stellar masses ($M_{\rm star}$, middle),
and  gas mass fraction ($M_{\rm gas}/(M_{\rm gas}+M_{\rm star}$, bottom) for the two models
with the `H-dependent' (blue) and `${\rm H_2}$-dependent' (red)  SF recipes.
The red line corresponds to the fiducial model.
The two models are exactly the same except the SF recipes.
It should be noted that the gas mass fraction  is estimated for all gas,
including warm gas in the halo and well above the stellar disk: the gas mass fraction
is fairly large ($\sim 0.6$). The gas mass fraction is $0.38$ in the stellar disk
($R\le 7.5$ kpc and $|z| \le 1$ kpc), which is more consistent with the observed
gas mass fraction of less luminous disk galaxies like the LMC 
($\sim 30$\%; e.g., van den Bergh 2000).
}
\label{Figure. 5}
\end{figure}

\subsection{Star formation}

We adopt the following two star formation (SF) recipes and compare between
the results of  models with the two recipes. One is  the `H-dependent' recipe in which
star formation rate (SFR) around $i$th gas particle  
depends simply on the total gas density ($\rho_i$). This has been a standard recipe
since Katz (1992) implemented this in galaxy formation simulations.
The other is `${\rm H_2}$-dependent',  in which the SFR depends on ${\rm H_2}$ gas
density rather than $\rho_i$. This is more realistic and reasonable,
given that star formation is ongoing within GMCs composed of ${\rm H_2}$ gas 
(e.g., K12).  We mainly investigate models with the ${\rm H_2}$-dependent SF recipe
in the present study.

In the H-dependent SF recipe, 
a gas particle is converted
into a new star if (i) the local dynamical time scale is shorter
than the sound crossing time scale (mimicking
the Jeans instability) , (ii) the local velocity
field is identified as being consistent with gravitationally collapsing
(i.e., div {\bf v}$<0$),
and (iii) the local density exceeds a threshold density for star formation ($\rho_{\rm th}$).
Recent different models adopted different $\rho_{\rm th}$,
for example, 0.1 cm$^{-3}$ (RJ12), 1 cm$^{-3}$ (RK12), 10 cm$^{-3}$ (R08),
and $5-500$ cm$^{-3}$ (K12).
We mainly investigate the models with $\rho_{\rm th}=1$ cm$^{-3}$,
and briefly discuss how the present results depend on $\rho_{\rm th}$
later.
Owing to the adopted SF model with $\rho_{\rm th}$,
  gas densities can not become very high.
For example, the maximum gas density and minimum dynamical time scale 
obtained for the fiducial model
(described later) with a gas mass resolution of $10^5 {\rm M}_{\odot}$
and $\rho_{\rm th}=1$ cm$^{-3}$
are $\sim 20$ cm$^{-3}$ and $\sim 3 \times 10^6$ yr, respectively.
The maximum density can be as large as $\sim 8000$ cm$^{-3}$ in some models
in which ${\rm H_2}$ formation is severely suppressed.

Gas particles are converted into new stellar particles in the ${\rm H_2}$-dependent
SF recipe described as follows.
$f_{\rm H_2}$ is estimated for each gas particle at each time step.
A gas particle can be regarded as a `SF candidate' gas particle 
if the above three SF conditions (i)-(iii) are satisfied. 
It could be possible to convert some fraction ($\propto f_{\rm H_2}$)
of a SF candidate  gas particle 
into a new star at each time step until the mass of the gas particle
becomes very small. However, this SF conversion method can increase dramatically
the total number of stellar particles, which becomes  numerically very costly.
We therefore adopt the following SF conversion method.
A SF candidate gas
particle is regarded as having  a SF probability of 
$f_{\rm H_2}$. At each time step   random numbers ($R_{\rm 2}$; $0\le R_2 \le 1$)
are generated and compared with $f_{\rm H_2}$.
If $R_2 < f_{\rm H_2}$, then the gas particle can be converted into
a new stellar one.

Each $i$th stellar particle is born with a fixed IMF and an initial mass
$m_{0, i}$. The stellar mass decreases with  time owing to mass loss
by SNe Ia, SNe II, and, AGB stars and the mass at time $t$ ($m_i$) can be significantly
different from $m_{0, i}$.
The adopted IMF in number is  defined
as $\psi (m_{\rm I}) = C_i{m_{\rm I}}^{-\alpha}$,
where $m_{\rm I}$ is the initial mass of
each individual star and the slope $\alpha =2.35$
corresponds to the Salpeter IMF.
The normalization factor $C_i$ is a function of $m_{0, i}$,
$m_{\rm l}$ (lower mass cut-off), and $m_{\rm u}$ (upper mass cut-off):
\begin{equation}
C_i=\frac{m_{0, i}
\times (2-\alpha)}{{m_{\rm u}}^{2-\alpha}-{m_{\rm l}}^{2-\alpha}}.
\end{equation}
where $m_{\rm l}$ and $m_{\rm u}$ are  set to be   $0.1 {\rm M}_{\odot}$
and  $100 {\rm M}_{\odot}$, respectively.
We adopt $\alpha=2.35$ for all models in the present study.

\subsection{Gravitational dynamics, hydrodynamics, and feedback}

 Since the present  code is a revised version of our GRAPE-SPH code
(Bekki 2009; Bekki \& Couch 2011),
the calculation of gravitational interaction between particles
is based on the direct summation of gravitational force of the  particles (i.e., no usage
of tree codes). The detail and performance (on GPU clusters) of the new
code will be  given in our future papers (Bekki 2013).
The direct gravitational calculation is done by the GPU whereas
other calculations related to gas dynamic, star formation, and chemical evolution
are  done by the CPU in the GPU clusters used by the present study.

One of key ingredients of the code is that 
the gravitational softening length ($\epsilon$) is chosen for each
component in a galaxy (i.e.,
multiple gravitational softening lengths).
Thus the gravitational softening length ($\epsilon$) 
is different between dark matter ($\epsilon_{\rm dm}$)
and gas ($\epsilon_{\rm g}$) and $\epsilon_{\rm dm}$ is determined by the initial
mean separation of dark matter particles. 
Furthermore,  when two different components interact gravitationally,
the mean softening length for the two components
is applied for the gravitational calculation.
For example, $\epsilon = ({\epsilon}_{\rm dm}+{\epsilon}_{\rm g})/2$
is used for gravitational interaction between 
dark matter and gas  particles.
The total numbers for dark matter ($N_{\rm dm}$)  and   gas particles ($N_{\rm g}$)
is 900000 and 100000, respectively, in the fiducial model described later.

In the present study, $\epsilon_{\rm dm}$ is relatively large ($\sim 1 kpc$ for
a model with $M_{\rm h}=10^{11} M_{\odot}$), which might severely suppress the
formation of substructures in baryonic components. However, we think that
owing to (i) the adopted multi-softening lengths and (ii) initially virialized
dark matter halo,  such severe suppression of substructure formation,
which could be regarded as a numerical artifact, is highly unlikely to occur. 
Indeed, clumpy structures of baryonic components can be formed in the present
models, in particular, in ${\rm H_2}$ distributions.

We consider that the ISM in galaxies can be modeled as an ideal gas with
the ratio of specific heats ($\gamma$) being 5/3. 
The basic methods to implement SPH in the present study are essentially
the same as those proposed by Hernquist \& Katz (1989). 
We adopt the predictor-corrector algorithm (that is accurate to second order
in time and space) in order to integrate the equations 
describing  the time  evolution of a system.
Each particle is allocated an individual time step width ($\Delta t$) that is determined
by physical properties of the particle.
 The maximum time step width ($\Delta t_{\rm max}$)
is $0.01$ in simulation units, which means that  $\Delta t_{\rm max}=1.41 \times 10^6$ yr
in the present study. Although a gas particle is allowed to have a minimum time step
width of $1.41 \times 10^4$ in the adopted individual timestep scheme,
no particle actually has such a short time step width owing to conversion
from gas to star in high-density gas regions.

Each SN is assumed to eject the feedback energy ($E_{\rm sn}$) 
of $10^{51}$ erg and 90\% and 10\% of $E_{\rm sn}$ are used for the increase
of thermal energy (`thermal feedback') 
and random motion (`kinematic feedback'), respectively.
The energy-ratio of thermal to kinematic feedback is consistent with
previous numerical simulations by Thornton et al. (1998) who investigated
the energy conversion processes of SNe in  detail.
The way to distribute $E_{\rm sn}$ of SNe among neighbor gas particles
is the same as described in B12.
The radiative cooling processes 
are properly included  by using the cooling curve by
Rosen \& Bregman (1995) for  $100 \le T < 10^4$K 
and the MAPPING III code 
for $T \ge 10^4$K 
(Sutherland \& Dopita 1993).

\begin{table*}
\centering
\begin{minipage}{175mm}
\caption{A range of model parameters for the models with $N=10^6$.}
\begin{tabular}{ccccccc}
{Parameters}
& { $M_{\rm h}$ (${\rm M}_{\odot}$)
\footnote{We mainly investigate the five representative models
with $M_{\rm h}=10^{10}, 3\times 10^{10}, 10^{11}, 3\times 10^{11}$, 
and $10^{12} {\rm M}_{\odot}$ for different $\lambda$. 
We also investigate a model with $M_{\rm h}=3 \times 10^9 {\rm M}_{\odot}$
and $\lambda=0.038$ to search for  a lowest possible $D$ in low-mass disk galaxies. } }
& { $\lambda$ \footnote{ $\lambda=0.02, 0.038$, and 0.06 are investigated.} } 
& { $\tau_{\rm acc}$ (Gyr) \footnote{$\tau_{\rm dest}$ is fixed at 0.5 Gyr for all models.}}
& {SF recipe}
& {$\rho_{\rm th}$ (atom cm$^{-3}$)} 
& { Dust yield \footnote{D98-type  yield is used for most models. For uniform dust yield
model,  $F_{\rm dust}$ (dust mass fraction among metals from SNe and AGB stars) is
set to be  0.1. } }\\
Range &  $3 \times 10^{9}-10^{12}$ 
& $0.02-0.06$  & $0.13-0.5$  & H- or ${\rm H_2}$-dependent  & $1-10$
& D98-type or uniform \\
\end{tabular}
\end{minipage}
\end{table*}

\begin{figure*}
\psfig{file=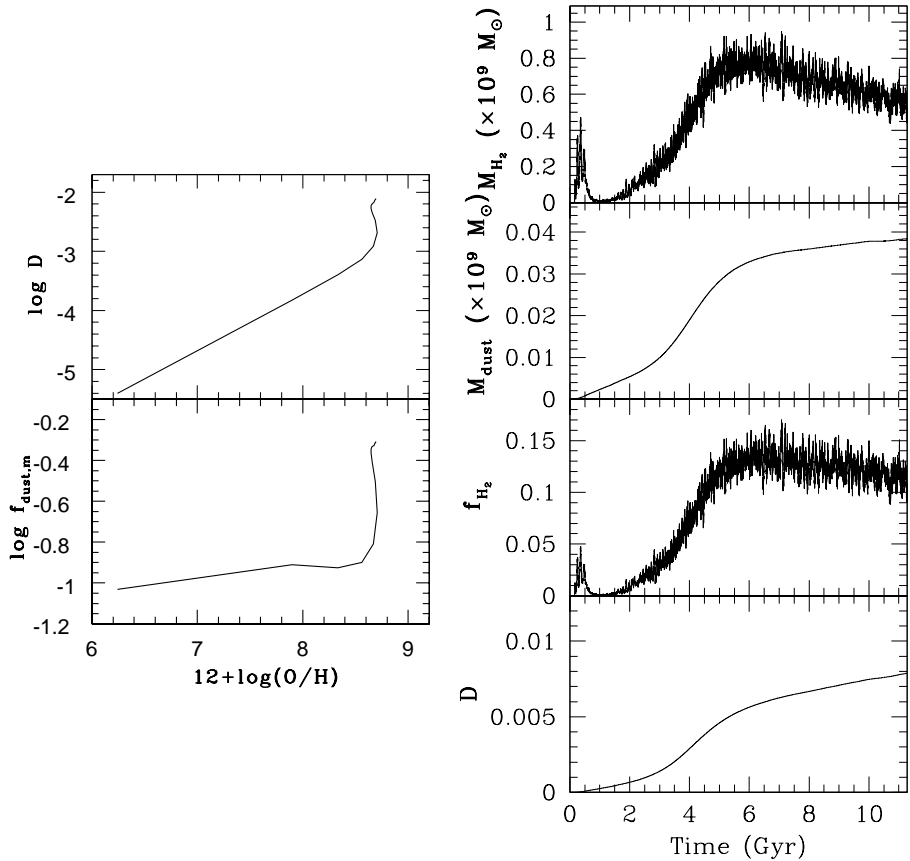,width=12.0cm}
\caption{
The left two panes are
for the time evolution of the forming disk galaxy in the fiducial model
 on the $A_{\rm O}-D$ and
$A_{\rm O}-f_{\rm dust,m}$ planes,
and the right four are for the time evolution  of
$M_{\rm H_2}$ (top), $M_{\rm dust}$ (the second from top),
$f_{\rm H_2}$ (the second from bottom), and $D$ (bottom).
$f_{\rm dust, m}$ is the mass fraction of metals (ejected from stars) that are locked up
in dust.
}
\label{Figure. 6}
\end{figure*}

\begin{figure*}
\psfig{file=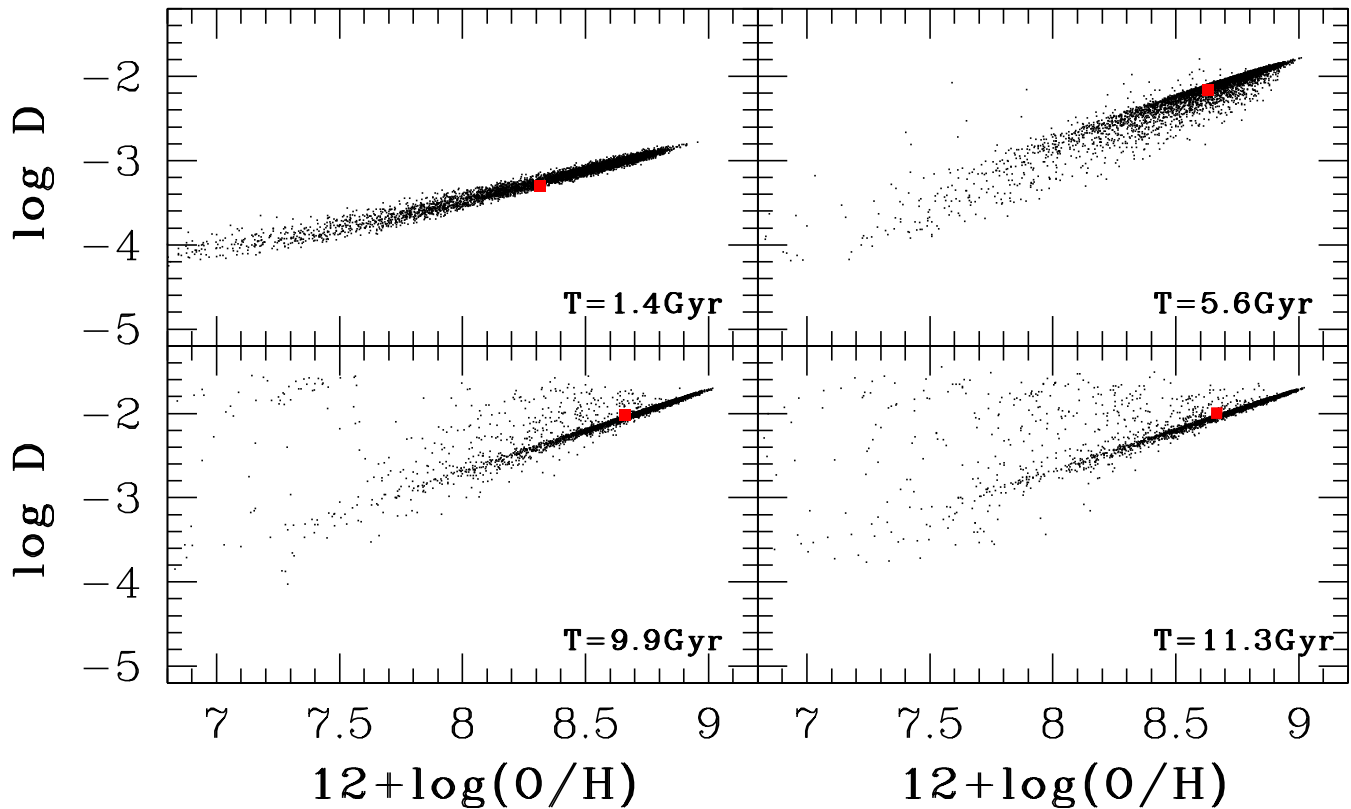,width=18.0cm}
\caption{
The locations of gas particles on the $A_{\rm O}-D$ plane for four representative
times steps in the fiducial model: $T=1.4$ Gyr (upper left), $T=5.6$ Gyr (upper right),
$T=9.9$ Gyr (lower left), and $T=11.3$ Gyr (lower right). The red square in each panel
indicates the mean values of $A_{\rm O}$ and $D$.
One of every 10 gas particles is shown in each panel.
}
\label{Figure. 7}
\end{figure*}

\begin{figure*}
\psfig{file=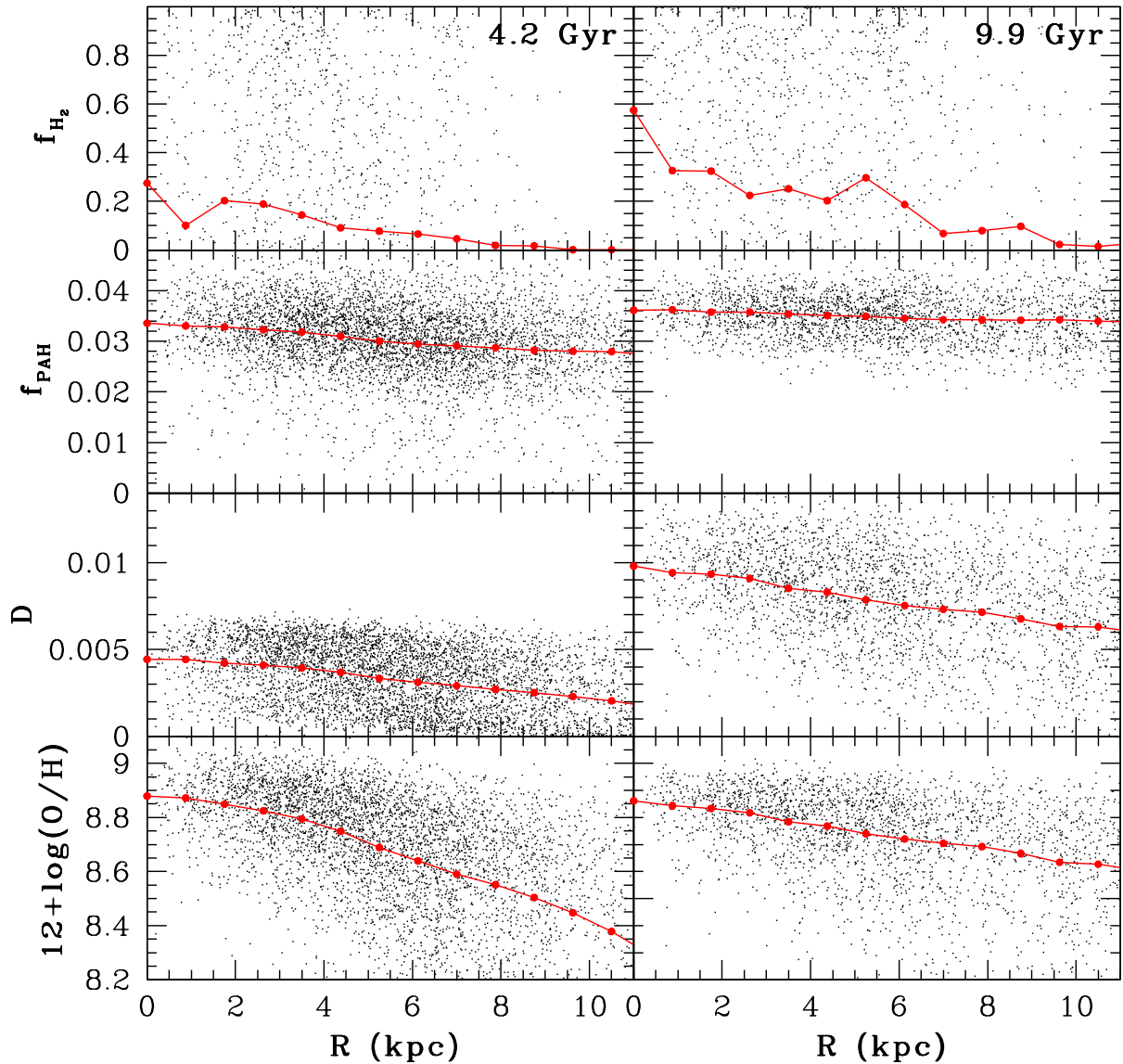,width=16.0cm}
\caption{
The radial gradients of $f_{\rm H_2}$ (top), $f_{\rm PAH}$ (the second from top),
$D$ (the second from bottom), $A_{\rm O}$ (bottom) for the two representative
time steps, $T=4.2$ Gyr (left) and $T=9.9$ Gyr (right) in the fiducial model.
Small black dots indicate  gas particles and red circles represent
the  mean values
for each radial bin.
One of every 10 gas particles is shown  in each panel.
}
\label{Figure. 8}
\end{figure*}

\begin{figure*}
\psfig{file=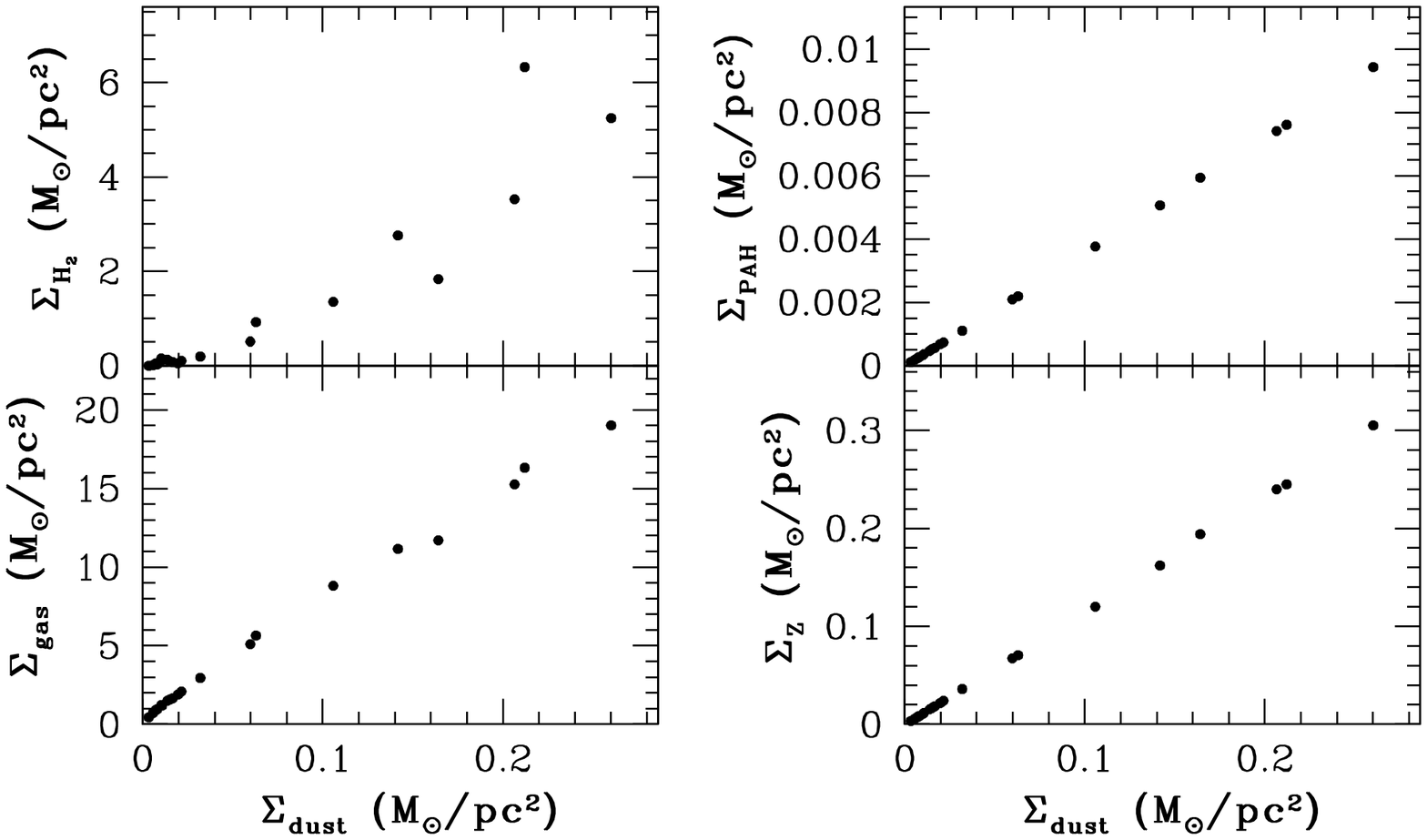,width=18.0cm}
\caption{
 Correlations of dust surface  densities ($\Sigma_{\rm dust}$) with
${\rm H_2}$ surface densities ($\Sigma_{\rm H_2}$, upper left),
PAH  surface densities ($\Sigma_{\rm PAH}$, upper right),
gas surface densities ($\Sigma_{\rm gas}$, lower left),
and metal surface densities ($\Sigma_{\rm Z}$, lower left)
at $T=11.3$ Gyr (i.e., final time step)
in the fiducial model.
 Each black dot represents the azimuthally averaged value
at each radial bin.
}
\label{Figure. 9}
\end{figure*}

\subsection{Initial conditions of galaxy formation}

Although we mainly investigate the evolution of forming disk galaxies embedded in
massive dark mater halos,
we also investigate physical properties of merger remnants that are morphologically
similar to early E/S0 galaxies.
This is because we discuss the observed differences in dust and ${\rm H}_2$ properties
between different Hubble morphological  types.

\subsubsection{Isolated disk galaxies}

Disk galaxies form from slow accretion of halo gas embedded in virialized massive
dark matter halos in the present study. The halo gas of
a forming disk galaxy is initially in hydrostatic
equilibrium determined by the mass distribution of the dark matter halo;
the gas thereafter collapses and accretes onto the central region
of the dark halo  owing to radiative cooling processes.
This disk formation model is essentially similar to those adopted by Kaufmann et al. (2007)
for the detailed investigation of angular momentum transfer in disk galaxy formation
and by RJ12 for dwarf galaxy formation.

The ratio of the dark matter halo mass ($M_{\rm dm}$)  to gaseous halo mass ($M_{\rm g}$)
in a forming disk galaxy is
fixed at 9 for most  models. The initial total halo mass is denoted
as $M_{\rm h}$ ($=M_{\rm dm}+M_{\rm g}$).
We adopt the NFW profile for the density distribution of the dark  matter 
halo suggested from CDM simulations:
\begin{equation}
{\rho}(r)=\frac{\rho_{0}}{(r/r_{\rm s})(1+r/r_{\rm s})^2},
\end{equation}
where  $r$, $\rho_{0}$, and $r_{\rm s}$ are
the spherical radius,  the characteristic  density of a dark halo,  and the
scale
length of the halo, respectively.
The $c$-parameter ($r_{\rm vir}/r_{\rm s}$, where $r_{\rm vir}$ is the virial radius)
is chosen for a given $M_{\rm h}$ according to the predicted $c$-$M_{\rm h}$ relation
in the $\Lambda$CDM simulations (e.g., Neto et al. 2007).

The gaseous halo initially has the same spatial distribution as the dark
matter and is assumed to be initially in hydrostatic equilibrium. 
The initial temperature of a halo gas particle is therefore determined by
the gas
density, total mass, and gravitational potential at the location of the
particle via Euler's equation for hydrostatic equilibrium (e.g., the
equation
1E-8 in Binney \& Tremaine 1987).
Therefore  gaseous temperature $T_{halo}(r)$ at radius $r$ from the center
of a disk galaxy can be described as:
\begin{equation}
T_{\rm halo}(r)=
\frac{m_{\rm p}}{k_{\rm B}}
\frac{1}{{\rho}_{\rm halo}(r)}
\int_{r}^{\infty}
{\rho}_{\rm halo}(r)
\frac{GM(r)}{r^2}
dr,
\end{equation}
where $m_{\rm p}$ $G$, and $k_{\rm B}$ are the proton mass,
the gravitational constant, and the Boltzmann constant, respectively,
and $M(r)$ is the total mass within $r$ determined by
the adopted mass distributions of dark matter and baryonic
components in the forming disk galaxy.

The dark matter halo is dynamically supported by isotropic velocity dispersion
and has no net angular momentum in the present study. The gas has a net 
angular momentum  with the radial distribution of specific
angular momentum ($j$) described as  $j \propto r$ 
that is consistent with the
prediction of CDM simulations (e.g., Bullock et al. 2001). 
The spin parameter ($\lambda = \frac{J|E|^{1/2}}{GM^{5/2}}$) is a free parameter
that controls the initial  total amount of rotation
and we mainly investigate $\lambda=0.038$ in the present study.
The angular rotation of a gas particle  (with the rotating axis being the $z$-axis)
is determined by the adopted $\lambda$, $j(r)$, and the position of the particle.

\subsection{The choice of dust parameters}

Since  no chemodynamical simulations have so far investigated variously 
different dust properties
of galaxies, it is not so clear what a reasonable set of dust model parameters 
(i.e., $\tau_{\rm acc}$ and $\tau_{\rm dest}$)
is for explaining the observed dust properties. 
We accordingly investigate many `low-resolution' ($N \approx 10^5$)
 models with different parameter values 
of $\tau_{\rm acc}$ and $\tau_{\rm dest}$ and  thereby try to  derive  a reasonable set of
the parameters that can explain observations.
We then perform  `high-resolution' ($N=10^6$) simulations and investigate physical properties
of gas, dust, and stars in galaxies by using the derived set of dust model
parameters. The low-resolution simulations have $N_{\rm dm}=10^5$ and 
$N_{\rm g}=2 \times 10^4$.

In determining reasonable $\tau_{\rm acc}$ and $\tau_{\rm dest}$,
we use the observed $D$ and $f_{\rm PAH}$ of the LMC, because a growing number
of observational data sets have been accumulated for the LMC (e.g., Meixner et al. 2010).
We adopt the LMC-type models with $M_{\rm h}=10^{11} {\rm M}_{\odot}$,  $\lambda=0.038$
and variously different $\tau_{\rm acc}$ and $\tau_{\rm dest}$
and investigate the time evolution of $D$ and $f_{\rm PAH}$ for 11.3 Gyr
and compare the final $D$ and $f_{\rm PAH}$ with the observations.
Fig. 2 show the results for some of the models as well as the observed $D$ and $f_{\rm PAH}$
with observational error bars. 
The observational data sets are from Meixner et al. (2010) and Leroy et al. (2011),
and the error bars of 0.2 dex are shown for $A_{\rm O}$, as done in Leroy et al. (2011)
for possible spatial variation of $A_{\rm O}$.
In these models, $\tau_{\rm dest}=0.5$ Gyr
and $R_{\rm PAH}=0.05$. 
The models with $\tau_{\rm acc}=0.13$ Gyr and 0.25 Gyr can better explain the
observed location of the LMC on the $A_{\rm O}-D$ plane than other ones.
The model with $\tau_{\rm acc}=0.13$ Gyr, however, shows $f_{\rm PAH}$ that is
significantly lower than the observed value. The model with $\tau_{\rm acc}$=0.25 Gyr
shows $f_{\rm PAH}$ that is consistent with the observed value.

Thus it is reasonable and realistic  for the present study to adopt 
($\tau_{\rm acc}$,$\tau_{\rm dest}$)=(0.25,0.5) Gyr in high-resolution simulations.
It is confirmed that as long as $\tau_{\rm dest}/\tau_{\rm acc} \approx 2$
(for $\tau_{\rm dest}=0.13-0.5$ Gyr),
then the observed $D$ and $f_{\rm PAH}$ can be reasonably well reproduced.
Given that the above set of model parameters
($\tau_{\rm acc}$,$\tau_{\rm dest}$)=(0.25,0.5) Gyr)  is consistent with those used 
for D98 that explains dust properties of the Galaxy self-consistently,
we mainly investigate the models with ($\tau_{\rm acc}$,$\tau_{\rm dest}$)=(0.25,0.5) Gyr
in the high-resolution simulations.

\subsection{Parameter study}
We mainly describe the results of  the `fiducial' model for which the model parameters
are described in Table 3. This fiducial model has a total mass of 
$M_{\rm h}=10^{11} {\rm M}_{\odot}$ that is similar to the mass of the LMC
before its first passage of the Galaxy's virial radius 
(i.e., before the loss of its initial mass; Bekki 2011 and 2012).
We adopt this fiducial model, because recent observational studies have provided
a rich amount of information on dust and ${\rm H_2}$ properties of the LMC
including 2D distributions of $D$, $f_{\rm PAH}$, and $f_{\rm H_2}$ across
the LMC (e.g., Kawamura et al. 2009; Meixner et al. 2010), which can be compared
with the fiducial model in detail.
We also present the results of some representative  models  with different parameters
and the range of model parameters investigated in the present study are shown
in Table 4.
For all models, the initial iron abundances ([Fe/H]) and dust-to-metal ratios
for gas particles
are set to be $-3$ and 0.1,  respectively. Owing to the adopted  non-zero initial dust mass
in gas,  ${\rm H_2}$ formation is possible from the very early stage of
disk galaxy formation. 

We investigate the time evolution of dust and ${\rm H_2}$ properties
for the last $\sim 11$ Gyr, which corresponds to the period of
disk growth via gas accretion after virialization of the dark matter halos. 
The present simulations are therefore somewhat idealized in the sense that
they can not describe the very early merging of sub-galactic clumps that
formed stellar halos of disk galaxies. However, we consider that these simulations
enable us to grasp some essential gradients of long-term dust and ${\rm H_2}$ evolution
in galaxies. We will adopt more realistic initial conditions of galaxy
formation based on a $\Lambda$CDM cosmology and thereby investigate physical properties
of dust and ${\rm H}_2$ 
in our forthcoming papers.
In the following,  $T$ in a simulation  represents the time that has elapsed since 
the simulation started.

\section{Results}

\subsection{The fiducial model}

\subsubsection{Time evolution of dust and ${\rm H_2}$}

 Figs. 3 and 4 show the time evolution of 2D distributions of gas, new stars, 
${\rm H_2}$, and dust 
($\mu_{\rm g}$, $\mu_{\rm s}$, $\mu_{\rm H_2}$, and $\mu_{\rm dust}$, respectively)
for the last 11.3 Gyr  in the fiducial model.
For clarity, the time evolution of surface mass densities 
(in a logarithmic scale) is shown for each component.
As a   gaseous disk forms through early rapid accretion of halo gas,  new stars can 
form 
from the central high-density gaseous regions of the disk ($T=1.4$ Gyr). 
Multiple explosions of SNe in the early burst phase can blow some fraction of the early
gas disk so that numerous  giant (kpc-sized) gaseous holes can be developed 
($T=2.8$ and 5.6 Gyr). 
As the disk grows by further  gas accretion,  
star formation can occur in the outer part of the disk
so that the disk size becomes larger ($T=5.6$ and 11.3 Gyr). The giant holes 
can be still  conspicuous  in the outer part of the final disk
owing to the presence of SN explosions ($T=11.3$ Gyr).
The final stellar disk has a disk-like structure with a higher density
surrounded by a diffuse halo-like or thick disk-like stellar component.
A strong stellar bar can not be formed during the $\sim 11$ Gyr evolution in this model.

 The formation of ${\rm H_2}$ is possible from high-density gaseous regions 
of the central part of the disk in the early phase of the disk formation ($T=1.4$ Gyr).
The simulated ${\rm H_2}$ gas shows clumpy structures, particularly, in later epochs
(e.g., $T=2.8$ and 5.6 Gyr),
and a more compact distribution in comparison with the total gas distribution.
As stars form from   ${\rm H_2}$ gas across the forming disk, 
a larger amount of dust can be produced so that the formation
efficiency of ${\rm H_2}$ can increase. As a result of this, the ${\rm H_2}$ distribution
becomes more widespread and a larger amount of ${\rm H_2}$ gas can be contained in the disk
($T=2.7$ and 5.6 Gyr). 
Owing to the lack of a strong bar,  the efficient gas-transfer to the central region
of the disk through dynamical effects of the bar on gas does not occur.
Consequently,
a very strong concentration of ${\rm H_2}$ gas
can not be seen in this model, though $f_{\rm H_2}$ is higher in inner regions of the disk.
The simulated distribution of dust does not look so clumpy as that of ${\rm H_2}$
gas throughout the disk formation and evolution.
Clearly, the gas disk has significantly smaller surface 
dust densities in the outer regions.

As shown in Fig. 5,  SFR can rapidly increase during the first 1 Gyr evolution to reach
as high as $1 {\rm M}_{\odot}$ yr$^{-1}$ in the disk. The SFR then decreases slowly to 
finally become less than $0.1 {\rm M}_{\odot}$ yr$^{-1}$.
In this lower mass disk model,  SN feedback effects play a vital role in suppressing
star formation within the disk. 
Comparison between the fiducial model and the one with H-dependent SF recipe
suggests that there is no remarkable difference in the long-term SF histories between
the two models. However, there are some significant differences in SF histories
between  low-mass models
($M_{\rm h} \le 10^{10} {\rm M}_{\odot}$) with H- and ${\rm H_2}$-dependent SF recipes,
as described later in Appendix C.
SFR can more violently change in the H-dependent SF model for the first few Gyr 
in comparison with the fiducial one.
This violent change is due to the fact that stars can form  
in small high-density
clumps in a bursty manner and then 
star formation can be truncated by strong supernova feedback effects.
In the fiducial model with the ${\rm H_2}$-dependent SF,
the SFRs in small high-density regions can not so steeply
rise, because the ${\rm H_2}$ densities are not so high
(in spite of the high total gas densities).
Slower consumption of gas in the clumps therefore ends up with the less
violent change of  the total SFR in
the fiducial model. 
There are no major differences in
the final total stellar mass ($M_{\rm star}$) and gas mass fraction
($M_{\rm g}/(M_{\rm g}+M_{\rm s}$)) between the models with
H- and ${\rm H_2}$-dependent  
SF recipes.

Fig. 6 shows the time evolution of the disk galaxy on the $A_{\rm O}-D$ 
and $A_{\rm O}-f_{\rm dust,m}$ planes and the time evolution of
the total dust ($M_{\rm dust}$) and ${\rm H_2}$ masses ($M_{\rm H_2}$) and
mean $D$ and $f_{\rm H_2}$. Here
the mean $D$, $A_{\rm O}$, and $f_{\rm dust,m}$ values
averaged among {\it all gas particles} are shown. 
$D$ of the disk  evolves  along a straight line ($D \propto A_{\rm O}$)
until $A_{\rm O} \approx 8.6$ and then very steeply increases to become $D \approx 0.01$.
The origin of this steep rise in $D$ is explained as follows.
In the late evolution phase of the disk (12+$\log {\rm O/H} > 8.6$),  
the star formation rate can drop significantly owing to the lower gas density and
thus the number of SN explosions becomes small.
As a result of this,
dust destruction by SNe can become much less efficient.  
On the other hand,
the dust can still grow  rapidly owing to accretion of gas-phase metals
onto the pre-existing grains (i.e., at the expense of the metals). 
Accordingly the gas-phase abundance ($A_{\rm O}$) can  slightly decrease during
this rapid $D$ increase.  On the other hand,  $f_{\rm dust,m}$ can very gradually
increase until $A_{\rm O} \approx 8.6$ and then very steeply increases
almost without changing $A_{\rm O}$.

Although ${\rm H_2}$ can be rapidly produced
during the first 0.5 Gyr evolution, the ${\rm H}_2$ gas can be efficiently consumed 
by the first burst of star formation in the forming disk. $M_{\rm H_2}$ can 
slowly increase after the early high SFR phase ($T<2$ Gyr) and takes a peak value
($\sim 8 \times 10^8 {\rm M}_{\odot}$) 
around $T=6$ Gyr. After the peak, both $M_{\rm H_2}$ and $f_{\rm H_2}$ decrease
slowly owing to gas consumption by star formation.
The decrease for the last 5 Gyr is  more clearly seen in $M_{\rm H_2}$
in comparison with $f_{\rm H_2}$.
Both $M_{\rm dust}$ and $D$ can slowly increase even after the peak formation phase
of ${\rm H}_2$ owing to dust mass growth via accretion of gas-phase metals in ISM.

Fig.7 shows the distributions of gas particles on the $A_{\rm O}-D$ plane
at four different time steps ($T=1.4$, 5.6, 9.9, and 11.3 Gyr).
Only gas particles that are within the  disk ($z \le 1$ kpc)
are plotted so that $A_{\rm O}-D$ relations for the disk can be
investigated.
The particles initially form a narrow line,  the slope of which 
is significantly shallower than the observed slope  of the $A_{\rm O}-D$ relation
($D \propto A_{\rm O}$; $T=1.4$ Gyr).
As time passes by,  the $A_{\rm O}-D$ relation becomes steeper 
and gas particles become more widely spread on the $A_{\rm O}-D$ plane
($T=5.6$ Gyr).  A larger number of metal-poor particles with $A_{\rm O} <8$ can 
have higher $D$ ($>0.01$), because they are located where SN explosions are rare
and therefore dust can more efficiently grow by accretion of gas-phase metals.
The majority of stars can be finally located within  a  narrow line
($D \propto A_{\rm O}$) on the plane at $T=11.3$ Gyr, though dispersions of $D$ for a given
$A_{\rm O}$ is large.

\begin{figure}
\psfig{file=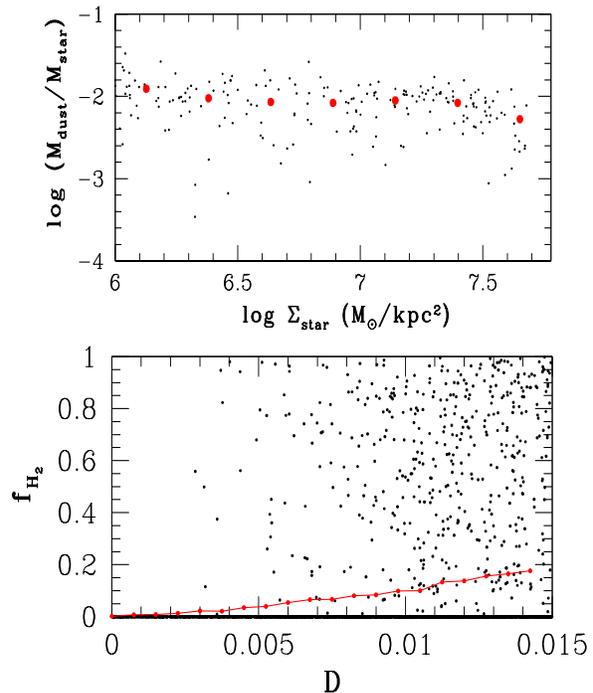,width=8.0cm}
\caption{
{\it Upper panel:}
 A correlation of the  stellar surface densities ($\Sigma_{\rm star}$)
with the dust-to-star mass ratios
($R_{\rm dust} =M_{\rm dust}/M_{\rm star}$) in individual local regions 
at $T=11.3$ Gyr in
the fiducial model.
Smaller black dots are for individual local regions and large red
dots indicate the mean $R_{\rm dust}$ at each radial bin.
{\it Lower panel:}
Plots of gas particles on the $D-f_{\rm H_2}$ plane at $T=11.3$ Gyr in the
fiducial model. Red dots indicate the mean $f_{\rm H_2}$ at each $D$ bin.
The red line connecting red dots means that ${\rm H_2}$ is higher for
gas with higher $D$.
One of every 10 gas particles is shown for clarity.
About 81\% of gas particles shown here have $f_{\rm H_2}=0$ (i.e.,
no molecular hydrogen).
}
\label{Figure. 10}
\end{figure}

\begin{figure}
\psfig{file=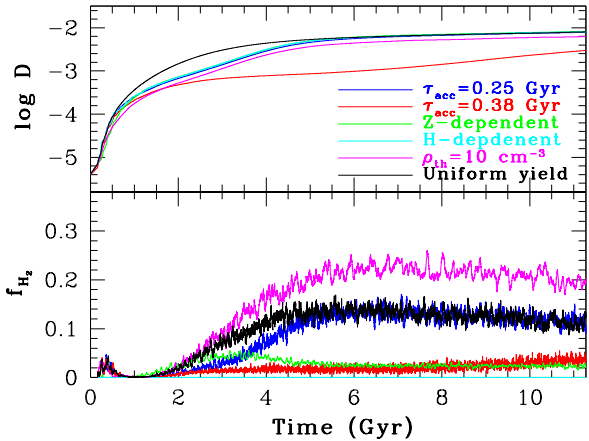,width=8.0cm}
\caption{
The time evolution of $D$ (upper) and $f_{\rm H_2}$ (lower) for six models with 
$\tau_{\rm acc}=0.25$ Gyr (blue), $\tau_{\rm  acc}=0.38$ Gyr (red),
Z-dependent SF recipe (green), H-dependent one (cyan), $\rho_{\rm th}=10$ cm$^{-3}$ (magenta),
and uniform dust yield (black).
The blue line corresponds to the fiducial model.
The basic parameters for initial conditions of galaxy formation
(e.g., $M_{\rm h}$ and $\lambda$)  are exactly the same between
the six models.
}
\label{Figure. 11}
\end{figure}

\subsubsection{Radial gradients}

Fig. 8 shows that the radial gradients of $f_{\rm H_2}$, $f_{\rm PAH}$,
$D$, and $A_{\rm O}$ evolve significantly with time.
Dust abundances can more rapidly increase in the inner disk regions owing
to their more rapid star formation and chemical evolution histories, which end up with
more rapid increases of ${\rm H_2}$ formation rates in the inner regions.
Consequently, the radial gradient of $f_{\rm H_2}$ becomes steeper between $T=4.2$ and 9.9 Gyr,
and the ${\rm H_2}$ gradients are always negative (i.e., higher in the inner regions). 
A  weak negative radial gradient of $f_{\rm PAH}$ 
($\sim -6 \times 10^{-4}$ kpc$^{-1}$) can be seen 
at $T=4.2$ Gyr, and this gradient becomes even weaker  ($\sim -2 \times 10^{-4}$ kpc$^{-1}$)
at $T=9.9$ Gyr. The $f_{\rm PAH}$ dispersion for a given $R$ appears to be larger
at $T=4.2$ Gyr than at $T=9.9$ Gyr. The almost lack of $f_{\rm PAH}$ gradient 
is discussed later in terms of the recent observational results of PAH dust in the LMC.

The disk galaxy shows a negative radial $D$ gradient at $T$=4.2 and 9.9 Gyr
($\sim -2\times 10^{-4}$ kpc$^{-1}$ and $-4 \times 10^{-4}$ kpc$^{-1}$, respectively)
and the gradient becomes steeper at later epochs. 
A negative radial $A_{\rm O}$ gradient (i.e., gas-phase metallicity gradient)
can be clearly seen at the two epochs; the gradient becomes shallower at later epochs.
Dispersions of four properties, $f_{\rm H_2}$, $f_{\rm PAH}$, $D$, and $A_{\rm O}$
are large for a given radius,
which reflects the fact that chemical evolution and star formation histories
can be quite different between different local regions.
The derived characteristic negative radial gradients of $f_{\rm H_2}$,
$D$, and $A_{\rm O}$ can be seen in other models, though the amplitudes
of the gradients  depend on model parameters.

\subsubsection{Dust-gas-star correlations}

Fig. 9 shows that there are clearly positive 
correlation of $\Sigma_{\rm gas}$ and $\Sigma_{\rm H_2}$ with
$\Sigma_{\rm dust}$ in the final disk ($T=11.3$ Gyr), 
though the $\Sigma_{\rm H_2}-\Sigma_{\rm dust}$ correlation
appears to be weaker. The almost constant $\Sigma_{\rm gas}/\Sigma_{\rm dust}$ ($\sim 100$)
reflects the mean $D$ of the disk and therefore the slope of
the $\Sigma_{\rm gas}-\Sigma_{\rm dust}$ relation depends strongly on the dust abundances
of simulated disk galaxies (and thus on their chemical and dust evolution histories).
 Leroy et al. (2011) investigated correlations of surface 
densities of dust ($\Sigma_{\rm dust}$) with those
of total gas ($\Sigma_{\rm gas}$) and
${\rm H}_2$ gas derived from CO for galaxies in the Local Group.
They found that  a stronger $\Sigma_{\rm dust}-\Sigma_{\rm gas}$ correlation
can be more clearly in the LMC in comparison with
$\Sigma_{\rm dust}-\Sigma_{\rm H_2}$ correlation.
These observational results
are consistent at least
qualitatively with the present results shown in Fig. 9.

Fig. 10 shows that there is no/little correlation between $\Sigma_{\rm star}$
(local stellar density)
and $R_{\rm dust}=M_{\rm dust}/M_{\rm star}$ (dust-to-star mass ratio)
in local regions of the final disk: this could
be regarded as 
a very weak correlation in the sense that local regions with higher $\Sigma_{\rm star}$
have smaller $M_{\rm dust}/M_{\rm star}$.
This very weak correlation for $\log (\Sigma_{\rm star}) < 7.5$ ${\rm M}_{\odot}$ kpc$^{-1}$
can be seen in other disk models.
However, the remnants of gas-rich major mergers shows a slightly  different 
$\Sigma_{\rm star}-R_{\rm dust}$ relation for higher $\Sigma_{\rm star}$ (Bekki 2013),
which could be an important difference between late- and early-types
galaxies.

Fig. 10 also shows that there is a positive correlation
between $f_{\rm H_2}$ and $D$ in local regions of the final disk, though dispersion in
$f_{\rm H_2}$ is large for a given $D$. The derived larger $f_{\rm H_2}$ for higher
$D$ (or higher $D$ in more ${\rm H_2}$-rich ISM)
is reasonable, given that a larger amount of ${\rm H_2}$ 
can be produced owing to higher $R_{\rm gr}$ for higher $D$ in the present ${\rm H_2}$
formation model.
This result has an important implication on the evolution of radial $D$ gradients
of galaxies in clusters of galaxies, which is discussed later.

\begin{figure*}
\psfig{file=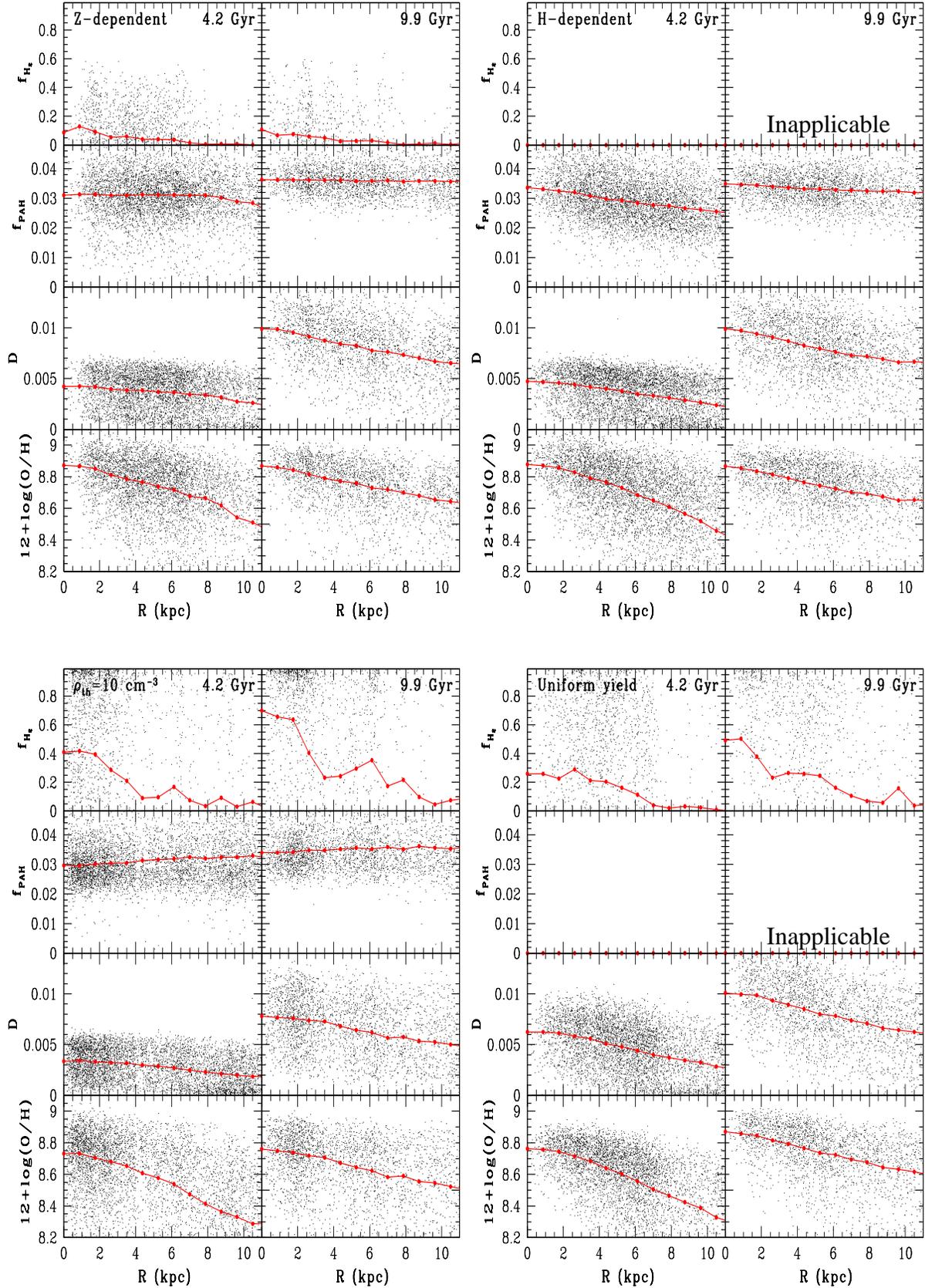,width=16.0cm}
\caption{
The same as Fig. 8 but for different models with
Z-dependent SF recipe (upper left),
 H-dependent one (upper right), $\rho_{\rm th}=10$ cm$^{-3}$ (lower left),
and uniform dust yield (lower right).
In the H-dependent model,  $f_{\rm H_2}$ is not estimated at all
(it is set to be 0).
The  PAH dust evolution is not included in the uniform dust model
(i.e., $f_{\rm PAH}=0$). Thus `inapplicable'  
is shown for the relevant panels of these models for clarity.
}
\label{Figure. 12}
\end{figure*}

\subsection{Parameter dependences}

Here we briefly summarize the governing dependences of dust and ${\rm H}_2$ properties,
and their correlations with other  key model
parameters (e.g., $\tau_{\rm acc}$ and $M_{\rm h}$).

\subsubsection{Dust parameters}

The time evolution and final values of $D$ and $f_{\rm H_2}$ depend strongly
on $\tau_{\rm acc}$. As shown in Fig. 11, although the very early evolution
of $D$ and $f_{\rm H_2}$ ($T<1$ Gyr) is not different between the two models
with $\tau_{\rm acc}=0.25$ and 0.38 Gyr,  the model with shorter $\tau_{\rm acc}$
(i.e., more rapid dust growth) shows more efficient production of dust and 
${\rm H_2}$ and thus has larger values of $D$ and $f_{\rm H_2}$. 
For these two models, a factor of $\sim 0.5$ difference in $\tau_{\rm acc}$
ends up with a factor of $\sim 3$ difference in $D$ and $f_{\rm H_2}$.
This strong $\tau_{\rm acc}$-dependence of $D$ and $f_{\rm H_2}$ implies
that the  modeling of dust growth is very important for the formation and evolution
of dust and ${\rm H_2}$: $\tau_{\rm acc}$ needs to be carefully chosen
when $f_{\rm H_2}$ in galaxies is investigated by using numerical simulations with dust.
D98 suggested that  ISM-phase averaged dust accretion
timescale can be as long a $\sim (1-2) \times 10^8$ yr, which is roughly consistent
with the adopted reasonable value of $\tau_{\rm acc}$.

Fig. 11 shows  that there are no significant differences in
the evolution of  $D$ and $f_{\rm H_2}$
between the fiducial models with D98-type 
and the model in which uniform dust yields are adopted (i.e., no PAH dust production),
as long as $F_{\rm dust}=0.1$
is adopted as  in  previous one-zone models.
Fig. 12 also shows that the final disks in the two models have very similar radial
profiles of ${\rm H_2}$, $D$, and $A_{\rm O}$ (See Fig. 8 for comparison).
These results imply that the evolution of  ${\rm H_2}$ and dust
is not  sensitive to the inclusion of of dependence to dust production rate
and dust composition  on stellar masses. A reasonable value
of $F_{\rm dust}$, however, should be carefully chosen in a simpler model
with uniform dust yields.

\subsubsection{SF recipes}

The models with  $M_{\rm h}=10^{11} {\rm M}_{\odot}$ yet different
SF recipes and SF threshold gas densities  $\rho_{\rm th}$
are investigated. The three principle results on 
the dependences of the time evolution
and final values of $D$ and $f_{\rm H_2}$
on  SF recipes, which are shown in Figs. 11 and 12,
are summarized as follows.
First,  there are no major differences in the $D$ evolution between the models
with the D-dependent (i.e., fiducial model) and H-dependent SF recipes,
which suggests that $D$ evolution is relatively insensitive to the modeling of global
star formation in galaxies.

Second,  although $D$ evolution is essentially the same between
the models with the D-dependent (i.e., fiducial model) and Z-dependent
SF recipes, the $f_{\rm H_2}$ evolution is significantly different between the two
in the sense that $f_{\rm H_2}$ is significantly lower in the model
with the Z-dependent SF recipe.  The origin of the 
lower $f_{\rm H_2}$ is discussed later in \S 4.1. 
Third, a  factor of 10 difference in  $\rho_{\rm th}$
($1 \le \rho_{\rm th} \le 10$  atom cm$^{-3}$) can yield a factor of $\sim 2$
difference in $f_{\rm H_2}$ in the present study.  
However,
the evolution and the final value of $D$ do not depend so strongly on $\rho_{\rm th}$:
$D$ is only slightly smaller in the model with higher $\rho_{\rm th}$ owing
to stronger suppression of star formation that produces dust.

\begin{figure}
\psfig{file=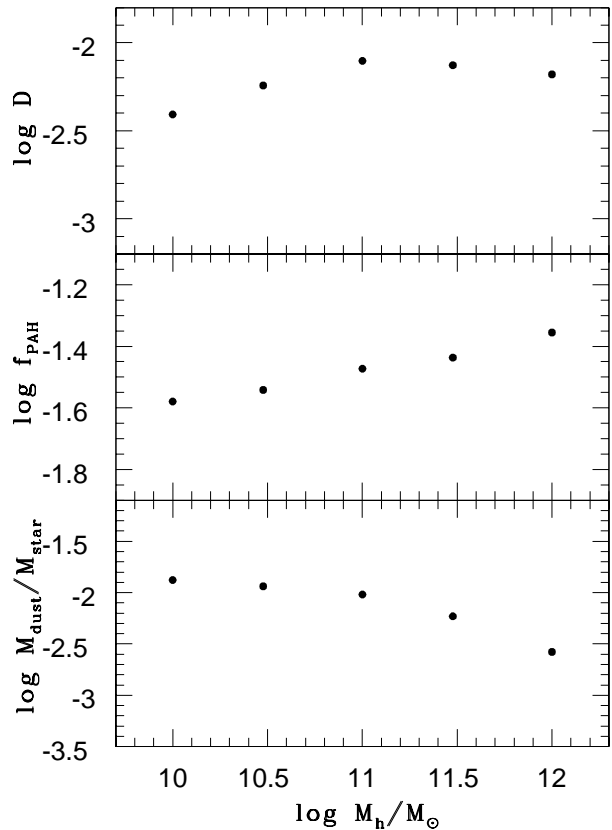,width=8.0cm}
\caption{
The dependences of $D$ (top), $f_{\rm PAH}$ (middle), 
and $R_{\rm dust}$ ($=M_{\rm dust}/M_{\rm star}$; bottom)
on $M_{\rm h}$ for $\lambda=0.038$.
}
\label{Figure. 13}
\end{figure}

\subsubsection{Galaxy mass and spin parameter}

Fig. 13 shows that  $D$ is larger in more massive galaxies (i.e., larger $M_{\rm h}$)
for $M_{\rm h} \le 10^{11} {\rm M}_{\odot}$. This $M_{\rm h}-D$ relation is not clearly
seen in more luminous disk galaxies with $M_{\rm h}>10^{11} {\rm M}_{\odot}$: the
$M_{\rm h}-D$ relation is quite flat.
The final $f_{\rm PAH}$ depends weakly on $M_{\rm h}$ such that $f_{\rm PAH}$ is larger for
larger $M_{\rm h}$. The derived $M_{\rm h}-f_{\rm PAH}$ relation may well be 
approximated as $f_{\rm PAH} \propto M_{\rm h}^{0.12}$. Since more massive galaxies
have higher chemical abundances in the present study,
this result means that galaxies with higher $A_{\rm O}$ can have higher $f_{\rm PAH}$.
High-mass galaxies have higher metallicities so that $f_{\rm H_2}$ can be higher.
The time evolution of $f_{\rm H_2}$ for high-mass ($M_{\rm h}=10^{12} {\rm M}_{\odot}$)
and low-mass ($M_{\rm h}=10^{10} {\rm M}_{\odot}$) disk galaxies is given and briefly discussed
in Appendix C.

The dust-to-star mass ratio ($R_{\rm dust}$) depends on $M_{\rm h}$
such that it is smaller for larger $M_{\rm h}$. Also, the $M_{\rm h}-R_{\rm dust}$
relation appears to be different below and above $M_{\rm h}=10^{11} {\rm M}_{\odot}$:
$R_{\rm dust}$ depends more strongly on $M_{\rm h}$ for $M_{\rm h}>10^{11} {\rm M}_{\odot}$.
The final disks in the models 
with higher  $\lambda$ ($0.06$) show lower $M_{\rm H_2}$, $M_{\rm dust}$, $f_{\rm H_2}$,
and $D$, mainly because larger disks with lower mass densities can be formed in the models.
Negative radial gradients of dust and ${\rm H_2}$ properties and dust-${\rm H_2}$ 
correlations clearly seen in the fiducial model can be also seen in the models with different
$M_{\rm h}$ and $\lambda$.

 Fig. 14 shows that irrespective of galaxy halo masses ($M_{\rm h}$),
final $\Sigma_{\rm gas}$ positively correlate with $M_{\rm dust}$ 
(i.e., $\Sigma_{\rm gas} \propto \Sigma_{\rm dust}$) in disk galaxies.
The slopes of the correlations are steeper in galaxies with lower $M_{\rm h}$  owing to
their lower dust-to-gas ratios, and the maximum $\Sigma_{\rm dust}$ is larger
for larger $M_{\rm h}$. 
Positive correlations between $\Sigma_{\rm dust}$ and $\Sigma_{\rm H_2}$ can be
seen, though they are significantly weaker in comparison with 
the $\Sigma_{\rm dust}-\Sigma_{\rm gas}$ relations. The derived stronger
correlations in 
the $\Sigma_{\rm dust}-\Sigma_{\rm gas}$ relations are consistent with
observations by Leroy et al. (2011). The comparative model
with $\tau_{\rm acc}=0.38$ Gyr, in which dust growth is slower,
shows a steeper slope 
in the $\Sigma_{\rm dust}-\Sigma_{\rm gas}$ relation
and a smaller maximum value of $\Sigma_{\rm dust}$.
These results for the comparative model imply that dust accretion processes
can be a key parameter that controls
$\Sigma_{\rm dust}-\Sigma_{\rm gas}$ and
$\Sigma_{\rm dust}-\Sigma_{\rm H_2}$ relations in disk galaxies.

\begin{figure*}
\psfig{file=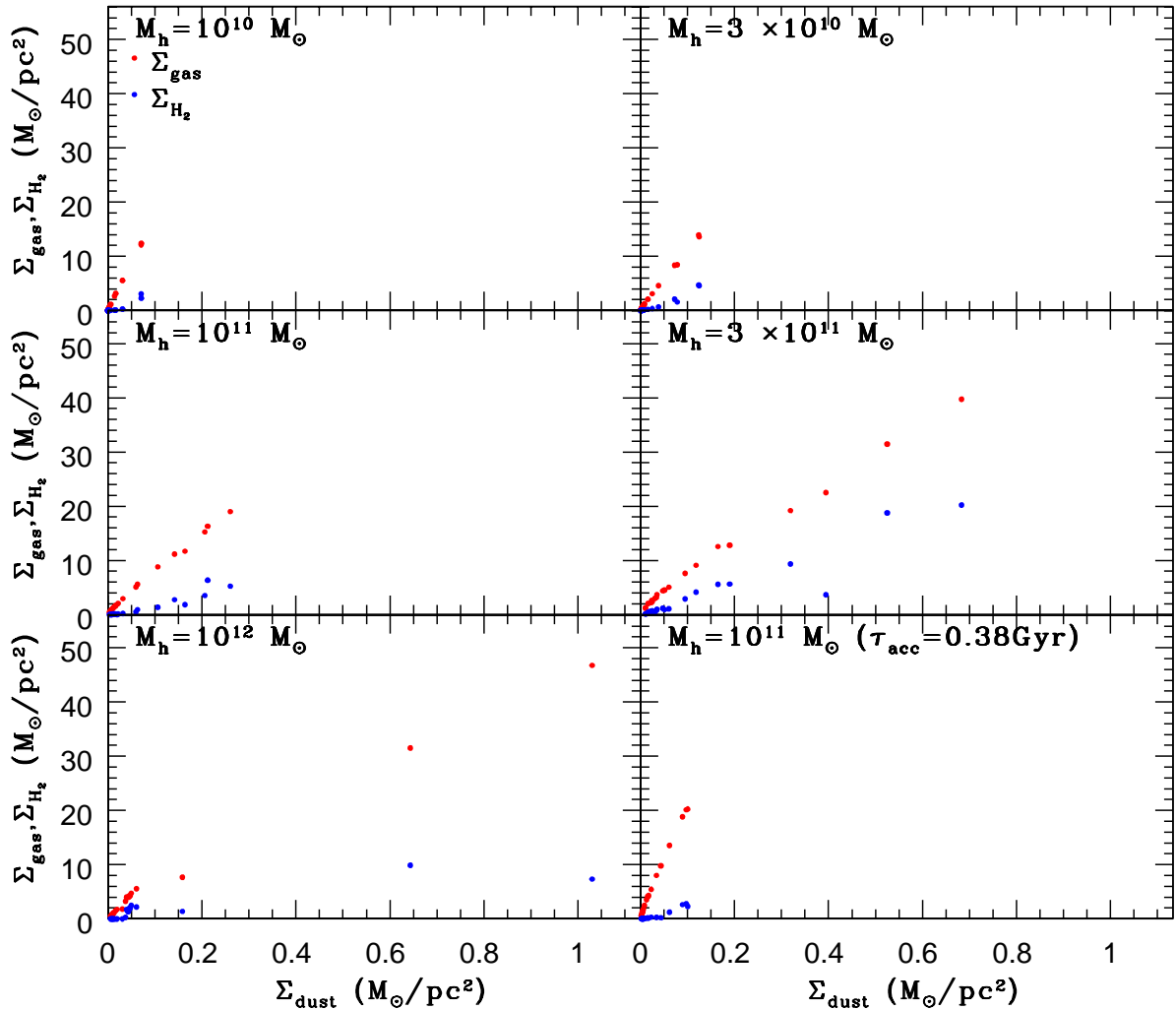,width=16.0cm}
\caption{
Correlations of dust surface densities ($\Sigma_{\rm dust}$) 
with total gas  ($\Sigma_{\rm gas}$, red dots)
and ${\rm H_2}$ ones ($\Sigma_{\rm H_2}$, blue dots) for  five models
with different $M_{\rm h}$ for $\lambda=0.038$.  For comparison,  
the model with $M_{\rm h}=10^{11} {\rm M}_{\odot}$ and $\tau_{\rm acc}=0.38$ Gyr
is shown.
}
\label{Figure. 14}
\end{figure*}

\section{Discussion}

\subsection{Necessary to model dust evolution in predicting star formation
histories ?} 

Previous chemical evolution models clearly showed that dust-to-metal
ratios ($D_{\rm z}$) of galaxies  evolve
with time especially  in the early evolution phases of galaxies
(e.g., Hirashita 1999;  Inoue 2003; Calura et al. 2008). These previous
models are one-zone,  in which star formation models are rather idealized
(e.g., non-inclusion of  ${\rm H}_2$ formation models) and dynamical evolution controlling
star formation processes is not explicitly included.
Therefore it is important for the present study to investigate how the  mass fraction
of metals that are locked up in dust  ($f_{\rm dust,m}$) can evolve with
time in the present 
chemodynamical model.
Fig. 14,  showing the $f_{\rm dust, m}$ evolution for four representative models 
with different $\tau_{\rm acc}$ and $M_{\rm h}$, confirms that $f_{\rm dust,m}$ can
rapidly change in the early evolution of galaxies ($T<4$ Gyr), though 
the means by which   $f_{\rm dust,m}$ changes
depends on the model parameters. This result strongly suggests that 
the total amount of dust in galaxies can not be linearly  proportional
to their  metallicities ($Z$) especially  in early galaxy formation phases.

Recent numerical simulations of galaxy formation
and evolution  with ${\rm H}_2$-regulated star formation (e.g., P06 and K12)
assumed that dust abundances ($D$) of galaxies are linearly proportional to $Z$
and the proportionality constant does not vary with time and location in galaxies.
The results shown in Fig. 14 suggests that this assumption is not realistic, especially,
in the early formation phases of galaxies. 
An important question here is  how  different
the predicted properties of galaxies are between numerical simulations with Z-dependent
(e.g., P06 \& K12) and the present 
D-dependent star formation models.
Fig. 15 shows the time evolution of star formation rates and dust properties
($D$ and $f_{\rm PAH}$) in the two models with $M_{\rm h}=10^{11} {\rm M}_{\odot}$
and $D$- and $Z-$dependent star formation
recipes.  Although the SFR can more violently change with time in the first
$\sim 2$ Gyr evolution for the model with the  Z-dependent SF recipe, 
the overall SFR is very
similar  between these  two models. As a result of this, the time evolution
of dust properties is very similar between the two models.  
This similarity in the time evolution of SFR, D, and $f_{\rm PAH}$ can be 
seen in models with greater  $M_{\rm h}$ ($ \ge 10^{11} {\rm M}_{\odot}$).

These results imply that the Z-dependent SF recipe is reasonable and realistic enough
to investigate the time evolution of SFRs (regulated by ${\rm H}_2$ formation and
evolution) in galaxies,
given that additional new parameters such as $\tau_{\rm acc}$ and $\tau_{\rm dest}$
should be considered in the $D$-dependent SF recipe.
One might not have to adopt this recipe, 
unless one wishes to
investigate dust properties such as radial gradients of $D$ and $f_{\rm PAH}$
in detail. 
One of advantages of the present chemodynamical model is that we can discuss
the observed
gas-phase metallicities, dust properties, and dynamical properties of galaxies
in a fully self-consistent manner.
Thus we can make the most of the present chemodynamical model when we investigate correlations 
between physical properties of dust, gas, and stars in galaxies.

 However, as shown in Fig. 11,
the final $f_{\rm H_2}$ is significantly lower in the Z-dependent SF model
than in the D-dependent one. The derived low $f_{\rm H_2}$ ($ \sim 0.05$) is 
not  consistent with $f_{\rm H_2}$ fractions observed in less luminous
disk galaxies like the LMC (e.g., $f_{\rm H_2} \approx  0.09-0.16$; van den Bergh 2000).
Dust-to-metal ratios can increase
during chemical evolution of galaxies
so that ${\rm H_2}$ production on dust grains can become more efficient.
Since the Z-dependent model does not include this evolutionary effect,
it can underpredict $f_{\rm H_2}$ in the present SPH simulations.
This does not mean that the Z-dependent model has a problem in predicting
$f_{\rm H_2}$.
The time evolution of $f_{\rm H_2}$ depends
on the adopted $\tau_{\rm c}$ (i.e., optical depth) and the 
fixed initial dust-to-metal ratio in
the Z-dependent SF model (KMT09).
If different values of these two parameters are chosen in the Z-dependent model,
the final $f_{\rm H_2}$
might well become  more similar to that derived  in the D-dependent model
and thus to the observed one.
Accordingly, we do not regard  the significant $f_{\rm H_2}$ difference
between the two models as a serious problem: depending on which model we adopt,
we need to carefully chose the model parameters (e.g., $\tau_{\rm dust}$ and
$\tau_{\rm c}$ etc).

\subsection{Origin of dust scaling relations}
We here discuss the observed three key dust scaling relations: correlations
between  $A_{\rm O}$ and $D$ (e.g., Galametz et al. 2011; Leroy et al. 2011),
between $M_{\rm star}$ and $R_{\rm dust}=M_{\rm dust}/M_{\rm star}$ 
(e.g., Corbelli et al. 2012;
Cortese et al. 2012), and between $M_{\rm H_2}$ and $M_{\rm dust}$ (Corbelli et al. 2012). 

\subsubsection{$A_{\rm O}-D$ relation}

Recent observational studies have confirmed that galaxies with higher $A_{\rm O}$
are more likely to show higher $D$,
which is approximated roughly as $D \propto A_{\rm O}$
(e.g., Galametz et al. 2011; Leroy et al. 2011). 
Although the observed $A_{\rm O}-D$ relation
is not strong and the observational estimation of dust masses of galaxies
has some uncertainties, 
it is important for the present study to confirm whether the relation  
can be well reproduced by the present new model with dust formation and evolution.
Fig. 17 shows the locations of six simulated disk galaxies different $M_{\rm h}$ 
($=3\times 10^9, 10^{10}, 3\times 10^{10}, 10^{11}, 3\times 10^{11}, 10^{12}
{\rm M}_{\odot}$) on the $A_{\rm O}-D$ plane.
For comparison, 
the location of
the Galaxy in the one-zone model by D98 
and those of dwarf galaxies at $A_{\rm O}=8.2$, 8.4, and 8.6 in the best one-zone model
with the $\chi$ parameter being 30 by Lisenfeld \& Ferrara (1998)
are plotted in this figure.

Clearly, the simulated galaxies have  a $A_{\rm O}-D$ relation very similar 
to the observed relation, though the simulated $D$ is slightly smaller than
the observed one for a given $A_{\rm O}$.  Given the possible observational
errors in the estimation of $D$,  this similarity suggests that
the present new model can explain very well the observed $A_{\rm O}-D$ relation.
The present simulations predict systematically higher $D$ for a given $A_{\rm O}$
for lower $A_{\rm O}$ ($<8.4$) in comparison with the one-zone models. 
The observed large dispersion 
in $D$ at lower $A_{\rm O}<8.4$ for a given $A_{\rm O}$  has not been well reproduced
by the present models and thus needs to be investigated in our future studies.

In the present model,  supernova feedback effects play a vital role in 
determining global star formation histories of galaxies
and thus the  total amount of gas that is converted into new stars in galaxies.
Supernova feedback effects can less strongly suppress star formation in more massive
galaxies owing to deeper gravitational potential wells. As a result of this
chemical enrichment,  dust production can 
proceed more efficiently in more massive galaxies.
Therefore, more massive galaxies can finally have higher metallicities, $D$, and  
mass-ratios  of final stars  to initial gas.
The simulated $A_{\rm O}-D$ relation due to this effectiveness of
supernova feedback is strongly dependent on $M_{\rm h}$. Thus the present study
is the first chemodynamical simulation that has successfully  reproduced the observed
$A_{\rm O}-D$ relation.

\begin{figure}
\psfig{file=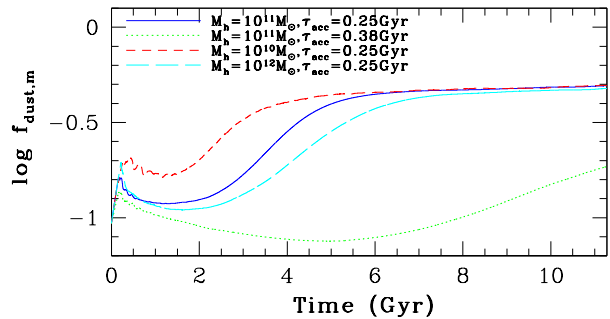,width=8.0cm}
\caption{
The time evolution of $f_{\rm dust,m}$ for different models with
$M_{\rm h}=10^{11} {\rm M}_{\odot}$ and $\tau_{\rm acc}=0.25$ Gyr (solid, blue),
$M_{\rm h}=10^{11} {\rm M}_{\odot}$ and $\tau_{\rm acc}=0.38$ Gyr (dotted, green),
$M_{\rm h}=10^{10} {\rm M}_{\odot}$ and $\tau_{\rm acc}=0.25$ Gyr (short-dashed, red),
and $M_{\rm h}=10^{12} {\rm M}_{\odot}$ and $\tau_{\rm acc}=0.25$ Gyr (long-dashed,  cyan).
}
\label{Figure. 15}
\end{figure}

\begin{figure}
\psfig{file=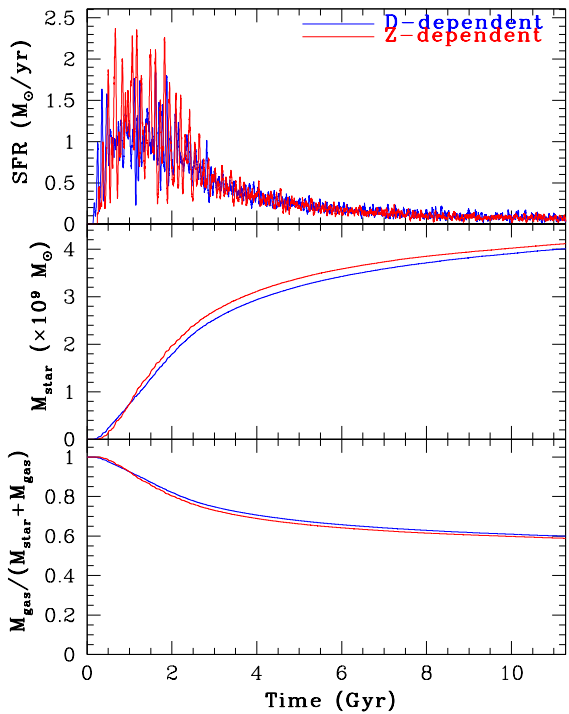,width=8.0cm}
\caption{
The same as Fig. 5 but for the models with
with `D-dependent' (blue) and `Z-dependent' (red) SF recipes.
The blue lines correspond to the fiducial model.
In the two models, the SFR at each local region is estimated from local ${\rm H_2}$
densities. However, the way to estimate ${\rm H_2}$ mass fraction at each local region
is different between the two.  The Z-dependent SF recipe is
described in the main text.
}
\label{Figure. 16}
\end{figure}

\begin{figure}
\psfig{file=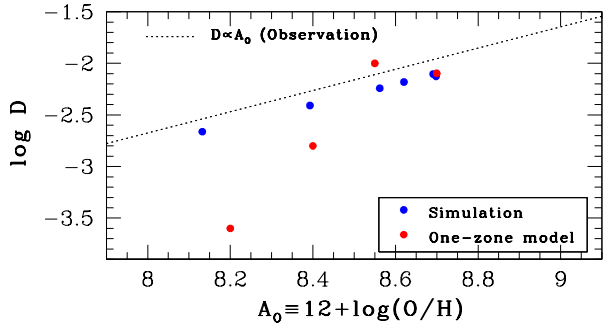,width=8.0cm}
\caption{
The locations of six models with different $M_{\rm h}$ yet same $\lambda$ (=0.038)
from the present simulation (blue circles)
on the $A_{\rm O}-D$ plane.
For comparison, the results of one-zone models (D98; Lisenfeld \& Ferrara 1998)
are shown by red circles.
The dotted line is the observed relation by Leroy et al. (2011).
}
\label{Figure. 17}
\end{figure}

\begin{figure}
\psfig{file=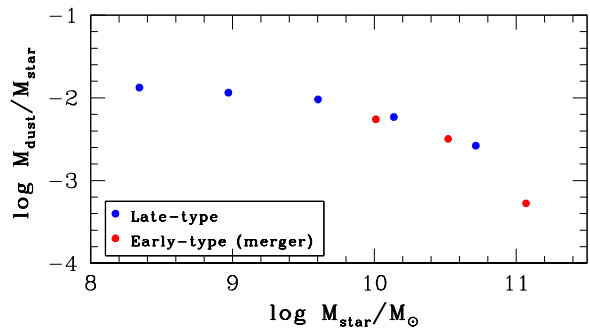,width=8.0cm}
\caption{
The location of five late-type disk galaxies  with different $M_{\rm h}$ (blue)
and three luminous ($M_{\rm h} \ge 10^{11} {\rm M}_{\odot}$)
early-type galaxies formed from major merging (red) on the $M_{\rm star}-R_{\rm dust}$
plane. The results of major mergers are taken from Bekki (2013).
}
\label{Figure. 18}
\end{figure}

\subsubsection{$R_{\rm dust}-M_{\rm star}$ relation}

Recently Cortese et al. (2012) have found an interesting dust scaling relation
that galaxies with larger $M_{\rm star}$ are more likely to have higher $R_{\rm dust}$.
They have also found that $R_{\rm dust}$ is systematically lower in early-type E/S0
galaxies in comparison with late-type disk galaxies. Although these correlations
of $R_{\rm dust}$ with $M_{\rm star}$ and galactic morphologies show larger dispersions,
they could have valuable physical meanings of galaxy formation and evolution.
We have thus investigated the $R_{\rm dust}-M_{\rm star}$ relation for a number of
representative models with different parameters. In order to discuss
the dust properties of early-type galaxies, 
we have taken the $R_{\rm dust}-M_{\rm star}$ relation
of early-type galaxies from our another work on E/S0 formation by gas-rich
mergers (Bekki 2013).
Fig. 18 shows that (i) $R_{\rm dust}$ is lower for larger $M_{\rm star}$ 
and (ii) this relation appears to be steeper for $\log (M_{\rm star}/{\rm M}_{\odot}$)
larger than 9.6.  These simulated trends are qualitatively similar to the observed
ones (Cortese et al. 2012), which implies that the present model is quite good at
grasping some essential ingredients of the formation process of this
$R_{\rm dust}-M_{\rm star}$ relation. However, the present disk formation models
can not reproduce
the observed galaxies with very low $R_{\rm dust}$ ($\log R_{\rm dust} < -3$), in particular,  
those with luminous galaxies with  $\log (M_{\rm star}/{\rm M}_{\odot}) > 10$.

This failure of the disk formation models
may reflect the limitation of the models in which violent merger
events are not explicitly included.  However,  the major merger models, 
in which two late-type disk galaxies can be transformed into early-type E/S0 ones,
can show significantly lower $R_{\rm dust}$ (Bekki 2013). 
Fig. 18 shows $R_{\rm dust}$ of the remnants of  three luminous major merger models
with prograde-prograde orbital configurations for 
$M_{\rm h} \ge 10^{11} {\rm M}_{\odot}$.
As shown in Fig. 18, the remnant  with $M_{\rm h}=10^{12} {\rm M}_{\odot}$
shows $\log R_{\rm dust}<-3$, which  implies that early-type galaxies
should  have lower  $R_{\rm dust}$, if they are formed from major merging.
It should be here noted that recent simulations of gas-rich major galaxy   mergers
by Hayward et al. (2011)
suggested that the low $R_{\rm dust}$ of a galaxy
is  due simply to consumption of metal-enriched
gas.
A minor fraction of galaxies have extremely low $R_{\rm dust}$ ($\log R_{\rm dust}<-4$),
which can not be reproduced by any model in the present study.
These galaxies might have experienced some environmental effects (e.g., dust stripping by
hot intra-cluster medium) or evaporation of dust by some hot radiation sources.
The origin of these galaxies will be explored  in our forthcoming papers.

\begin{figure*}
\psfig{file=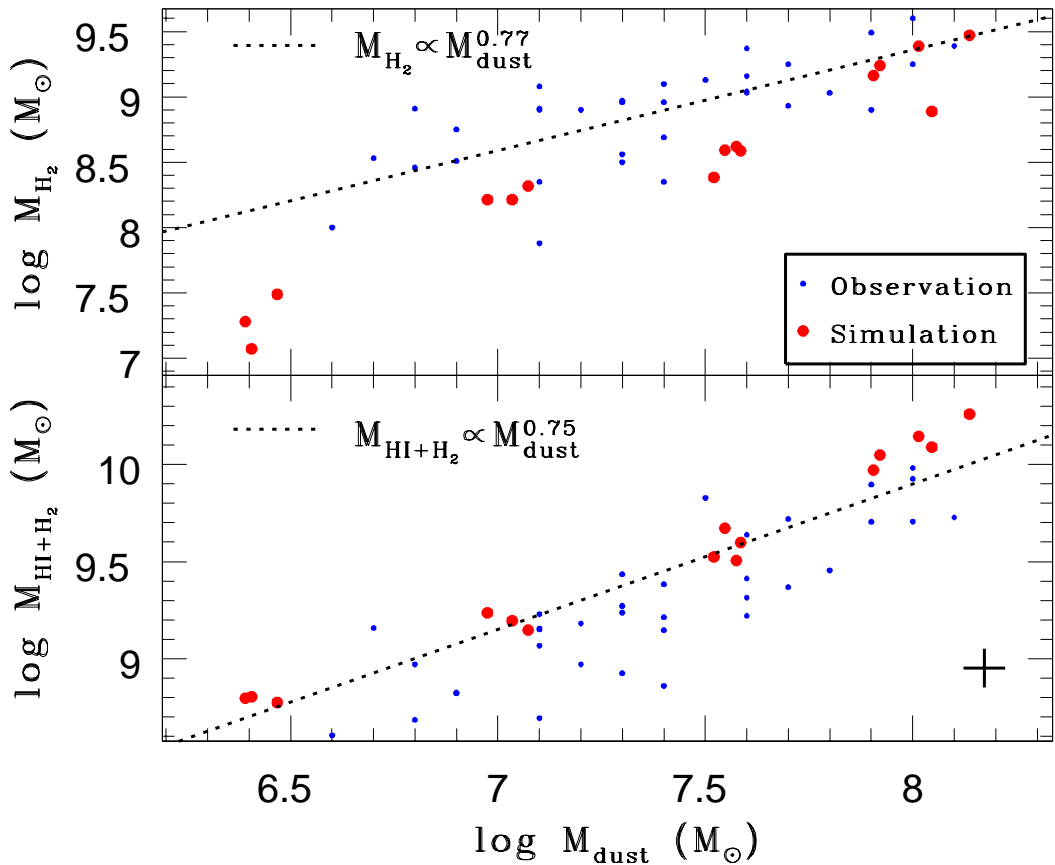,width=14.0cm}
\caption{
Big red dots show
the locations of 15 simulated disk galaxies with different $M_{\rm h}$ and $\lambda$
(=0.02, 0.038, and 0.06) on the $M_{\rm dust}-M_{\rm H_2}$ (upper)
and $M_{\rm dust}-M_{\rm HI+H_2}$ planes (lower).
For comparison,
observational results of 35 galaxies from Corbelli et al. (2012)
are shown by small black dots.
The dotted lines are the observationally derived
relations from Corbelli et al. (2012) and observational error bars ($\sim 0.1$ dex)
are shown by a large cross.
}
\label{Figure. 19} 
\end{figure*}

\subsubsection{$M_{\rm H_2}-M_{\rm dust}$ relation}

Corbelli et al. (2012) have recently investigated the total masses of dust,
${\rm H_2}$, and ${\rm H_I}$ in late-type Virgo cluster galaxies and found
$M_{\rm dust}-M_{\rm H_2}$ and $M_{\rm dust}-M_{\rm gas}$ relations
(where $M_{\rm gas}=M_{\rm HI}+M_{\rm H_2}$)  for
the galaxies. The best-fit relation between $M_{\rm H_2}$ and $M_{\rm dust}$
and between $M_{\rm gas}$ and $M_{\rm dust}$
are described as
\begin{equation}
\log M_{\rm H_2}=0.77 (\pm 0.12) \times \log M_{\rm dust} + 3.2 (\pm 0.9)
\end{equation}
and
\begin{equation}
\log M_{\rm gas}=0.75 (\pm 0.09) \times \log M_{\rm dust} + 3.9 (\pm 0.6),
\end{equation}
respectively. 
The metallicity-dependent CO-to-${\rm H_2}$ conversion factor is used for
the total ${\rm H_2}$ mass estimation of 35 galaxies in deriving the above
relations. 
Fig. 19 shows these two observational relations as well as the results from
the present models with different $M_{\rm h}$ and $\lambda$ for comparison.

Clearly, the simulated $M_{\rm dust}-M_{\rm gas}$ relation is quite similar to
the observed relation, although the locations of the simulated massive disk galaxies 
($M_{\rm dust} \ge  8 \times 10^8 {\rm M}_{\odot}$ or 
$M_{\rm h} \ge 3 \times 10^{11} {\rm M}_{\odot}$) are slightly above the observed relation.
The locations of the simulated less massive galaxies with 
$M_{\rm dust} <  4 \times 10^8 {\rm M}_{\odot}$ on the $M_{\rm dust}-M_{\rm H_2}$ plane
are appreciably  ($\sim 0.5$ dex) 
below the observed relation. However, given the observed large error (0.9 dex)
for the $M_{\rm dust}-M_{\rm H_2}$ relation,  the discrepancy between the simulation
and the observation can not be serious. Instead, the present model appears to do a good job
in reproducing the $M_{\rm dust}-M_{\rm H_2}$ relation  as well as 
$M_{\rm dust}-M_{\rm gas}$.
The present results also suggest that the $M_{\rm dust}-M_{\rm H_2}$ relation
can become steeper for $M_{\rm dust} \le 10^7 {\rm M}_{\odot}$.
The slightly under-abundant $M_{\rm H_2}$ for a given $M_{\rm dust}$ 
derived in the present study
would need to be re-investigated in our future simulations  with more realistic
initial conditions of galaxy formation.

\begin{figure*}
\psfig{file=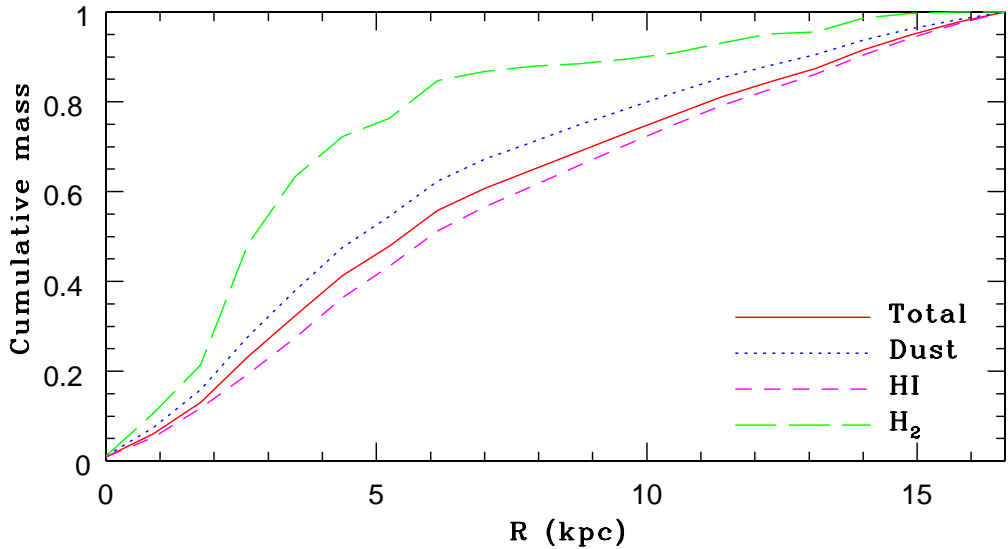,width=14.0cm}
\caption{
The cumulative mass distributions of total gas (solid, red), dust (dotted, blue),
H~{\sc i} (short-dashed, magenta), and ${\rm H_2}$ (long-dashed, green) at $T=11.3$ Gyr
in the fiducial model.
}
\label{Figure. 20}
\end{figure*}

\subsection{Radial gradients of $D$ and $f_{\rm H_2}$}

Recent observational studies have revealed radial gradients of $A_{\rm UV}$
(UV attenuation), dust surface densities, and $D$ in galaxies 
(e.g., Boissier et al. 2004; Pappalardo et al. 2012; P12). Recent theoretical works
have investigated the time evolution of radial gradients of $D$ and $D_{\rm z}$
for different models of dust growth and destruction (e.g., Mattsson et al. 2012).
The present study has clearly shown that luminous disk galaxies can have
negative $D$ gradients (i.e., higher $D$ in inner regions), which is qualitatively
consistent with the observational results by P12 for
non-HI-deficient galaxies in the Virgo cluster of galaxies.
The slightly steeper gradients of ${\rm H}_2$  in comparison
with $D$ in P12 appears to be also consistent with
our model predictions.

One of interesting recent results on radial $D$ gradients (P12) is that
the $D$ gradients appear to be different between galaxies with different
HI-deficiency parameters ($def_{\rm HI}$):
The parameter $def_{\rm HI}$ is defined
as the difference between the observed H~{\sc i}
mass (in logarithmic units) and 
that expected for a galaxy of the same linear diameter 
and morphological type in a comparison sample of isolated objects
(Roberts \& Haynes 1994).
Galaxies with $def_{\rm HI}<0.4$ (i.e., non-deficient) show negative $D$ gradients
whereas those with $def_{\rm HI}>0.7$ (i.e., strongly HI-deficient) 
show slightly positive ones (Fig. 13 in P12). Given that no model in the present study
shows such positive $D$ gradients,  this observation  needs to be discussed
in the context of environmental effects on galaxies.
If H~{\sc i} gas in the outer parts of galaxies
can be more efficiently stripped by some environmental effects
(e.g., ram pressure or tidal stripping) in comparison with dust, then
$D$ might well can increase in the outer parts after this stripping process.
On the other hand,
if the inner parts remain intact during the stripping process,
$D$ of the inner parts would not change.
Therefore,
this  more efficient stripping of H~{\sc i} gas in the outer parts of
galaxies could end up with positive radial $D$ 
gradients.

It appears to be  theoretically unclear
whether and why  this more efficient stripping of
H~{\sc i} gas could be possible. However,  Fig. 10,  which demonstrates a larger
amount of dust in gas with higher ${\rm H_2}$ mass fractions,  suggests the following
scenario for the more efficient H~{\sc i} stripping.
A larger amount of dust in the outer parts of disks
can be locked up  in GMCs in comparison with H~{\sc i} gas (as shown in Fig. 10).
Therefore, if gas stripping process is more
efficient for H~{\sc i} gas, then
$D$ could increase after gas stripping. 
Fig. 20 shows the cumulative mass distributions for H~{\sc i}, ${\rm H_2}$,
dust, and total gas (i.e., H~{\sc i}+${\rm H_2}$) in the fiducial model.
The more compact ${\rm H_2}$ distribution in Fig. 20
strongly suggests that H~{\sc i} can be more efficiently stripped by tidal or
ram pressure stripping in the outer parts of disks.
Therefore, the above scenario seems to be reasonable and realistic.
Given that no previous studies have so far investigated the above scenario,
it is our intention in future studies to  confirm that
the origin of the observed positive $D$ gradients is really related to
more efficient H~{\sc i} stripping of disk galaxies  in clusters of galaxies.


\subsection{Origin of PAH dust properties}

The present study has shown that the PAH-to-dust mass ratio ($f_{\rm PAH}$)
is higher for luminous
(or massive) disk galaxies with higher metallicities ($A_{\rm O}$).
This correlation is qualitatively consistent with recent observational results
by Draine et al. (2007), which shows $f_{\rm PAH}$
(which is equivalent to their $q_{\rm PAH}$)
is 1\% for metal-poor galaxies with  $A_{\rm O}<8.1$ and 3.55\% for metal-rich
galaxies with $A_{\rm O} \ge 8.1$.
Interestingly, their results show a lack of galaxies with higher $f_{\rm PAH}>0.02$
and $A_{\rm O}<8.1$.  This is also consistent with the present PAH formation models which show
a rapid rise of $f_{\rm PAH}$ only after $A_{\rm O}>8.1$ (i.e., very steep
slope of the simulated $A_{\rm O}-f_{\rm PAH}$ relations for $A_{\rm O}>8.1$; See Fig. 2).
It should be noted, however, that the present models can not show very low $f_{\rm PAH}$
($<0.005$) for the {\it present} galaxies with $A_{\rm O}<8.1$ (i.e., the simulated galaxies
show  low $f_{\rm PAH}$ {\it only} in  their early evolution).

The present study has shown that  
the radial $f_{\rm PAH}$ gradients of the present disk galaxies with total masses
similar to that of the LMC
can be only slightly negative ($\sim -2\times 10^{-4}$ kpc$^{-1}$). 
This almost flat radial gradients
appears to be inconsistent with the recent observation  by Meixner et al. (2010), 
which has revealed a possible enhancement of $f_{\rm PAH}$ in the central region of
the LMC.  The LMC could have experienced central starburst events owing to
the LMC-SMC-Galaxy tidal interaction in the last several Gyr
(e.g., Bekki \& Chiba 2005), which might enhance the $f_{\rm PAH}$ in the central
regions and consequently enhance the radial $f_{\rm PAH}$ gradient. 
The simulated disk galaxies of the present study do not experience  tidal interaction
with other galaxies and thus keep the almost flat $f_{\rm PAH}$ gradients.
Accordingly, the above apparent inconsistency could be due largely to 
differences in the evolution of PAH gradients between isolated and interacting galaxies
(rather than some problems of PAH modeling).
The present study thus implies that radial $f_{\rm PAH}$ gradients 
of disk galaxies could have some fossil information about the past interaction
histories with other galaxies.

\subsection{Comparison with other models}

The present chemodynamical model
incorporated, for the first time,  the evolution of  different chemical
components, dust formation and destruction, and
${\rm H_2}$ formation on dust grains  in a  self-consistent manner.
Previous simulations, however,   included ${\rm H_2}$ formation 
(but not dust) with different
models. By assuming that dust abundances are proportional to metals ($Z$) in galaxies,
P06 incorporated ${\rm H_2}$ formation on dust grains and
${\rm H_2}$ destruction by ISRF in SPH hydrodynamical
simulations of galaxies for the first time.
RK08 adopted (i)  a different model for ${\rm H_2}$ formation
and (ii) a more sophisticated
ISRF model based on the photoionization code CLOUDY in their SPH simulations of galaxies,
and discussed global star formation histories of galaxies.
Cosmological simulations by Gnedin et al. (2009) and Christensen et al. (2012)
adopted a model of ${\rm H_2}$ formation that is essentially similar to that by
P06.

Chemical evolution of variously different components (e.g., Mg and Fe) and 
dust evolution are not included in these previous models.
The dust-to-gas ratios ($D$), which are 
key parameters in ${\rm H_2}$ formation on dust grains, are simply assumed to
be proportional to gas-phase metallicities ($Z$) in previous models ($D_{\rm z}=D/Z
=$ constant).
As both observational and theoretical
studies showed  that $D$ is not simply proportional to $Z$ (i.e., there is large
dispersion in $D$ for a given $Z$; Galametz et al. 2011),
the adopted assumption  in previous models is neither reasonable nor realistic
in a strict sense. 
However, as shown in the present study,  the time evolution of SFRs and dust properties
does not depend strongly on whether the evolution of dust is fully self-consistently
included or not. 
Thus, the present study suggests that adopting a constant $D_{\rm z}$  
in numerical simulations  can be a good approximation
for investigating what really occurs in real galaxy evolution.

\subsection{Future directions}

Theis \& Orlova (2004) incorporated dust-gas interaction in hydrodynamical
simulations of gas-rich disk galaxies and thereby investigated how
pressure-free cold dust components, which are assumed to be coupled to
the gas by a drag force, can influence the gas dynamics of disk galaxies.
They found that if the dust mass fraction exceeds 2\% in disk galaxies,
then the disk can become destabilized by the dust-gas coupling.
Although their results suggest dust-gas coupling through a drag force
could influence the gas dynamics of disk galaxies,
such a possibly important effect is not included in the present
study. Accordingly, it is our intention  to incorporate the dust-gas
coupling through a drag force in more sophisticated chemodynamical
simulations. 

The destruction processes of dust grains by sputtering in the reverse
shock of a single SN remnant have been 
investigated by hydrodynamical models  in detail (e.g., Silvia et al. 2010).
In comparison with these  models on the detailed physical
processes of dust destruction by a single SN,
the present model for dust destruction could be fairly crude,
because only the dust destruction time scale is a key parameter
for the dust destruction processes. More sophisticated parameterization
for dust destruction processes will be necessary so that the time
evolution of dust contents can be more precisely predicted in our
future chemodynamical simulations.


\section{Conclusions}

We have investigated the time evolution of dust, gas, and star formation rates
in galaxies by using our new chemodynamical simulations with 
a self-consistent model for the formation and evolution of  dust and 
molecular hydrogen (${\rm H}_2$).
In this first  of a  series of papers, 
we have focused particularly on spatial distributions of dust, gas-to-dust ratios
($D$), molecular hydrogen fraction ($f_{\rm H_2}$),
gas-phase abundances ($A_{\rm O}\equiv$12+log(O/H)),  PAH-to-dust ratios ($f_{\rm PAH}$),
surface mass densities for dust ($\Sigma_{\rm dust}$), total gas ($\Sigma_{\rm gas}$),
and stars ($\Sigma_{\rm stars}$),
and dust-to-stellar mass ratios ($R_{\rm dust}=M_{\rm dust}/M_{\rm star}$).
We have also investigated
correlations and scaling-relations  between these properties and 
the dependences of dust and ${\rm H}_2$
properties on the halo and stellar masses ($M_{\rm h}$ and $M_{\rm star}$,
respectively) of galaxies.
The principle results are summarized as follows. \\

(1) Dust can play a vital role in regulating global star formation histories of galaxies,
mainly because the time evolution  of ${\rm H}_2$ formation efficiencies depends strongly
on dust evolution. Galactic star formation histories thus can depend on model
parameters for  dust formation and evolution,
in particular, the accretion time scale of dust ($\tau_{\rm acc}$).
The observed $D$ ($0.005 \sim 0.01$) in luminous disk galaxies can be reproduced
if $\tau_{\rm acc} \sim 0.25$ Gyr, which is similar to that used in previous one-zone
chemical evolution models with dust.  The adopted dust-dependent star formation model
is particularly important for low-mass disk galaxies with 
$M_{\rm h} \le 10^{10} {\rm M}_{\odot}$. 

(2) Different local regions of a disk galaxy can have significantly different $D$
and $A_{\rm O}$ owing to their different star formation and chemical evolution histories
within the galaxy. The local regions, however, show a $A_{\rm O}-D$ correlation
($D \propto A_{\rm O}$) and the $A_{\rm O}-D$ correlation evolves with time.
The spatial distributions of dust show negative radial gradients of $D$ and $A_{\rm O}$ 
(i.e., higher $D$ and $A_{\rm O}$ in inner regions) in
the simulated disk galaxies.

(3) $f_{\rm H_2}$ rapidly  increases owing to more efficient production
of dust in the first several Gyr of disk galaxy formation.
The distributions of $f_{\rm H_2}$ show negative radial gradients
(i.e., larger $f_{\rm H_2}$ in inner regions) and the gradients evolve
with time.
Local regions with higher $D$ are more likely to  show higher $f_{\rm H_2}$.
The spatial distributions of ${\rm H_2}$ gas in disk galaxies are more compact than  
those of dust, and the distributions of dust 
are more compact than those of H~{\sc i}  gas.

(4) $\Sigma_{\rm gas}$ can strongly correlate with $\Sigma_{\rm dust}$
($\Sigma_{\rm gas} \propto \Sigma_{\rm dust}$)  in 
local regions of a  disk  galaxy.
However, $\Sigma_{\rm dust}-\Sigma_{\rm H_2}$ correlations are less clearly seen
in comparison with $\Sigma_{\rm dust}-\Sigma_{\rm H_2}$.
$\Sigma_{\rm dust}$ is also well correlated with $\Sigma_{\rm PAH}$ and $\Sigma_{\rm Z}$.
$R_{\rm dust}$ 
very weakly correlates with $\Sigma_{\rm star}$
in a galaxy such that $R_{\rm dust}$ is higher for lower $\Sigma_{\rm star}$.
$R_{\rm dust}$ is higher in early-type E/S0s formed by merging in comparison
with isolated late-type disks.

(5) If about 5\% of dust formed from
gaseous ejecta of C-rich AGB stars can become PAH dust, then
the present models can reproduce the observed typical fraction $f_{\rm PAH}$ 
($ \sim 0.03$) in luminous disk galaxies.  The time evolution of $f_{\rm PAH}$
is more rapid than that of $D$, and $f_{\rm PAH}$ does not change significantly
in the last several Gyr. 
Weak radial gradients of $f_{\rm PAH}$ can be seen in the simulated disk galaxies
only for the first several Gyr evolution,
 which means that the present galaxies are highly likely
to have flat or only slightly negative  $f_{\rm PAH}$ gradients.

 (6) Time evolution of SFRs and  $D$ is 
investigated for the fiducial  model with 
dust-dependent and metallicity-dependent star formation recipes. It is found
that there are no major differences in the time evolution 
of SFRs and $D$ between the models.
These results imply that the evolution of SFRs and dust properties
are not  sensitive to whether a dust model is self-consistently included  
or not for luminous
galaxies with $M_{\rm h} \sim 10^{11} {\rm M}_{\odot}$.
It should be stressed, however, that the metallicity-dependent  
star (and ${\rm H_2}$) formation model
could possibly underestimate $f_{\rm H_2}$, because it does not include the increase
of dust-to-metal ratios ($D_{\rm z}$) 
(thus the higher probability of conversion from H~{\sc i} to ${\rm H_2}$)
during chemical evolution. \\

Preliminary results on the dependences of dust and ${\rm H}_2$ properties
on model parameters (e.g., $M_{\rm h}$ and $M_{\rm star}$) 
are summarized as follows. \\

(1) The present chemodynamical model can reproduce reasonably well the observed
$A_{\rm O}-D$ relation ($D \propto A_{\rm O}$),
though the simulated $D$ for a given $A_{\rm O}$ is slightly 
smaller than the observed one. Galaxies with larger $M_{\rm h}$ ($M_{\rm star}$) 
can have higher 
$A_{\rm O}$, $D$, and $f_{\rm PAH}$ owing to more efficient chemical enrichment
(i.e, more efficient dust production) in more massive galaxies.

(2) The final (i.e., present) values of $R_{\rm dust}$ 
depend strongly on $M_{\rm h}$ (and $M_{\rm star}$)
such that galaxies with larger $M_{\rm h}$ (and larger $M_{\rm star}$)
have lower $R_{\rm dust}$. 
Early-type galaxies
transformed from late-type disks via galaxy merging can have lower $R_{\rm dust}$
in comparison with late-type disks. Thus, the present study
suggests that  morphological transformation of galaxies
can cause significant  evolution of $R_{\rm dust}$.

(3) Disk galaxies with larger $M_{\rm h}$ are more likely to have
higher $f_{\rm H_2}$.  
The final $f_{\rm H_2}$  can be very low 
in low-mass disk galaxies with  $M_{\rm h}$ for $M_{\rm h} < 10^{10} {\rm M}_{\odot}$,
because star formation is more severely suppressed in these low-mass disks
(i.e., a smaller amount of dust is produced and used for ${\rm H_2}$ formation).

(4) Disk galaxies with larger $M_{\rm dust}$ have larger $M_{\rm gas}$ 
($=M_{\rm HI} + M_{\rm H_2}$) and the simulated $M_{\rm dust}-M_{\rm gas}$
relation is consistent reasonably well with the observed one
($M_{\rm gas} \propto M_{\rm dust}^{0.75}$). 
The $M_{\rm dust}-M_{\rm H_2}$  relation derived for the simulated disk galaxies
is slightly steeper than the observed relation ($M_{\rm H_2} \propto M_{\rm dust}^{0.77}$).

\section{Acknowledgment}
I (Kenji Bekki; KB) am   grateful to the referee  for  constructive and
useful comments that improved this paper.
Numerical simulations  reported here were carried out on
the three GPU clusters,  Pleiades, Fornax,
and gSTAR kindly made available by International Center for radio astronomy research
(ICRAR) at  The University of Western Australia,
iVEC,  and the Center for Astrophysics and Supercomputing
in the Swinburne University, respectively.
This research was supported by resources awarded under the Astronomy Australia Ltd's ASTAC scheme on Swinburne with support from the Australian government. gSTAR is funded by Swinburne and the Australian Government's
Education Investment Fund.
KB is grateful to Cameron Yozin-Smith for his reading this paper and giving  
useful comments to me.
KB acknowledges the financial support of the Australian Research Council
throughout the course of this work.

\appendix

\section{Numerical tests for subgrid physics}

\begin{figure*}
\psfig{file=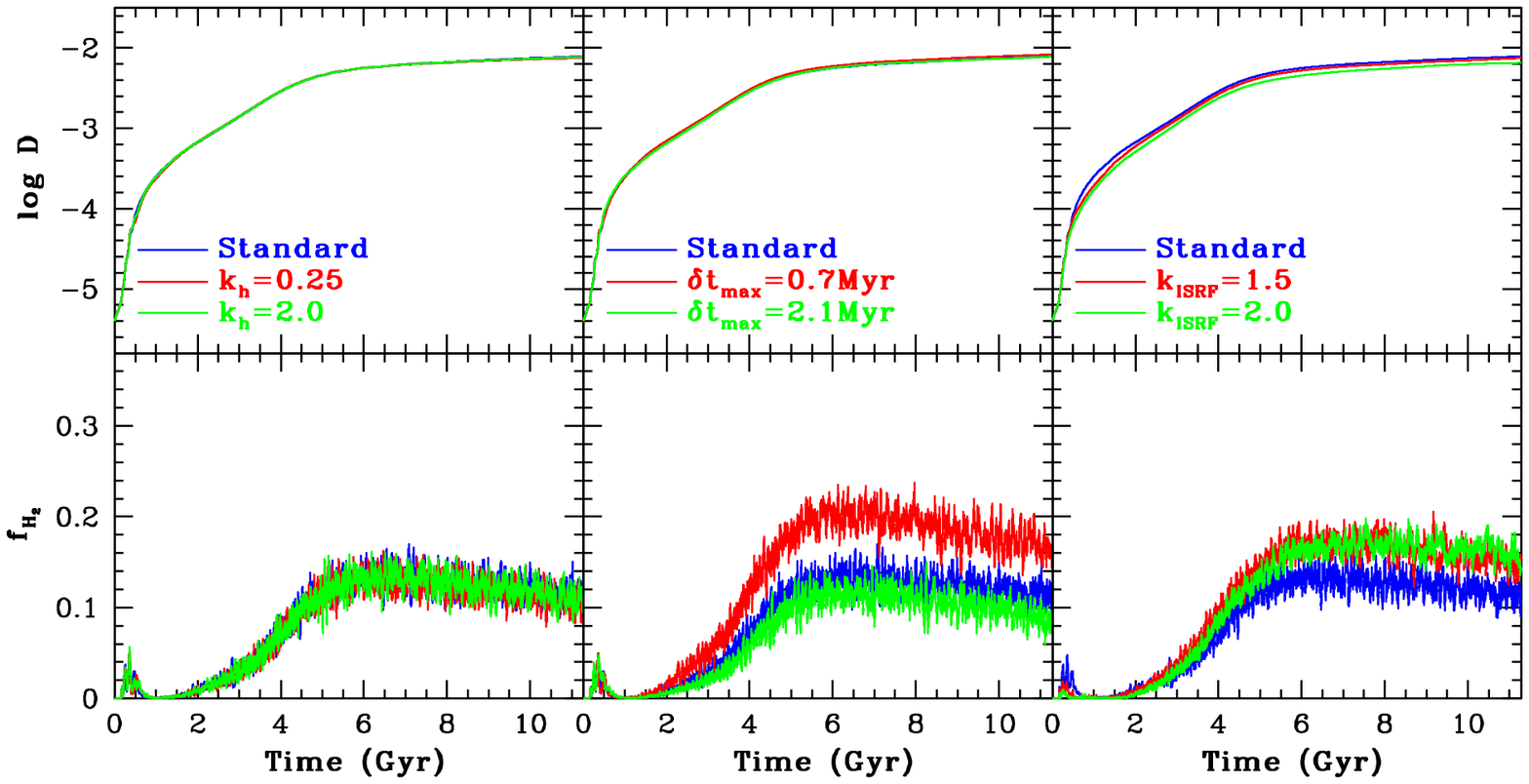,width=16.0cm}
\caption{
The same as Fig. 11 but for the seven comparative models (including the fiducial model):
comparison between different $k_{\rm h}$ (left),  $\delta t_{\rm max}$ (middle),
and $k_{\rm ISRF}$ (right).
}
\label{Figure. 21}
\end{figure*}

\begin{table}
\centering
\begin{minipage}{80mm}
\caption{A summary of parameters of subgrid physics  for
additional test models.}
\begin{tabular}{cccc}
{ model }
& { $k_{\rm h}$  \footnote{ The parameter $k_{\rm h}$ is used for determining
column densities of neutral and molecular hydrogen:
$N_{1,i}=\int_{-k_{\rm h} h_i}^{ k_{\rm h} h_i} n_{1,i} dr \approx 2k_{\rm h}n_{1,i}h_i$
for neutral hydrogen.}} 
& { $k_{\rm ISRFF}$  \footnote{ The parameter $k_{\rm ISFR}$ is used for determining
the local volume $V_{\rm ISRF}$ ($\propto k_{\rm ISRF}^3 \epsilon_{\rm g}^3$)
of the surrounding region of a gas particle.}} 
& { $\delta t_{\rm max}$ 
\footnote{ The maximum time step width in units of $10^6$ yr.}} \\
fiducial  & 1.0 & 1.0 & 1.4  \\
T1 & 0.5 & 1.0 & 1.4  \\
T2 & 2.0 & 1.0 & 1.4  \\
T3 & 1.0 & 1.0 & 0.7 \\
T4 & 1.0 & 1.0 & 2.1 \\
T5 & 1.0 & 1.5 & 1.4  \\
T6 & 1.0 & 2.0 & 1.4  \\
\end{tabular}
\end{minipage}
\end{table}

In the present study,  ${\rm H_2}$ formation on dust grains and ${\rm H_2}$ destruction
by ISRF are estimated for each SPH particle at a given time step by using
the smoothing length ($h_i$) 
and the gravitational softening length ($\epsilon_{\rm g}$).
For example, we estimate the column density of H~{\sc i}  gas 
(which is necessary to estimate ${\rm H_2}$ formation) as follows:
\begin{equation}
N_{1,i}=\int_{-k_{\rm h} h_i}^{ k_{\rm h} h_i} n_{1,i} dr \approx 2k_{\rm h}n_{1,i}h_i,
\end{equation}
where $k_{\rm h}$ is set to be 1 in all models described in the main text.
Accordingly we need to demonstrate that the present results do not depend on the resolution
of simulations (i.e., $k_{\rm h}$).

ISRF for a gas particle is estimated for the local volume $V_{\rm ISRF}$ that is defined
as;
\begin{equation}
V_{\rm ISRF}=k_{\rm ISRF}^3\epsilon_{\rm g}^3, 
\end{equation}
where $\epsilon_{\rm g}$ is the gravitational softening length for 
the gas particle and $k_{\rm ISRF}$ is set to be 1
for all models described in the main text. Such local estimation of  ISRF implies
that stellar radiation from stars that are located outside 
$\epsilon_{\rm g}$ (for $k_{\rm ISFR}=1$) is completely ignored.
Although, the radiation density would not change significantly even if
a larger volume is used,  we need to demonstrate that the present results
do not depend on $k_{\rm ISRF}$.
Hopkins et al. (2011) have discussed a similar resolution problem related to
the estimation of the total amount of radiation from massive stars in their simulations.
Also, the present results could possibly depend on
the maximum time step width ($\delta t_{\rm max}$), though timescales of key
physical processes (e.g., $\tau_{\rm dust}$) are much longer than the adopted
$\delta t_{\rm max}$ for models described  in the main text.

We thus investigate the models in which model parameters are the same as those
adopted in the fiducial  model except $k_{\rm h}$, $k_{\rm ISRF}$, and 
$\delta t_{\rm max}$. The parameter values for these test models
(T1$-$T6 as well as the fiducial  model) are summarized in Table A1.
Key results
on the time evolution of $D$ and $f_{\rm H_2}$,
which are shown in Fig. A1, 
are summarized as follows.
First, the time evolution and final values of $D$ and $f_{\rm H_2}$  
do not depend on $k_{\rm h}$ for $0.25 \le k_{\rm h} \le 2.0$.
This result means that the present results are insensitive to the method
to estimate  column densities of neutral and molecular hydrogen.
Second, the final $f_{\rm H_2}$ value in the model with $\delta t_{\rm max}=0.7$ Myr
(a finer time step width) is 34\% larger than that in the fiducial model,
though there are no significant differences in $D$ between the two.
The final $f_{\rm H_2}$ in
the model with  $\delta t_{\rm max}=2.1$ Myr is  25\% smaller 
than that in the fiducial  model.

Gas densities can become higher in the model with smaller $\delta t_{\rm max}$
so that the probability of conversion from neutral to molecular hydrogen
can become higher. As a result of this,
$f_{\rm H_2}$ can become larger. The derived dependences of $f_{\rm H_2}$
on $\delta t_{\rm max}$ imply that $f_{\rm H_2}$ can be better predicted
by using models with smaller  $\delta t_{\rm max}$.  However, a larger amount
of calculation time is required for completing
numerical simulations  with smaller $\delta t_{\rm max}$.
Thus, if one has a limited amount of allocated time for 
performing  numerical simulations (on GPU clusters or supercomputers),
one would need to compromise on the accuracy of the predicted $f_{\rm H_2}$.

Fig. A1 also shows that the final $f_{\rm H_2}$ and $D$ do not depend so strongly
on $k_{\rm ISRF}$: $f_{\rm H_2}$ is
slightly higher in the models with larger $k_{\rm ISRF}$,
probably because ISRF (i.e., radiation density)  can be somewhat 
weaker for a larger volume (i.e., more efficient
$f_{\rm H_2}$ formation due to weaker ISRF).
These results imply that the present results are not so sensitive to the adopted method
to estimate ISRF  (i.e., non-inclusion of the contribution of distant stellar particles to 
ISRF).

\section{SF histories of low-mass disks}

Fig. B1 shows the time evolution of SFRs in the low-mass model with
$M_{\rm h}=10^{10} {\rm M}_{\odot}$ for the ${\rm H}-$ and ${\rm H_2}-$dependent
SF recipes. Although the fiducial model shows only slight differences in SFRs between the
two different SF recipes, this model shows more remarkable differences in the early
SF histories. The model with the ${\rm H}-$dependent SF shows systematically higher
SFR for $T=1 \sim 2$ Gyr in comparison  with that with the ${\rm H_2}-$dependent SF.
This suppression of SF in early phase of the disk formation in the ${\rm H_2}$-dependent
SF model
is due largely to the lower ${\rm H_2}$ fraction in the
low-mass disk model.
Such suppression  can be seen in other low-mass models  
with the ${\rm H_2}-$dependent SF recipe
($M_{\rm h} \le 10^{10} {\rm M}_{\odot}$).  These results imply that 
the adoption of ${\rm H_2}-$dependent SF recipe is important for SF histories
of low-mass disk galaxies. Severe suppression of SF has been already
reported by K12 in which  ${\rm H}_2$-regulated SF models are adopted.

\begin{figure}
\psfig{file=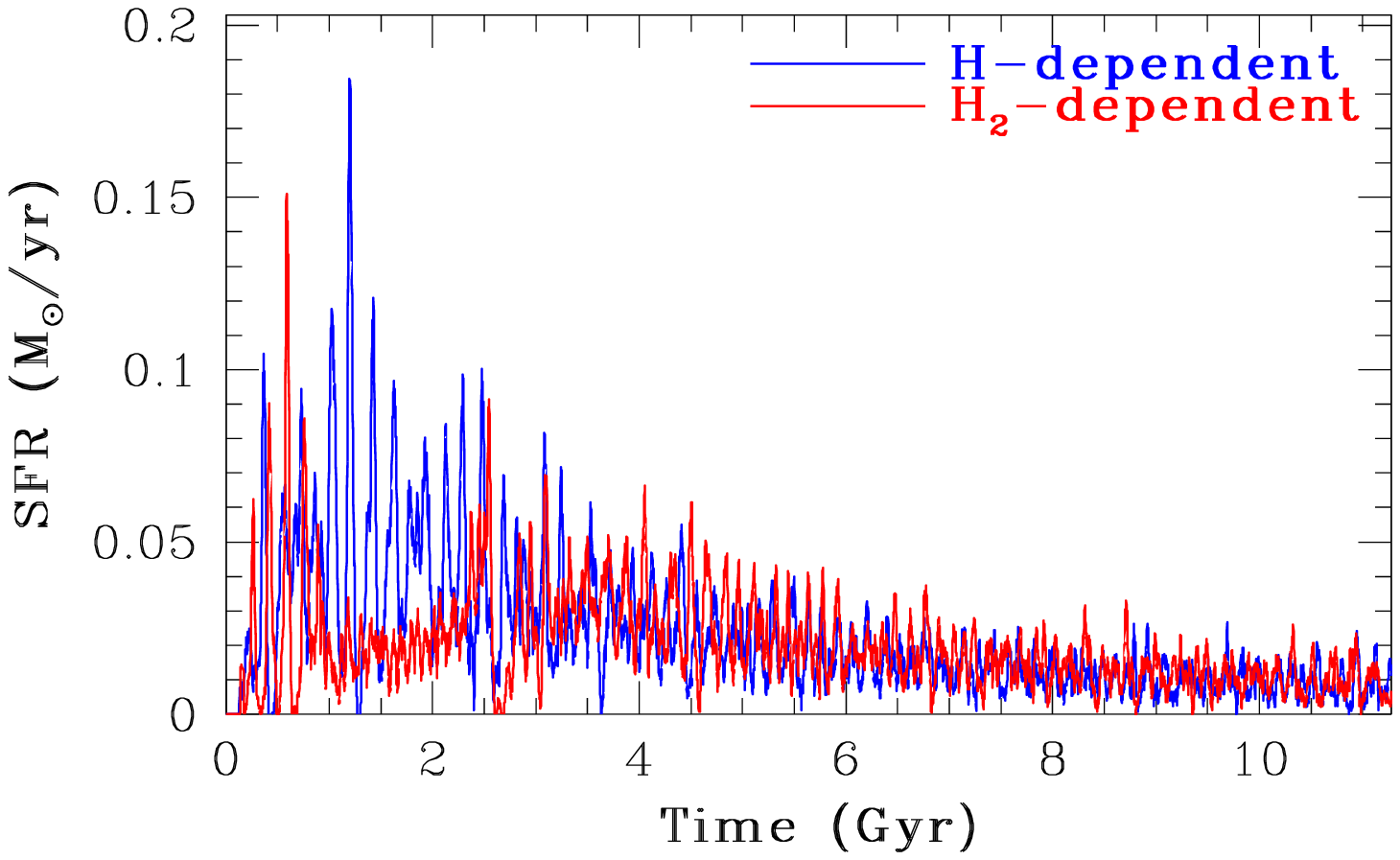,width=8.0cm}
\caption{
The time evolution of SFRs in the two models with H-dependent (blue) and ${\rm H_2}$
dependent (red) SF recipes. These results are 
for the low-mass disk models with $M_{\rm h}=10^{10} {\rm M}_{\odot}$
and $\lambda=0.038$.
}
\label{Figure. 22}
\end{figure}

\section{Lower $f_{\rm H_2}$ in galaxies with lower total masses}

\begin{figure}
\psfig{file=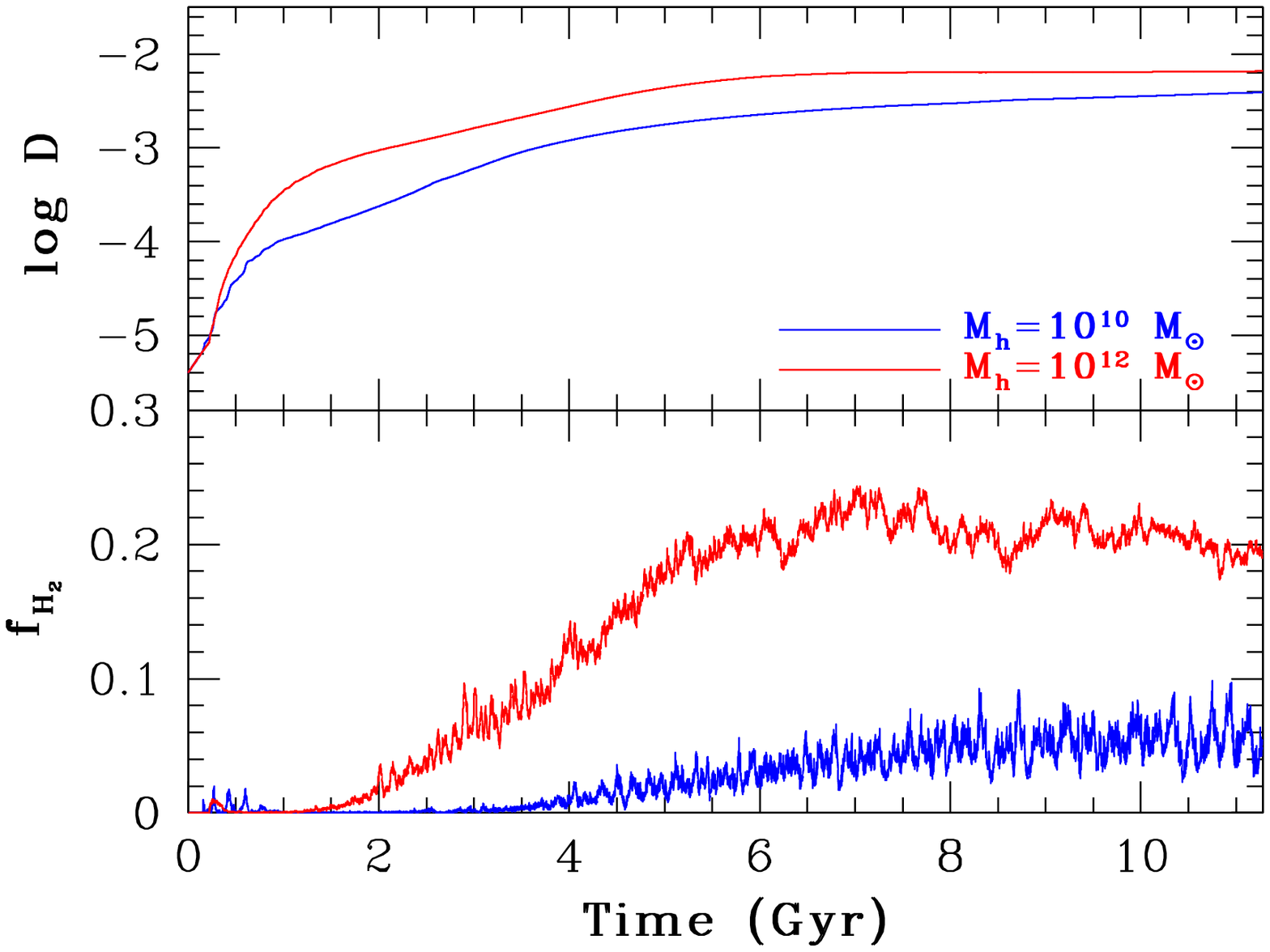,width=8.0cm}
\caption{
The same as Fig. 11 but for the low-mass halo model
with $M_{\rm h}=10^{10} {\rm M}_{\odot}$  (blue) 
and the high-mass model  with $M_{\rm h}=10^{12} {\rm M}_{\odot}$  (red).
}
\label{Figure. 23}
\end{figure}

In the present model, ${\rm H_2}$ formation efficiencies depend strongly on dust
abundances ($D$). Therefore, it is expected that more massive galaxies, where 
chemical enrichment proceeds more efficiently, can have larger $D$ and thus larger
$f_{\rm H_2}$. Fig. C1 shows the time evolution of $D$ and $f_{\rm H_2}$ in
the low-mass disk model with $M_{\rm h}=10^{10} {\rm M}_{\odot}$ and
the high-mass one with $M_{\rm h}=10^{12} {\rm M}_{\odot}$. 
Clearly, both final $D$ and $f_{\rm H_2}$ are higher in the high-mass model,
and  the evolution of $f_{\rm H_2}$ is more rapid in the high-mass model.
In the present ${\rm H_2}$ formation model,
the final $f_{\rm H_2}$ in disks  with $M_{\rm h} \le 10^{10} {\rm M}_{\odot}$
can be very low ($<0.05$), which has important implications of gas contents in
dwarf galaxies. We will discuss such implications in detail in our forthcoming papers.


\begin{thebibliography}{}

\bibitem[)]{}
Bekki, K., 2009, MNRAS, 399, 2221

\bibitem[)]{}
Bekki, K., 2011, MNRAS, 416, 2359

\bibitem[)]{}
Bekki, K., 2013, in preparation

\bibitem[)]{}
Bekki, K., Shioya, Y., 1999, ApJ, 513, 108

\bibitem[)]{}
Bekki, K., \& Chiba, M.
2005, MNRAS, 356, 680

\bibitem[)]{}
Bekki, K., Couch, W. J., 2011, MNRAS, 415, 1783

\bibitem[)]{}
Bekki, K.,   Tsujimoto, T., 2012,  ApJ, in press

\bibitem[)]{}
Bekki, K., Shigeyama, T., Tsujimoto, T., 2012,  MNRAS, in press (B12)

\bibitem[)]{}
Bigiel, F.,  Leroy, A., Walter, F., Brinks, E.,
de Blok, W. J. G.,  Madore, B.,  Thornley, M. D., 2008, ApJ, 136, 2846

\bibitem[)]{}
Blitz, L., 1993,
in Protostars and planets III, p125

\bibitem[)]{}
Blitz, L.,  Rosolowsky, E., 2004, ApJ, 612, L29

\bibitem[)]{}
Boissier, S., Boselli, A.,  Buat, V., Donas, J., Milliard, B.,
2004, A\&A, 424, 465

\bibitem[)]{}
Bruzual, G.,  Charlot, S., 2003, MNRAS, 344, 1000

\bibitem[)]{}
Bullock, J. S., Dekel, A., Kolatt, T. S., Kravtsov, A. V., Klypin, A. A.,
Porciani, C., Primack, J. R., 2001, ApJ, 555, 240

\bibitem[)]{}
Calura, F., Pipino, A., Matteucci, F., 2008, A\&A, 484, 107

\bibitem[)]{}
Cazaux, S.,  Tielens, A. G. G. M., 2002, ApJ, 575, 29

\bibitem[)]{}
Christensen, C., et al., 2012, MNRAS, 425,  3058

\bibitem[)]{}
Corbelli, E., 2012, A\&A, 542, 32

\bibitem[)]{}
Cortese, L., Catinella, B.,  Boissier, S.,  Boselli, A.,  Heinis, S.,
2011, MNRAS, 415, 1797

\bibitem[)]{}
Cortese, L., et al. 2012, A\&A, 540, 52

\bibitem[)]{}
Draine, B. T., 2009, Physics of the interstellar and intergalactic medium

\bibitem[)]{}
Draine, B. T.,  Bertoldi, F., 1996, ApJ, 468, 269

\bibitem[)]{}
Draine, B. T., et al., 2007, ApJ, 663, 866

\bibitem[)]{}
Dunne, L., et al. 2011, MNRAS, 417, 1510

\bibitem[]{}
Dwek, E., 1998, ApJ, 501, 643 (D98)

\bibitem[)]{}
Edmunds, M., 2001, MNRAS, 328, 223

\bibitem[)]{}
Elmegreen, B. G., 1993, ApJ, 411, 170

\bibitem[]{}
Ferrarotti, A. S.,  Gail, H.-P., 2006, A\&A, 447, 553

\bibitem[]{}
Fu, J.,  Guo, Q.,  Kauffmann, G.,  Krumholz, M. R., 2010, MNRAS, 409, 515

\bibitem[]{}
Galametz, M., Madden, S. C., Galliano, F., Hony, S.,  Bendo, G. J., Sauvage, M.,
2011, A\&A, 532, 56


\bibitem[]{}
Gnedin, N. Y., Tassis, K.,  Kravtsov, A. V., 2009, ApJ, 697, 55

\bibitem[)]{}
Gould, R. J.,  Salpeter, E. E., 1963, ApJ, 138, 393

\bibitem[)]{}
Goldshmidt, O.,  Sternberg, A., 1995, ApJ, 439, 256

\bibitem[)]{}
Hayward, C. C., Keres, Ds.,  Jonsson, P., Narayanan, D., Cox, T. J.,  Hernquist, L.,
2011, ApJ, 743, 159

\bibitem[)]{}
Hernquist, L., \&  Katz, N., 1989, 70, 419

\bibitem[]{}
Hirashita, H., 1999, ApJ, 522, 220

\bibitem[]{}
Hirashita, H., 2012, MNRAS, 422, 1263

\bibitem[]{}
Hirashita, H.; Ferrara, A., 2002, MNRAS, 337, 921


\bibitem[)]{}
Hollenbach, D.,  Salpeter, E. E., 1971, ApJ, 163, 155

\bibitem[)]{}
Hopkins, P. F., Quataert, E., Murray, N., 2011, MNRAS, 417, 950

\bibitem[]{}
Inoue, A. K., 2003, PASJ, 55, 901

\bibitem[]{}
Jura, M., 1975, ApJ, 197, 575

\bibitem[]{}
Kaneda, H. et al., 2011, PASJ, 63, 601

\bibitem[)]{}
Katz, N., 1992, ApJ, 391, 502

\bibitem[]{}
Kawamura, A., 2009, ApJS, 184, 1

\bibitem[)]{}
Kawata, D.,  2001, ApJ, 558, 598

\bibitem[]{}
Kaufmann, T., ; Mayer, L.,  Wadsley, J.,  Stadel, J.,  Moore, B.,
2007, MNRAS, 375, 53

\bibitem[]{}
Kennicutt, R. C., Jr., 1998, ApJ, 498, 541

\bibitem[]{}
Kozasa, T.; Hasegawa, H.; Nomoto, K.

\bibitem[]{}
Krumholz, M. R.,  McKee, C. F., Tumlinson, J.,
2009, ApJ, 693, 216 (KMT09)

\bibitem[]{}
Kuhlen, M., Krumholz, M. R., Madau, P.,  Smith, B. D.,  Wise, J.,
2012, ApJ, 749, 36 (K12)

\bibitem[]{}
Lagos, C. P.,  Lacey, C. G.,  Baugh, C. M., 2012, in prepring (arXiv1210.4974)

\bibitem[]{}
Leroy, A. K., Walter, F.,  Brinks, E., Bigiel, F.,  de Blok, W. J. G.,
Madore, B.,  Thornley, M. D., 2008, AJ, 136, 2782

\bibitem[]{}
Leroy, A. K.,  et. al., 2011, ApJ, 737, 12

\bibitem[]{}
Lisenfeld, U.,  Ferrara, A.,  1998, ApJ, 498, 145

\bibitem[]{}
Mckee, C. F., 1989, 
in IAU Symp. 135, Interstellar Dust, Edited by Louis J. Allamandola and A. G. G. M. Tielens,
p431

\bibitem[Mannucci et al.[]{}
Mannucci, F., Della Valle, M.,  Panagia, N.,  2006, MNRAS, 370, 773

\bibitem[Maoz et al.(2010)]{Maoz_10}
Maoz, D., Sharon, K.,  Gal-Yam, A.,  2010, ApJ, 722, 1879

\bibitem[]{}
Maoz, D., Mannucci, F.,  Li, W.,  Filippenko, A. V.,
Della Valle, M.,    Panagia, N.,
2011, MNRAS, 412, 1508

\bibitem[]{}
Mattsson, L., Andersen, A. C., Munkhammar, J. D., 2012, MNRAS, 423, 26

\bibitem[]{}
Meixner, M., et al. 2010, A\&A, 518, L71

\bibitem[]{}
Navarro, J. F., Frenk, C. S.,  White, S. D. M., 1996, ApJ, 462, 563

\bibitem[]{}
Neto A. F. et al., 2007, MNRAS, 381, 1450

\bibitem[]{}
Nozawa, T., Kozasa, T.,  Umeda, H.,  Maeda, K.,  Nomoto,  K., 2003, ApJ, 598, 785

\bibitem[]{}
Pappalardo, C., et al., 2012, A\&A, 545, 75 (P12)

\bibitem[]{}
Pelupessy, F. I.,  Papadopoulos, P.  P.,  van der Werf, P.,
2006, ApJ, 645, 1024 (P06)

\bibitem[]{}
Piovan, L., Chiosi, C., Merlin, E., Grassi, T., Tantalo, R.,
Buonomo, U., Cassara, L. P., 2011, in preprint (arXiv1107.4541)

\bibitem[]{}
Rahimi, A., Kawata, D., 2012, MNRAS, 422, 2609 (RK12)

\bibitem[]{}
Renzini, A. Buzzoni, A., 1986,
in  Spectral evolution of galaxies,
(Dordrecht, D. Reidel  Publishing Co.),  p.195


\bibitem[]{}
Revaz, Y., Jablonka, P., 2012, A\&A, 538, 822 (RJ12)

\bibitem[]{}
Roberts, M. S.,  Haynes, M. P. 1994, ARA\&A, 32, 115

\bibitem[]{}
Robertson, B. E.,  Kravtsov, A.  V., 2008, ApJ, 680, 1083 (R08)

\bibitem[]{}
Roman-Duval, J., et al. 2010, A\&A, 518, L74

\bibitem[]{}
Rosen, A.,  Bregman, J. N., 1995, ApJ, 440, 634

\bibitem[]{}
Rosolowsky, E., Engargiola, G., Plambeck, R., \&  Blitz, L., 2003, ApJ, 599, 258

\bibitem[]{}
Sandstrom, K. M. et al., 2012, ApJ, 744, 20

\bibitem[]{}
Schmidt, M., 1959, ApJ, 129, 243

\bibitem[]{}
Silvia, D. W., Smith, B. D.,  Shull, J. M., 2010, ApJ, 710, 1575

\bibitem[]{}
Skibba, R., et al.,  2012, ApJ, 761, 42

\bibitem[]{}
Smith, D. J. B., 2012, MNRAS, 427, 703

\bibitem[]{}
Spitzer, L. Jr., 1978,
Physical Processes in the Interstellar Medium

\bibitem[]{}
Sugimoto, D., Chikada, Y., Makino, J., Ito, T., Ebisuzaki, T.,
Umemura, M., 1990, Nat, 345, 33

\bibitem[]{}
Sutherland, R. S., Dopita, M. A., 1993, ApJS, 88, 253

\bibitem[]{}
Takagi, T., et al. 2010, A\&A, 514, 5

\bibitem[]{}
Theis, C., Burkert, A., Hensler, G., 1992, A\&A, 265, 465

\bibitem[]{}
Theis, C., Orlova, N., 2004, A\&A, 418, 959

\bibitem[]{}
Thornton, K., Gaudlitz, M., Janka, H.-Th.,  Steinmetz, M.,
1998, ApJ, 500, 95

\bibitem[]{}
Tielens, A. G. G. M., 2008, ARA\&A, 46, 289

\bibitem[]{}
Totani, T., Morokuma, T., Oda, T., Doi, M.,  Yasuda, N.,  2008, PASJ, 60, 1327

\bibitem[]{}
Tsujimoto, T., Nomoto, K., Yoshii, Y., Hashimoto, M., Yanagida, S.,
Thielemann, F.-K.,  1995, MNRAS, 277, 945 (T95)

\bibitem[]{}
Tsujimoto, T.,  Bland-Hawthorn, J.,  \& Freeman, K. C. 2010,  PASJ, 62, 447

\bibitem[]{}
van den Hoek, L. B.; Groenewegen, M. A. T., 1997, A\&AS, 123, 305 (VG97)

\bibitem[]{}
van der Bergh, S., Th Galaxies in the Local Group
\bibitem[]{}
Yamasawa, D.,  Habe, A., Kozasa, T.,  Nozawa, T.,  Hirashita, H.,  Umeda, H.,  Nomoto, K.,
2011, ApJ, 735, 44

\bibitem[]{}
Zhukovska, S.,  Gail, H.-P.,  Trieloff, M., 2008, A\&A, 479, 453

\bibitem[]{}
Zubko, V.,  Dwek, E.,  Arendt, R. G., 2004, ApJS, 152, 211

\end{thebibliography}
\end{document}